\documentclass[preprint,prd,amsmath,amssymb,superscriptaddress,nofootinbib]{revtex4-1}
\usepackage[dvips]{graphicx}
\usepackage{amssymb}
\usepackage{rotating}
\usepackage{mathtools}
\usepackage{color}
\usepackage{multirow}
\usepackage[unicode]{hyperref}  

\newcommand\op[1]{\hat{#1}}						  
\newcommand\vecl[1]{{\bf{#1}}}         		  
\newcommand\vecs[1]{\boldsymbol{#1}}     		  
\newcommand\sqb[1]{\left[#1\right]}     		  
\newcommand{\ket}[1]{\left| #1 \right>} 		  
\newcommand{\bra}[1]{\left< #1 \right|}  		  
\newcommand\ovbb[0]{0\nu\beta\beta}			  
\newcommand\nme[0]{\mathcal{M}}				  

\begin{document}

\title{Analysis of Light Neutrino Exchange \\and Short-Range Mechanisms in $0\nu\beta\beta$ Decay} 

\author{Frank F. Deppisch}
\email{f.deppisch@ucl.ac.uk}
\affiliation{Department of Physics and Astronomy, University College London, Gower Street, London WC1E 6BT, UK}

\author{Lukas Graf}
\email{lukas.graf@mpi-hd.mpg.de}
\affiliation{Max-Planck-Institut f\"{u}r Kernphysik, Saupfercheckweg 1,
69117 Heidelberg, Germany}

\author{Francesco Iachello}
\email{francesco.iachello@yale.edu}
\affiliation{Center for Theoretical Physics, Sloane Physics Laboratory, Yale University, New Haven, Connecticut 06520-8120, USA}

\author{Jenni Kotila}
\email{jenni.kotila@jyu.fi}
\affiliation{Finnish Institute for Educational Research, University of Jyv{\"a}skyl{\"a}, P.O. Box 35, FI-40014 Jyv\"askyl\"a, Finland}
\affiliation{Center for Theoretical Physics, Sloane Physics Laboratory, Yale University, New Haven, Connecticut 06520-8120, USA}

\begin{abstract}
Neutrinoless double beta decay ($0\nu\beta\beta$) is a crucial test for lepton number violation. Observation of this process would have fundamental implications for neutrino physics, theories beyond the Standard Model and cosmology. Focussing on so called short-range operators of $0\nu\beta\beta$ and their potential interplay with the standard light Majorana neutrino exchange, we present the first complete calculation of the relevant nuclear matrix elements, performed within the interacting boson model (IBM-2). Furthermore, we calculate the relevant phase space factors using exact Dirac electron wavefunctions, taking into account the finite nuclear size and screening by the electron cloud. The obtained numerical results are presented together with up-to-date limits on the standard mass mechanism and effective $0\nu\beta\beta$ short-range operators in the interacting boson model framework. Finally, we interpret the limits in the particle physics scenarios incorporating heavy sterile neutrinos, Left-Right symmetry and $R$-parity violating supersymmetry.
\end{abstract}

\maketitle


\section{Introduction}
\label{sec:intro}

The nature of neutrinos and especially the origin of their masses are a crucial open question. While the Standard Model (SM) successfully explains the masses of the charged fermions it must be extended to incorporate neutrino masses. It would either require the presence of sterile neutrino states or effective lepton number violating (LNV) interactions. The first scenario allows the generation of \emph{Dirac} neutrino masses analogous to those of the charged fermions. While certainly feasible, given the stringent upper limits $m_\nu \lesssim \mathcal{O}(0.1)$~eV on the absolute neutrino masses from Tritium decay \cite{Otten:2008zz, Aker:2019uuj} and cosmological observations \cite{Ade:2015xua}, tiny Higgs Yukawa couplings are required. Also, total lepton number $L$ will no longer be an accidental symmetry. Unless $L$ symmetry is imposed, the sterile neutrinos would acquire an LNV \emph{Majorana} mass. The most popular example for such a scenario is the Seesaw mechanism where the sterile neutrinos have such a large Majorana mass $M\approx 10^{14}$~GeV naturally leading to light neutrino masses $m_\nu\approx 0.1$~eV \cite{Minkowski:1977sc, mohapatra:1979ia, Yanagida:1979as, seesaw:1979, Schechter:1980gr}.

High-scale seesaw mechanisms, or more generally scenarios where $L$ is broken at very high scales, are not the only way to generate light Majorana neutrino masses; other possibilities include LNV at low scales in secluded sectors, at higher loop order and when allowing higher-dimensional effective interactions. If $L$-breaking occurs close to the electroweak (EW) scale, higher-dimensional LNV operators can be important. From a phenomenological point of view, searching for processes that violate total $L$ thus play a crucial role in neutrino and Beyond-the-SM (BSM) physics. We here focus on the search for $\ovbb$ decay as the most sensitive approach to probe Majorana neutrino masses. Currently, the most stringent limit on the $0\nu\beta\beta$ decay half life $T_{1/2}$ is set in the Germanium isotope ${}^{76}_{32}$Ge \cite{Agostini:2020xta},
\begin{align}
\label{eq:cur-best-limit}
	T_{1/2}(^{76}\text{Ge}) \equiv T_{1/2}\left({}^{76}_{32}\text{Ge} \to {}^{76}_{34}\text{Se} + e^- e^-\right) > 1.8\times 10^{26}~\text{yr}.
\end{align}
However, Majorana neutrino masses are not the only contribution from BSM physics to $0\nu\beta\beta$ decay. We can generally consider the $0\nu\beta\beta$ decay rate by expressing high scale new physics contributions in terms of effective low-energy operators \cite{Pas:1999fc, Pas:2000vn, delAguila:2011gr, delAguila:2012nu}. This only assumes that there are no exotic particles beyond the SM below the $0\nu\beta\beta$ energy scale of $m_F \approx 100$~MeV. In this paper, we concentrate on so called short-range operators and their interplay with the standard light Majorana neutrino mass mechanism. As context, we provide a brief overview of the possible mechanisms for $0\nu\beta\beta$ decay which can be categorized in two main classes:

\begin{figure}[t!]
	\centering
	\includegraphics[clip,width=0.31\textwidth]{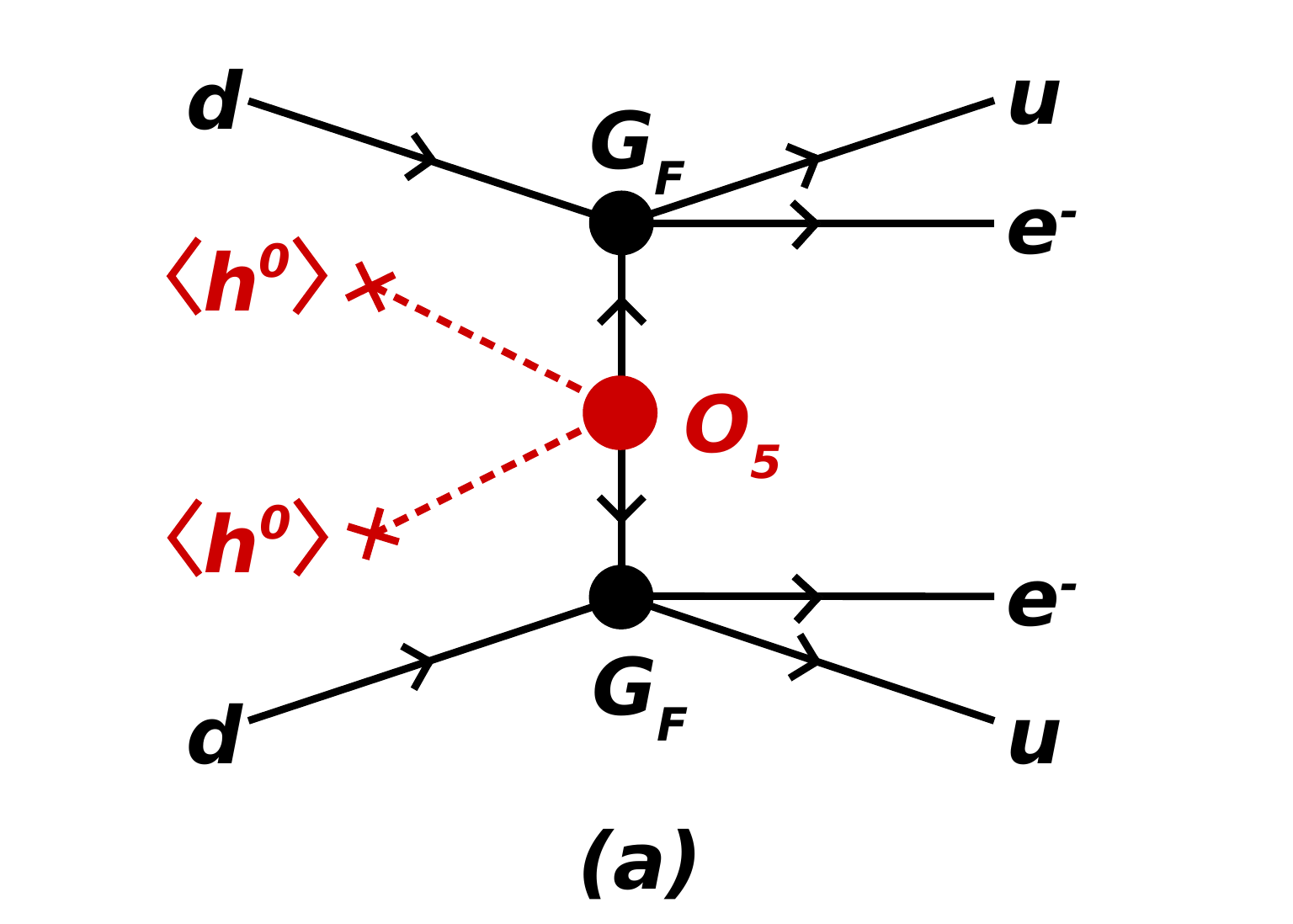}
	\includegraphics[clip,width=0.31\textwidth]{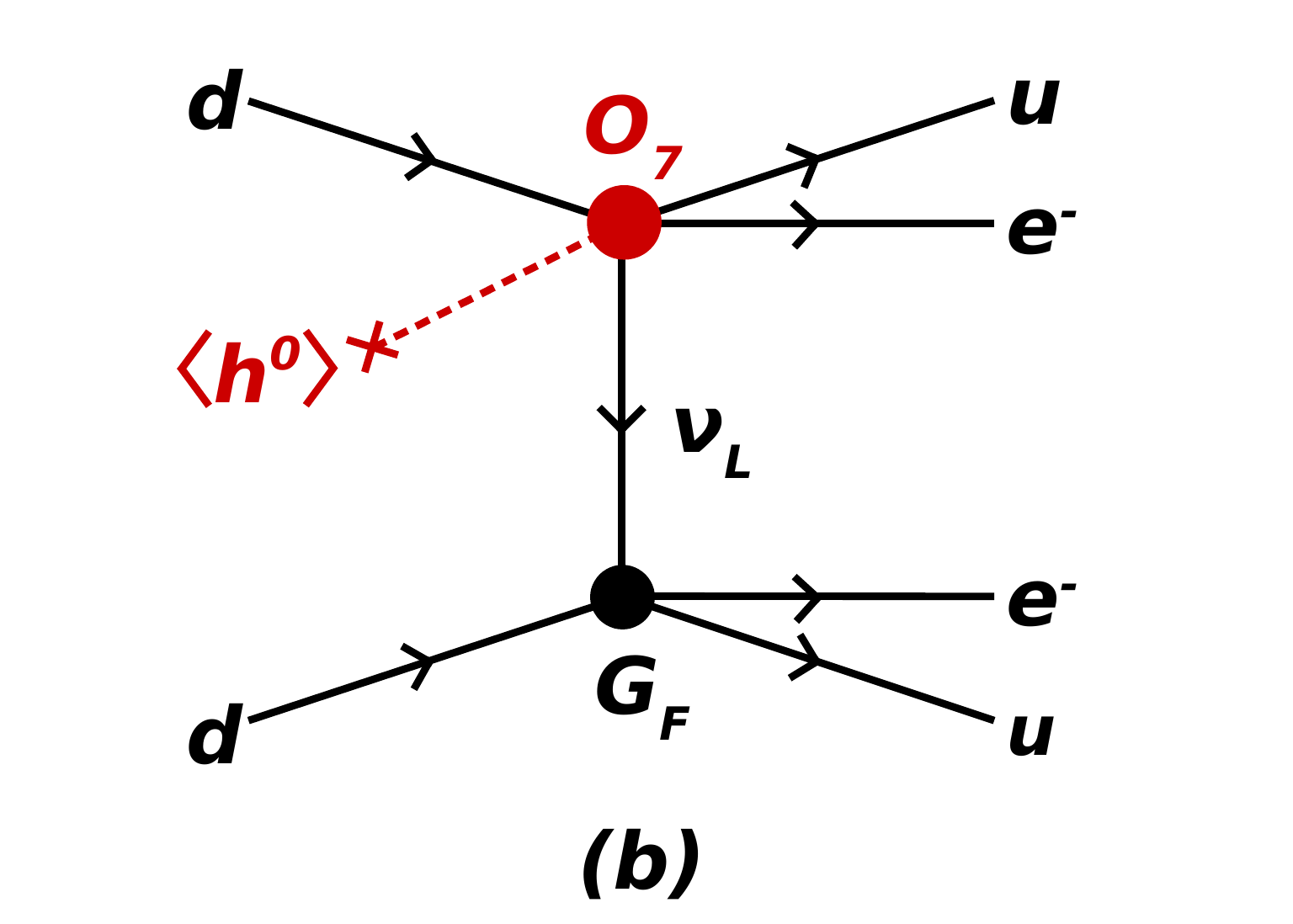}
	\includegraphics[clip,width=0.31\textwidth]{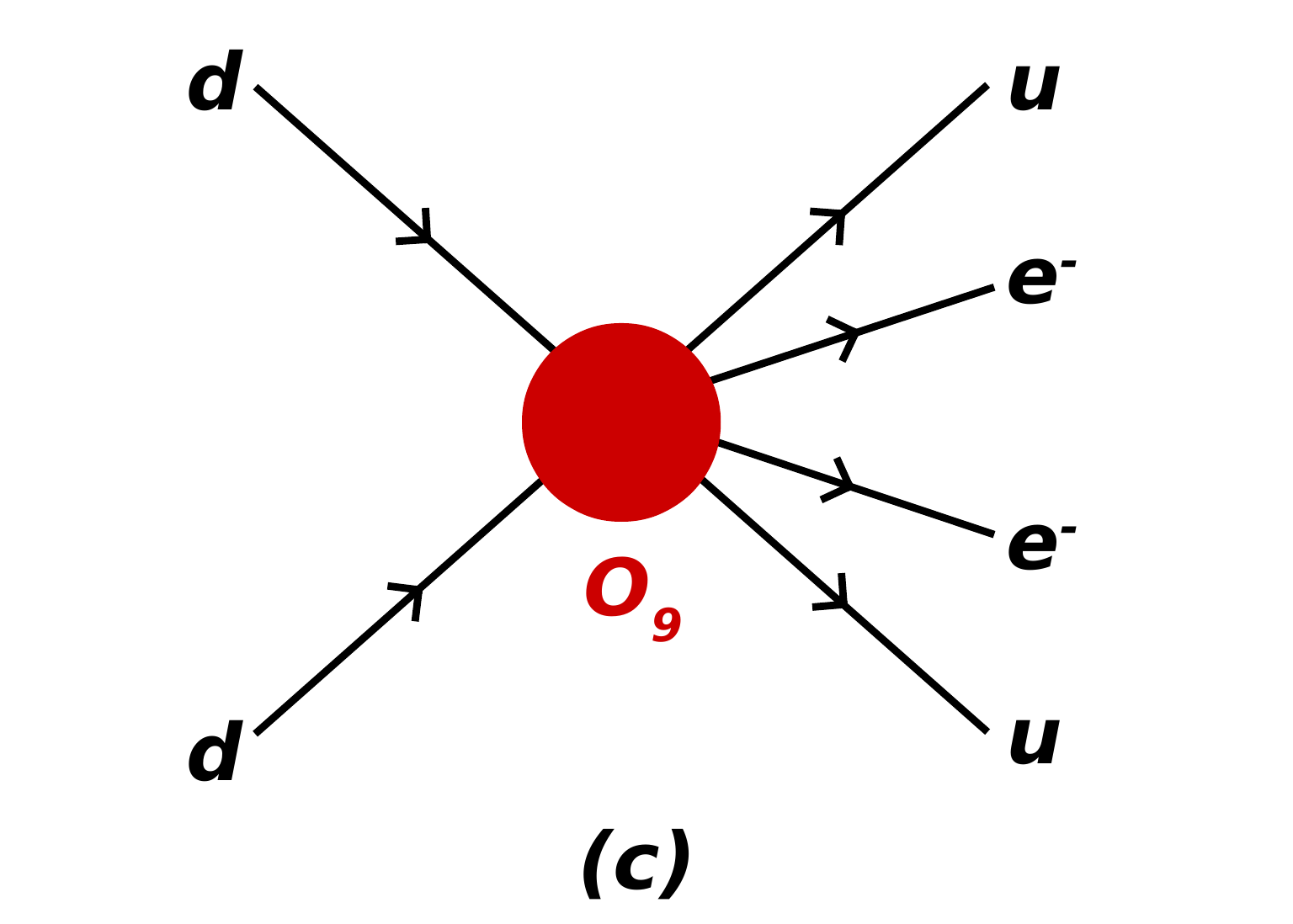}
	\caption{Contributions to $0\nu\beta\beta$ decay from effective LNV operators: (a) Standard light neutrino exchange via 5-dim operator; (b) Long--range contribution via 7-dim operator; (c) Short--range contribution via 9-dim operator. Adapted from \cite{Deppisch:2017ecm}.}
	\label{fig:contributions} 
\end{figure}
(i) Long-range transitions via exchange of a light neutrino. This includes the so-called standard mass mechanism in Fig.~\ref{fig:contributions}~(a) which is only possible if the neutrino is identical to its own antiparticle, i.e. if it is a Majorana fermion. The $0\nu\beta\beta$ decay rate can be estimated as $\Gamma^{0\nu\beta\beta}_{m_\nu} \sim m_\nu^2 G_F^4 m_F^2 Q_{\beta\beta}^5 \sim (m_\nu/0.1\,\text{eV})^2 (10^{26}\,\text{yr})^{-1}$. Here, $G_F$ is the SM Fermi coupling and the phase space scales as $Q_{\beta\beta}^5$ with the kinetic energy release $Q_{\beta\beta} = \mathcal{O}(1\,\text{MeV})$ for typical double beta decays. Specifically, the mass mechanism of $0\nu\beta\beta$ decay is sensitive to the effective neutrino mass $m_{\beta\beta} = \sum_i U_{ei}^2 m_{\nu_i}$, summing over the light Majorana neutrino masses $m_{\nu_i}$ weighted by the square of the charged-current leptonic mixing matrix elements $U_{ei}$. The inverse $0\nu\beta\beta$ decay half life in a given isotope is then conventionally expressed as
\begin{align}
\label{eq:halflife-intro}
	T_{1/2}^{-1} = \frac{|m_{\beta\beta}|^2}{m_e^2} G_\nu |\nme_\nu|^2,
\end{align}
with the phase space factor (PSF) $G_\nu$ and the nuclear matrix element (NME) $\nme_\nu$. The normalization with respect to the electron mass $m_e$ yields a small dimensionless parameter $|\epsilon_\nu| = |m_{\beta\beta}|/m_e$. The current bound in Eq.~\eqref{eq:cur-best-limit} sets a limit $|m_{\beta\beta}| \lesssim 79 - 180$~meV at 90\% confidence level (CL) for an unquenched axial coupling $g_A = 1.27$ \cite{Agostini:2020xta}, with the uncertainty mainly due to the NMEs in different nuclear models. Future experiments will probe $|m_{\beta\beta}| \approx 20$~meV \cite{Kim:2020vjv}, corresponding to the minimal value for inversely ordered neutrinos.

In BSM scenarios, a neutrino mass insertion is not necessarily required, cf. Fig.~\ref{fig:contributions}~(b). In such cases, the decay rate is estimated as $\Gamma^{0\nu\beta\beta}_\text{LR} \sim v^2\Lambda_{O_7}^{-6} G_F^2 m_F^4 Q_{\beta\beta}^5 \sim (10^5\,\text{GeV}/\Lambda_{O_7})^6 \times (10^{26}\,\text{yr})^{-1}$, with the SM Higgs vacuum expectation value (VEV) $v = 246$~GeV and the scale $\Lambda_{O_7}$ of the exotic dim-7 operator. Such long-range mechanisms via the exchange of light Majorana neutrinos with interactions beyond the SM have received considerable attention \cite{Doi:1981, Doi:1983, Tomoda:1990rs, Ali:2006iu, Ali:2007ec, Cirigliano:2017djv}, as the suppression at dim-7 is still fairly low and  $0\nu\beta\beta$ decay is sensitive to high scales. We note, though, that due to the neutrino helicity-flip intrinsic in the operator, typical mechanisms are suppressed by the light neutrino masses. It is generically difficult to have a dim-7 operator where the exotic long-range contribution dominates over the standard mass mechanism \cite{Helo:2016vsi} though it can be achieved in ultraviolet complete theories with a modestly suppressed standard contribution~\cite{Deppisch:2014zta, Deppisch:2017vne, Cirigliano:2018yza, Li:2020flq}.
	
(ii) Short-range contributions where all mediating particles are heavier than $m_F\approx 100$~MeV, cf. Fig.~\ref{fig:contributions}~(c), represented as contact interactions with six external fermions. These are the main focus of our analysis and they are generated by dim-9 and higher odd-dimensional operators. For a dim-9 operator, the decay rate can be estimated as $\Gamma^{0\nu\beta\beta}_\text{SR}\sim \Lambda_{O_9}^{-10} m_F^6 Q_{\beta\beta}^5 \sim (5\,\text{TeV}/\Lambda_{O_9})^{10} (10^{26}\,\text{yr})^{-1}$, with the operator scale $\Lambda_{O_9}$. The inverse $0\nu\beta\beta$ decay half life triggered by such a mechanism is expressed similarly to Eq.~\eqref{eq:halflife-intro} as $T_{1/2}^{-1} = |\epsilon_I|^2 G_I |\nme_I|^2$, with the PSF $G_I$ and NME $\nme_I$, both depending on the Lorentz structure of the effective operator. The coupling constant $\epsilon_I$ parametrizes the particle physics dynamics, i.e. the masses of the heavy states integrated out and their couplings. While such short-range contributions do not involve the exchange of light neutrinos at all, they still require the breaking of lepton number and the SM neutrinos will be of Majorana type. The short-range and standard mass mechanisms are thus expected to compete but the relative strength is highly dependent on the underlying model.

A detailed analytic derivation of the relevant NMEs for short-range operators was provided in our previous paper \cite{Graf:2018ozy}, where we included additional NMEs that become important when the latest values of the nucleon form factors are taken into account. Moreover, we calculated PSFs using the exact radial wave functions and we presented the single electron energy and angular correlation distributions for the exotic short-range $0\nu\beta\beta$ decay mechanisms. In the present paper, we numerically evaluate all relevant NMEs within the IBM-2 framework. This will allow us to set upper limits on the effective couplings $\epsilon_I$ where we will highlight the exchange of heavy sterile neutrinos as an important example. Within the same framework, we also provide updated NMEs for the standard light neutrino exchange and we analyse its interference with short-range mechanisms. The NMEs for the $\ovbb$ transitions are generally difficult to calculate and the limits derived are affected for any contribution. Detailed treatments using different nuclear structure model approaches can be found in \cite{Barea:2009zza, Barea:2013bz, Barea:2015kwa, Simkovic:2007vu, Simkovic:2013qiy, Suhonen:1991sk, Suhonen:2012wd, Caurier:2007xz, Menendez:2008jp, Rodriguez:2010mn}. Despite tremendous efforts to improve the nuclear theory calculation, the latest matrix elements obtained using various approaches differ in many cases by factors of $\sim(2-3)$.

The paper is organized as follows. We summarize the effective short-range Lagrangian at the quark level in Sec.~\ref{sec:effLagrangian} together with examples of underlying particle physics scenarios. The calculation of the $0\nu\beta\beta$ NMEs in the IBM-2 NME framework is outlined in Sec.~\ref{sec:nme} and that of the PSFs in Sec.~\ref{sec:PSFs}. We then present our numerical results in Sec.~\ref{sec:numresults} where we provide up-to-date limits on the standard mass mechanism and effective short-range $0\nu\beta\beta$ operators. Sec.~\ref{sec:summary} concludes our discussion with a summary and an outlook.

\section{Short-range LNV Operators and Neutrino Mass Models}
\label{sec:effLagrangian}

In general, new physics where lepton number is broken at a a high scale will induce SM effective operators of dimension-5, 7, 9 and higher \cite{Babu:2001ex, deGouvea:2007xp}. After EW symmetry breaking, this will give rise to long- and short-range contributions to $0\nu\beta\beta$ decay as outlined in the introduction, cf. Fig.~\ref{fig:contributions}. In this work we focus on short-range contributions and their potential interplay with the standard mass mechanism. 

\subsection{Effective Lagrangian}
The general effective short-range interaction Lagrangian can be written in terms of five different Lorentz-invariant classes of fermion current products \cite{Pas:2000vn},
\begin{align}
\label{eq:lagsr}
	\mathcal{L}_\text{SR} &= 
	\frac{G^2_F\cos^2\theta_C}{2m_p}\sum_{C_1,C_2,c}\left(
		  \epsilon_1^\chi J^{\phantom{\mu}}_{C_1} J^{\phantom{\mu}}_{C_2} j^{\phantom{\mu}}_c
		+ \epsilon_2^\chi J_{C_1}^{\mu\nu} J^{\phantom{\mu}}_{C_2,\mu\nu} j^{\phantom{\mu}}_c
		+ \epsilon_3^\chi J_{C_1}^\mu J^{\phantom{\mu}}_{C_2,\mu} j^{\phantom{\mu}}_c 
		\right.\nonumber\\
		&\qquad\qquad\qquad\qquad\,\,\left.
		+ \epsilon_4^\chi J_{C_1}^\mu J^{\phantom{\mu}}_{C_2,\mu\nu} j^\nu 
		+ \epsilon_5^\chi J_{C_1}^\mu J^{\phantom{\mu}}_{C_2} j^{\phantom{\mu}}_\mu
	\right) + \text{h.c.},
\end{align}
where the sum is over all unique combinations $\chi = \{C_1, C_2(,c)\}$ of chiralities $C_1, C_2, c = R, L$ of the quark and electron currents involved,
\begin{gather}
\label{eq:allcurrents}
	J_{R,L}          = \bar u_a(1\pm\gamma_5) d_a, \quad
	J^\mu_{R,L}      = \bar u_a\gamma^\mu(1\pm\gamma_5) d_a, \quad
	J^{\mu\nu}_{R,L} = \bar u_a\sigma_{\mu\nu}(1\pm\gamma_5) d_a, \\
	j_{R,L}          = \bar e(1\mp\gamma_5) e^c, \quad
	j^\mu            = \bar e\gamma^\mu\gamma_5 e^c.
\end{gather}
Here, the 4-component Dirac spinor operators representing the up-quark, down-quark and electron are denoted by $u$, $d$ and $e$, respectively. Quark $SU(3)_C$ colour indices are denoted by $a$, and each quark current forms a colour singlet in our parametrization.  As the lepton current must violate lepton number by two units, the charge conjugate electron field $e^c$ appears there. Note that the chirality assignment in $j_{R,L}$ is flipped, i.e. the index $L$ is associated with $1+\gamma_5$. This is due to the appearance of the charge-conjugated electron field and, for example, the operator $\bar e(1+\gamma_5) e^c$ describes the creation of two left-handed electrons. Furthermore, the usual definition $\sigma_{\mu\nu} = \frac{i}{2} \left[\gamma_\mu, \gamma_\nu\right]$ is used. The normalization of the Lagrangian by the factor $G_F^2\cos^2\theta_C/(2m_p)$ with the Fermi constant $G_F$, the SM Cabibbo angle $\theta_C$ and the proton mass $m_p$ is conventional and results in dimensionless couplings $\epsilon^\chi_i$. In principle, each unique current combination will be associated with a separate coupling, $\epsilon_i^\chi = \epsilon_i^{C_1 C_2 (c)}$. Note that in Ref.~\cite{Pas:2000vn}, the Lagrangian is defined without the factor $\cos^2\theta_C$. We chose to include it as the resulting PSFs can be defined in the same way as that for standard light neutrino exchange, cf. Sec.~\ref{sec:total-rate}.

Not all possible combinations of chiralities have to be considered in the Lagrangian Eq.~\eqref{eq:lagsr}, as redundancies and cancellations occur. First, the identity
\begin{align}
\label{eq:tensor-current}
	\sqb{\bar u\sigma^{\mu\nu}(1+\gamma_5)d} \sqb{\bar u\sigma_{\mu\nu}(1-\gamma_5)d} \equiv
	\sqb{\bar u\sigma^{\mu\nu}(1-\gamma_5)d} \sqb{\bar u\sigma_{\mu\nu}(1+\gamma_5)d} = 0
\end{align}
implies that terms corresponding to $\epsilon_2^{RLL}$, $\epsilon_2^{LRL}$, $\epsilon_2^{RLR}$ and $\epsilon_2^{LRR}$ trivially vanish. Second, the Pauli exclusion principle dictates that $\bar e\gamma^\mu e^c = 0$ and $\bar e\sigma_{\mu\nu}(1\pm\gamma_5) e^c = 0$, and thus any operator containing vector, tensor or axial-tensor electron currents can be omitted. Altogether, the short-range operators in Eq.~\eqref{eq:lagsr} contain 24 independent 9-dimensional operators invariant under the broken SM gauge group $SU(3)_C \times U(1)_Q$ \cite{Graf:2018ozy}.

\subsection{Example New Physics Scenarios with Short-Range Contributions}
\label{sec:newphysics}

To illustrate the generation of different short-range contributions, we consider three well know scenarios beyond the SM.

\subsubsection{Light and Heavy Neutrinos}
\label{sec:light-heavy-neutrinos}
As discussed in the introduction, the exchange of light active Majorana neutrinos is the most prominent mechanism for $0\nu\beta\beta$ decay. As a long-range contribution, it is not represented in the Lagrangian Eq.~\eqref{eq:lagsr} but arises from the SM charged current
\begin{align}
\label{eq:lagrangian}
	\mathcal{L} = 
	\frac{G_F\cos\theta_C}{\sqrt{2}}
	\left[\bar u\gamma^\mu(1-\gamma_5)d\right]
	\sum_{i=1}^3 U_{ei}\left[\bar e\gamma^\mu(1-\gamma_5)\nu_i\right]
	+ \text{h.c.}.
\end{align}
The sum is over the three SM neutrino mass eigenstates $\nu_i$, constructed as the Majorana spinors $\nu_i = \nu_{i,L} + \nu_{i,L}^c$ from the SM active left-handed neutrinos $\nu_{i,L}$ and their charge-conjugates. This gives rise to the mass mechanism of $0\nu\beta\beta$ decay sensitive to the effective Majorana neutrino mass
\begin{align}
\label{eq:bbmass}
	m_{\beta\beta} = \sum_{i=1}^3 U_{ei}^2 m_{\nu_i}.
\end{align}
The $0\nu\beta\beta$ decay half life in a given isotope is then conventionally expressed as in Eq.~\eqref{eq:halflife-intro}.

One of the most attractive extensions of the SM involves adding fermionic states $\nu_{i,S}$ ($i = 1,\dots, n_N$) that are sterile under the SM gauge interactions. They can thus acquire (Dirac or Majorana type) masses without spoiling the SM gauge invariance and eventually mix with the SM neutrinos after electroweak symmetry breaking. We can again form Majorana states by constructing $N_i = \nu_{i,S} + \nu_{i,S}^c$. The sterile states participate in the leptonic charged current due to mixing with the active neutrinos,
\begin{align}
\label{eq:lagrangian-sterile}
	\mathcal{L} = 
	\frac{G_F\cos\theta_C}{\sqrt{2}}
	\left[\bar u\gamma^\mu(1-\gamma_5)d\right]
	\sum_{i=1}^{n_N} V_{eN_i}\left[\bar e\gamma^\mu(1-\gamma_5)N_i\right]
	+ \text{h.c.},
\end{align}
where $V_{eN_i}$ are the elements of the active-sterile mixing matrix.

If the sterile neutrinos are much lighter than the nuclear physics scale $p_F \approx 100$~MeV, their contributions to $0\nu\beta\beta$ decay will be completely analogous to that of the active neutrinos and they can be included in Eq.~\eqref{eq:halflife-intro} by replacing
\begin{align}
	m_{\beta\beta} \to m_{\beta\beta} + \sum_{i=1}^{n_N} V_{eN_i}^2 m_{N_i},
	\qquad (m_{N_i} \ll 100~\text{MeV}).
\end{align}
Note that the $U_{ei}$, and hence $m_{\beta\beta}$, as well as the $V_{eN_i}$ are in general complex numbers and cancellations can occur. In fact, if the Majorana states $N_i$ are solely responsible for the light neutrinos masses in a Seesaw scenario, the active and sterile contributions cancel to zero.

If instead the sterile states are much heavier than the nuclear physics scale, $m_{N_i} \gg 100$~MeV, they can be integrated out, resulting in a contribution of the type $J_L^\mu J^{\phantom{\mu}}_{L,\mu}j^{\phantom{\mu}}_L$ and the associated coupling $\epsilon_3^{LLL}$ is matched with the underlying physics parameters as
\begin{align}
\label{eq:eps3LLL}
	\epsilon_3^{LLL} = \sum_{i=1}^{n_N} V_{eN_i}^2 \frac{m_p}{m_{N_i}},
	\qquad (m_{N_i} \gg 100~\text{MeV}).
\end{align}
Note that the above considerations apply for sterile neutrinos that are Majorana fermions. This includes quasi-Dirac states that can be described by pairs of Majorana neutrinos ($N_1$, $N_2$) with a small mass splitting $|m_{N_1} - m_{N_2}| \ll m_{N_{1,2}}$ and a relative CP phase of $\pi/2$, $V_{eN_2} = i V_{eN_1} \Rightarrow V_{eN_2}^2 = - V_{eN_1}^2$. In the limit of Dirac sterile neutrinos with $m_{N_1} = m_{N_2}$, the contributions to $0\nu\beta\beta$ decay cancel.

\subsubsection{Left-Right Symmetry}
\label{sec:lrsm}
The minimal Left-Right symmetric model (LRSM) is based on the extended gauge symmetry $SU(3)_C \times SU(2)_L \times SU(2)_R \times U(1)_{B-L}$ \cite{Pati:1974yy, Mohapatra:1974gc, Senjanovic:1975rk}. It has a rich neutrino and $0\nu\beta\beta$ decay phenomenology as it naturally contains right-handed Majorana neutrinos $N_i$ ($i=1,2,3$) that are charged under the $SU(2)_R$ part of the gauge group, forming a doublet together with the right-handed leptons. This gives rise to right-handed charged currents,
\begin{align}
\label{eq:lagrangianLR}
	\mathcal{L} = 
	\frac{g_R^2\cos\theta^R_C}{8m_{W_R}^2}
	\left[\bar u\gamma^\mu(1+\gamma_5)d\right]
	\sum_{i=1}^3 U^R_{ei}\left[\bar e\gamma^\mu(1+\gamma_5)N_i\right]
	+ \text{h.c.},
\end{align}
mediated by a right-handed $W_R$ boson with the gauge coupling strength $g_R$ of the $SU(2)_R$ group. The angle $\theta_C^R$ and the mixing matrix $U^R$ are the right-handed equivalents of the Cabibbo angle and the Pontecorvo–Maki–Nakagawa–Sakata matrix, respectively. The LRSM gauge group is understood to be spontaneously broken to that of the SM at a high scale giving masses to the right-handed $W_R$ boson and neutrinos $N_i$. In turn, the active SM neutrino acquire masses via mixing with the heavy neutrinos (Seesaw type I) as well as via the VEV of an electroweak triplet Higgs scalar present in the model (Seesaw type II).

Hence, the standard light neutrino and the sterile heavy neutrino contribution described above are generally present. In addition, the equivalent diagram with a heavy neutrino and two $W_R$ bosons contributes, giving rise to the short-range operator $J_R^\mu J^{\phantom{\mu}}_{R,\mu} j^{\phantom{\mu}}_R$ with $j_R = \bar{e}(1-\gamma_5) e^c$ associated with $\epsilon_3^{RRR}$ matched to the underlying physics parameters as
\begin{align}
\label{eq:lrsm-3RRR}
	\epsilon_3^{RRR} = \frac{g_R^2}{g^2}f_{LR}^2\sum_{i=1}^3 (U^R_{ei})^2 \frac{m_p}{m_{N_i}},\quad\text{with}\quad
	f_{LR} = \frac{g_R}{g}\frac{\cos\theta_C^R}{\cos\theta_C}
	\frac{m_W^2}{m_{W_R}^2},
\end{align}
where $g$ is the SM $SU(2)_L$ gauge coupling strength. Note that the contribution is not suppressed by the small light-heavy neutrino mixing but instead by the expectedly high $W_R$ mass $m_{W_R}$. The right-handed mixing matrix $U^R$ is approximately unitary with elements of order one, although cancellations due to complex phases can occur. 

The SM $W$ and the $W_R$ boson are also expected to mix with an angle as large as $\sin\theta_{LR}^W \lesssim g_R m_W^2/(g m_{W_R}^2)$. This permits the right-handed lepton current to couple with a left-handed quark current mediated by the SM $W$ giving rise to the contributions
\begin{align}
\label{eq:lrsm-3XXR}
	\epsilon_3^{LRR} = \epsilon_3^{RLR} = 
	\frac{\sin\theta_{LR}^W}{f_{LR}} \epsilon_3^{RRR}, \quad
	\epsilon_3^{LLR} = \frac{\sin^2\theta_{LR}^W}{f^2_{LR}} \epsilon_3^{RRR}.
\end{align}
With the $W$ mixing taking the generic value $\sin\theta_{LR}^W \approx g_R m_W^2/(g m_{W_R}^2) \approx f_{LR}$, all three effective couplings are of the same order. As mentioned, the LRSM also has the standard contribution from $m_{\beta\beta}$ and the sterile neutrino contribution $\epsilon_3^{LLL}$ in Eq.~\eqref{eq:eps3LLL}. In addition, the LRSM in principle also gives rise to the remaining contributions of type $\epsilon_3$, namely $\epsilon_3^{LRL} = \epsilon_3^{RLL}$ and $\epsilon_3^{RRL}$ but these are suppressed by both the light-heavy neutrino mixing and the high $W_R$ mass. Furthermore, the LRSM gives rise to additional long-range contributions that are not directly suppressed by the light neutrino masses.

Finally, the LRSM has contributions from the electroweak triplet scalars $\Delta_{L,R}$ that acquire VEVs $v_R$, $v_L \sim v^2/v_R$ during the spontaneous symmetry breaking, where $v_R$ is the breaking scale of the Left-Right symmetry. This gives rise to a diagram to $0\nu\beta\beta$ decay mediated by two $W_R$ bosons and the doubly-charged scalars $\Delta_{L,R}^{--}$. Taking into account the $W$ boson mixing, the contributions are
\begin{align}
\label{eq:doubly-charged-higgs}
	\epsilon_3^{RRR} = \frac{g_R^2}{g^2}f^2_{LR}
	\sum_{i=1}^3 (U^R_{ei})^2 \frac{m_p m_{N_i}}{m^2_{\Delta_R^{++}}}, \,\,
	\epsilon_3^{LRR} = \epsilon_3^{RLR} = 
	\frac{\sin\theta_{LR}^W}{f_{LR}} \epsilon_3^{RRR}, \,\,
	\epsilon_3^{LLR} = \frac{\sin^2\theta_{LR}^W}{f^2_{LR}} \epsilon_3^{RRR},   
\end{align}
analogous to Eqs.~\eqref{eq:lrsm-3RRR} and \eqref{eq:lrsm-3XXR}. Here, the heavy neutrino masses $m_{N_i}$ appear because the couplings of the triplet Higgs to the gauge boson and electrons are proportional to $v_R$ and the heavy neutrino Yukawa coupling, $m_N \sim y_N v_R$. Likewise, there are contributions from the left-handed $\Delta_L^{++}$ but they are additionally suppressed by the light neutrino masses (instead of $m_{N_i}$) and thus negligible. 

\subsubsection{$R$-Parity Violating Supersymmetry}

As the final example of an ultraviolet-complete theory, we consider the minimal supersymmetric Standard Model (MSSM) with $R$-parity violation \cite{Dimopoulos:1988jw, Hall:1983id}. Without explicitly imposing invariance under the discrete $R$ symmetry where each field carries the multiplicative quantum number $R = (-1)^{3B+L+2S}$, with the baryon number $B$, total lepton number $L$ and spin $S$, the MSSM allows for the  $R$-parity breaking terms
\begin{align}
\label{eq:rpv-superpotential}
	W \supset \lambda_{ijk}   L_i L_j \bar E_k
	        + \lambda'_{ijk}  L_i Q_j \bar D_k
	        + \lambda''_{ijk} \bar U_i \bar D_j \bar D_k,
\end{align}
in the superpotential. Here, the indices $i,j,k$ denote flavour generations of the superfields $L$, $\bar E$, $Q$, $\bar D$ and $\bar U$, associated with the SM weak lepton doublet $L$, the lepton singlet $e^c$, the quark doublet $Q$ and the quark singlets $d^c$, $u^c$. Short-range contributions to $0\nu\beta\beta$ are induced by the second term in Eq.~\eqref{eq:rpv-superpotential}, namely that associated with $\lambda'_{111}$ for the first lepton and quark generations \cite{Mohapatra:1986su}. They arise from diagrams with intermediate, heavy neutralinos, gluinos, squarks and sleptons. The corresponding short-range Lagrangian is \cite{Hirsch:1995ek}
\begin{align}
	\mathcal{L}_\text{SR} \supset \frac{G_F^2\cos^2\theta_C}{2m_p}
	\left(
		  \epsilon_1^{RRL} J_R J_R
		+ \epsilon_2^{RRL} J^{\mu\nu}_R J_{R,\mu\nu}^{\phantom{\mu}}
	\right)j_L,
\end{align}
i.e. a subset of the general short-range Lagrangian in Eq.~\eqref{eq:lagrangian} with scalar and tensor quark currents. The effective couplings $\epsilon_1^{RRL}$ and $\epsilon_2^{RRL}$ are generally functions of all supersymmetric particle masses and couplings involved. We here follow the assumptions of gluino dominance \cite{Hirsch:1995ek} where the diagrams involving gluinos and squarks contribute,
\begin{align}
\label{eq:rpv-epsilons}
	\epsilon_1^{RRL} = 
	\frac{8\pi\alpha_s\lambda'^{2}_{111}}{9\cos^2\theta_C}\frac{G_F^{-2}}{m^4_{\tilde q}}\frac{m_p}{m_{\tilde g}}, \qquad
	\epsilon_2^{RRL} = -\frac{1}{8}\epsilon_1^{RRL}.
\end{align}
Here we also assume degeneracy of squark masses $m_{\tilde q} = m_{\tilde u_L} = m_{\tilde d_R}$ in line with Ref.~\cite{Hirsch:1995ek}. In addition, $m_{\tilde g}$ is the gluino mass and $\alpha_s = 0.127$ is the strong fine structure constant at $m_W$. Note that the gluino dominance assumption is based on the relevant NME values and limits on supersymmetry particle masses from other sources and may thus not be appropriate in light of new results. We nevertheless adopt it for simplicity and to compare with  Ref.~\cite{Hirsch:1995ek}.   

\section{Determination of Nuclear Matrix Elements}
\label{sec:nme}

The NMEs for short-range mechanisms have been analytically derived in \cite{Graf:2018ozy}. We follow the approach therein and summarize the basic formalism using nucleon form factors.

\subsection{Nucleon Form Factors}
\label{subsec:quarktonucleons}
The nucleon matrix elements of the colour-singlet quark currents in Eq.~\eqref{eq:lagsr} have the structure~\cite{Adler:1975he}
\begin{align}
\label{eq:nucleoncurrents}
	\bra{p}\bar{u}(1\pm\gamma_5)d\ket{n} &= 
	\bar{N} \tau^+ 
	\sqb{F_S(q^2) \pm F_{P'}(q^2)\gamma_5} N', \\
	\bra{p}\bar{u}\gamma^\mu(1\pm\gamma_5)d\ket{n} &= 
	\bar{N} \tau^+ 
	\sqb{F_V(q^2)\gamma^\mu-i\frac{F_W(q^2)}{2m_p}\sigma^{\mu\nu}q_\nu}N' 
	\nonumber\\ 
\label{eq:vecNucleons}
	&\pm \bar{N} \tau^+ \sqb{F_A(q^2)\gamma^\mu\gamma_5 
	- \frac{F_P(q^2)}{2m_p}\gamma_5q^\mu}N', \\ 
	\bra{p}\bar{u}\sigma^{\mu\nu}(1\pm\gamma_5)d\ket{n} &= 
	\bar{N} \tau^+ \sqb{J^{\mu\nu} \pm \frac{i}{2} \epsilon^{\mu\nu\rho\sigma} J_{\rho\sigma}} N',
\label{eq:tenNucleons}
\end{align}
where $\tau^+$ denotes the isospin-raising operator which converts a neutron into a proton, and the tensor $J^{\mu\nu}$ in Eq.~\eqref{eq:tenNucleons} is defined as
\begin{align}
	J^{\mu\nu} = F_{T_1}(q^2)\sigma^{\mu\nu} + i\frac{F_{T_2}(q^2)}{m_p}(\gamma^\mu q^\nu 
	- \gamma^\nu q^\mu) + \frac{F_{T_3}(q^2)}{m_p^2}(\sigma^{\mu\rho} q_\rho q^\nu 
	- \sigma^{\nu\rho} q_\rho q^\mu).
\end{align}

The above matrix elements generally depend on the neutron and proton momenta $p_n = p_{N'}$ and $p_p = p_N$, respectively. The nucleon form factors are then functions of the momentum transfer $q = p_p - p_n$. The most general parametrization of the vector current in Eq.~\eqref{eq:vecNucleons} would include also induced scalar and axial-tensor terms --- these can be, however, safely neglected, since they vanish in the isospin-symmetric limit and they are not enhanced by any other effects \cite{PhysRev.112.1375}.

The momentum dependence in Eqs.~\eqref{eq:nucleoncurrents} - \eqref{eq:tenNucleons} is encoded in the nucleon form factors $F_X(q^2)$ with $X = S, P', V, W, A, P, T_1, T_2, T_3$, usually parametrized in the so-called dipole form, $F_X(q^2) = g_X/(1 + q^2 / m_X^2)^2$. Here, the so called charge $g_X$ represents the value of the form factor at zero momentum transfer, $g_X \equiv F_X(0)$, and the scale $m_X$ determines the shape of the form factor. We apply this parametrization to all form factors except for the pseudoscalar form factors $F_{P'}(q^2)$ and $F_{P}(q^2)$ which are enhanced by the pion resonance. The form factors with their corresponding parametrizations and charges are given by
\begin{align}
\label{eq:fs}
	F_S(q^2)    &= \frac{g_S}{(1 + q^2 / m_V^2)^2}, &  
	g_S &= 1.0~\text{\cite{Gonzalez-Alonso:2018omy}}, \\
\label{eq:fpprime}
	F_{P'}(q^2) &= \frac{g_{P'}}{(1 + q^2/m_V^2)^2}\frac{1}{1 + q^2/m^2_\pi}, & 
	g_{P'} &= 349~\text{\cite{Gonzalez-Alonso:2018omy}}, \\
	F_V(q^2)    &= \frac{g_V}{(1 + q^2 / m_V^2)^2}, & g_V &= 1.0, \\
	F_W(q^2)    &= \frac{g_W}{(1 + q^2 / m_V^2)^2}, & g_W &= 3.7, \\
	F_A(q^2)    &= \frac{g_A}{(1 + q^2 / m_A^2)^2}, & g_A &= 1.269, \\
	F_P(q^2)    &= \frac{g_P}{(1 + q^2 / m_A^2)^2}\frac{1}{1 + q^2 / m_\pi^2}, & 
	g_P &= 4 g_A \frac{m_p^2}{m_\pi^2}\left(1-\frac{m_\pi^2}{m_A^2}\right)
	= 231~\text{\cite{Simkovic:1999re}}, \\
\label{eq:fti}
	F_{T_i}(q^2)&= \frac{g_{T_i}}{(1 + q^2 / m_V^2)^2}, &
	g_{T_{1,2,3}} &= 1.0, -3.3, 1.34~\text{\cite{Adler:1975he}}.
\end{align}
The shape parameters are $m_V = 0.84$~GeV, $m_A = 1.09$~GeV \cite{Schindler:2006jq} and the pion mass is $m_\pi = 0.138$~GeV. The form factors $F_V(q^2)$, $F_W(q^2)$ and $F_A(q^2)$ can be determined experimentally and the parametrizations shown above provide a good description in the range $0 \leq |q| \leq 200$~MeV of interest in $0\nu\beta\beta$ decay. On the other hand, as it is not possible to directly obtain the induced pseudoscalar form factor from experiment, we use the parametrization suggested in Ref.~\cite{Simkovic:1999re}, which is based on the partially conserved axial-vector current (PCAC) hypothesis. The corresponding value of the free $g_P$ charge agrees with the recent chiral perturbation theory analysis \cite{Bernard:2001rs}, which yields the value $g_P = 233$. The value is also consistent with measurements of muon capture. With the muon mass $m_\mu = 0.105$~GeV, the resulting value of $F_{P}(-0.88 m_\mu^2) = 8.0$ agrees well with the measured value of $F_{P}(-0.88m_\mu^2) = 8.06 \pm 0.55$ \cite{Andreev:2012fj}. The scalar and pseudoscalar charges, $g_S$ and $g_{P'}$, come from recent lattice QCD calculations \cite{Gonzalez-Alonso:2018omy}. As there is not much information on the $q^2$-dependence of the corresponding form factors, we use the dipole parametrization, which, in the Breit frame, is the Fourier transform of the matter distribution. In the case of the pseudoscalar form factor we also include the monopole factor $1/(1 + q^2/m_\pi^2)$ used in chiral perturbation theory. As for the tensor form factors, only $F_{T_1}$ enters our calculations. The value of the corresponding charge $g_{T_1}$ quoted by Ref.~\cite{Gonzalez-Alonso:2018omy} reads $0.987\pm 0.055$. We emphasize that the charges in Eqs.~\eqref{eq:fs} - \eqref{eq:fti} are applicable at the free nucleon level. When calculating the $0\nu\beta\beta$ decay NMEs we will use an effective axial-vector charge $g_A = 1.0$ and, consequently, an induced pseudoscalar charge $g_P(g_A = 1.0) = 182$ to approximately account for quenching in the nuclear medium.

\subsection{Nuclear Matrix Elements}
\label{subsec:nmes}
The five different types of quark current products appearing in Eq.~\eqref{eq:lagsr} are mapped to the nucleon matrix elements according to Eqs.~\eqref{eq:nucleoncurrents} - \eqref{eq:tenNucleons}. By virtue of a non-relativistic expansion and the closure approximation, the resulting product of nucleon matrix elements is then mapped to the \emph{nuclear} matrix element between the final and initial $0^+$ nuclear states involved in the $0\nu\beta\beta$ decay. This procedure is described in Ref.~\cite{Graf:2018ozy} and we here summarize the definition of NMEs involved. One should note that in the following expressions the relative sign between GT and T terms is different than in our previous papers \cite{Barea:2009zza, barea12, Barea:2013bz, Barea:2015kwa} and other available literature taking into account tensor terms using the formulation in \cite{Simkovic:1999re}. The confusion about the relative sign arises from Eqs.~(13) and (22) in \cite{Simkovic:1999re}, where in Eq.~(13) a minus sign is used in front of the tensor term, while in Eq.~(22) the plus sign is used. The tensor term contributes very little to the standard long range mechanism, but, in the case of short range mechanisms, it has a notable effect. Thus we have checked the derivation and concluded that the following signs should be used.

The NMEs for the five short-range operators will generally depend on the chiralities of the two quark currents involved. For the first three operators associated with $\epsilon^\chi_1$, $\epsilon^\chi_2$ and $\epsilon^\chi_3$, the two quark currents are of the same type. Consequently, three possible combinations occur corresponding to the chiralities $RR$, $LL$ and $(RL + LR)/2$. It turns out that the resulting NMEs only depend on whether the quark chiralities are equal ($RR$, $LL$) or different $(RL + LR)/2$, represented by the upper and lower sign, respectively, in the expressions 
\begin{align}
\label{eq:nme1}
	\nme_1 &=   g_S^2\nme_F 
    	        \pm \frac{g_{P'}^{2}}{12}
    	        \left(\nme_{GT}^{'P'P'} + \nme_T^{'P'P'}\right), \\
\label{eq:nme2}        	    
	\nme_2 &= -2g_{T_1}^2 \nme_{GT}^{T_1T_1}, \\
\label{eq:nme3}
	\nme_3 &=   g_V^2 \nme_F
	            + \frac{(g_V + g_W)^2}{12}
	            \left(-2\nme^{\prime WW}_{GT} + \nme^{\prime WW}_T\right) 
				\nonumber\\
		   &    \mp \left[ g_A^2 \nme_{GT}^{AA}
		        - \frac{g_A g_{P}}{6} 
		        \left(\nme^{\prime AP}_{GT} + \nme^{\prime AP}_T\right)
				+ \frac{g_{P}^2}{48} 
				\left(\nme^{\prime\prime PP}_{GT} 
				      + \nme^{\prime\prime PP}_T\right)\right].
\end{align}
For the operators associated with $\epsilon^\chi_4$ and $\epsilon^\chi_5$, the two quark currents involved have different Lorentz structures and thus all four possible combinations of chiralities have to be considered in principle: $RR$, $LL$, $RL$ and $LR$. Again, it turns out that the NMEs only distinguish between the case where the quark chiralities are the same ($RR$, $LL\to$ upper sign) or different ($RL$, $LR\to$ lower sign),  
\begin{align}
\label{eq:nme4}
	\nme_4 &= \mp i \left[
			  g_A g_{T_1} \nme_{GT}^{AT_1}
			  - \frac{g_{P} g_{T_1}}{12}
			  \left(\nme^{\prime PT_1}_{GT} + \nme^{\prime PT_1}_T\right)
			  \right], \\
\label{eq:nme5}
	\nme_5 &=  g_Vg_S\nme_F
			   \pm \left[
			   \frac{g_A g_{P'}}{12}
			   \left(\tilde{\nme}^{AP'}_{GT} + \tilde{\nme}^{AP'}_T\right)
			   - \frac{g_{P}g_{P'}}{24}
			   \left(\nme^{\prime q_0 PP'}_{GT} + \nme^{\prime q_0 PP'}_T\right)
			   \right].
\end{align}
In the above expressions, we have explicitly factored the form factor charges $g_X = F_X(0)$. The $q$-dependence arising from the product of the reduced form factors $F_X(q^2)/g_X$ is still to be included in the various matrix elements appearing in Eqs.~\eqref{eq:nme1}-\eqref{eq:nme5}. The individual Fermi ($\mathcal{M}_F$), Gamow-Teller ($\mathcal{M}_{GT}$) and tensor ($\mathcal{M}_T$) NMEs along with the associated reduced form factor products $\tilde h(q^2)$ are given in Table~\ref{nme_ff}. The numerical values of these NME will be given in Sec.~\ref{sec:nme_numeric} but we would like to note that the so called recoil NMEs $\tilde{\mathcal{M}}^{AP}_{GT}$ and $\tilde{\mathcal{M}}^{AP}_{T}$, and the NMEs explicitly depending on the temporal momentum transfer $q_0$, $\mathcal{M}_{GT}^{\prime q_0 PP}$, $\mathcal{M}_{T}^{\prime q_0 PP}$ are difficult to evaluate exactly. We instead assume that the sum of nucleon spatial momenta is $\vecl{Q} = \vecl{p}_a + \vecl{p}_b \approx \vecl{q}$ \cite{Doi:1981, Doi:1983, Tomoda:1990rs}, approximately applicable in an average sense considering that the NME is calculated summing over all nucleons in the nucleus. Similarly, we take the average value $q_0 \sim \vecl{q}^2/m_p \approx 10$~MeV \cite{Tomoda:1990rs} for the temporal component of the momentum transfer. This allows to reduce the corresponding NMEs as indicated in Table~\ref{nme_ff}.
\setlength{\tabcolsep}{5pt}
\begin{table}[t!]
\centering
\begin{tabular}{l|l}
\hline
NME & $\tilde{h}_\circ(q^2)$ \\
\hline
$\mathcal{M}_F = \langle h_{XX}(q^2) \rangle$		
& $\tilde{h}_{XX}(q^2) = \frac{1}{(1+q^2/m_V^2)^4}$  \\
$\mathcal{M}_{GT}^{'P'P'} = 
\left\langle \frac{\vecl{q}^2}{m_p^2} h_{PP}(q^2) 
(\vecs{\sigma}_a\cdot\vecs{\sigma}_b) \right\rangle$ 
& $\tilde{h}_{PP}(q^2) = \frac{1}{(1 + q^2/m_A^2)^4}
\frac{1}{(1 + q^2/m_\pi^2)^2}$                       \\
$\mathcal{M}_T^{'P'P'} =
\left\langle \frac{\vecl{q}^2}{m_p^2}  h_{PP}(q^2) S_{ab} \right\rangle$	
& $\tilde{h}_{PP}(q^2)$                                         \\
\hline
$\mathcal{M}^{T_1T_1}_{GT} = \langle h_{XX}(q^2) (\vecs{\sigma}_a\cdot\vecs{\sigma}_b) \rangle$	
& $\tilde{h}_{XX}(q^2)$                                         \\
\hline
$\mathcal{M}^{\prime WW}_{GT} = 
\left\langle \frac{\vecl{q}^2}{m_p^2} h_{XX}(q^2) 
(\vecs{\sigma}_a\cdot\vecs{\sigma}_b) \right\rangle$		
& $\tilde{h}_{XX}(q^2)$                                         \\
$\mathcal{M}^{\prime WW}_T = 
\left\langle \frac{\vecl{q}^2}{m_p^2} h_{XX}(q^2) S_{ab} \right\rangle$	
& $\tilde{h}_{XX}(q^2)$                                         \\
$\mathcal{M}_{GT}^{AA} = \langle h_{AA}(q^2) (\vecs{\sigma}_a\cdot\vecs{\sigma}_b) \rangle$	
& $\tilde{h}_{AA}(q^2) = \frac{1}{(1+q^2/m_A^2)^4}$  \\
$\mathcal{M}^{\prime AP}_{GT} =  
\left\langle \frac{\vecl{q}^2}{m_p^2} h_{AP}(q^2) 
(\vecs{\sigma}_a\cdot\vecs{\sigma}_b) \right\rangle$ 
& $\tilde{h}_{AP}(q^2) = \frac{1}{(1 + q^2/m_A^2)^4}
\frac{1}{1 + q^2/m_\pi^2}$                                      \\
$\mathcal{M}^{\prime AP}_T = 
\left\langle \frac{\vecl{q}^2}{m_p^2}  h_{AP}(q^2) S_{ab} \right\rangle$	
& $\tilde{h}_{AP}(q^2)$                                         \\
$\mathcal{M}^{\prime\prime PP}_{GT} = \left\langle \frac{\vecl{q}^4}{m_p^4}
h_{PP}(q^2) (\vecs{\sigma}_a\cdot\vecs{\sigma}_b) \right\rangle$	
& $\tilde{h}_{PP}(q^2)$                                         \\
$\mathcal{M}^{\prime\prime PP}_{T} = \left\langle \frac{\vecl{q}^4}{m_p^4}  
h_{PP}(q^2) S_{ab} \right\rangle$ 
& $\tilde{h}_{PP}(q^2)$                                         \\
\hline
$\mathcal{M}_{GT}^{AT_1} = \langle h_{AX}(q^2) (\vecs{\sigma}_a\cdot\vecs{\sigma}_b) \rangle$	
& $\tilde{h}_{AX}(q^2) = \frac{1}{(1+q^2/m_V^2)^2} \frac{1}{(1+q^2/m_A^2)^2}$                           \\
$\mathcal{M}^{\prime PT_1}_{GT} = 
\left\langle \frac{\vecl{q}^2}{m_p^2} h_{XP}(q^2) 
(\vecs{\sigma}_a\cdot\vecs{\sigma}_b) \right\rangle$	
& $\tilde{h}_{XP}(q^2) = \frac{1}{(1 + q^2/m_V^2)^2}
\frac{1}{(1 + q^2/m_A^2)^2}\frac{1}{1 + q^2/m_\pi^2}$ \\
$\mathcal{M}^{\prime PT_1}_{T} = 
\left\langle \frac{\vecl{q}^2}{m_p^2} h_{XP}(q^2) S_{ab} \right\rangle$ 
& $\tilde{h}_{XP}(q^2)$                                         \\
\hline
$\tilde{\mathcal{M}}^{AP'}_{GT} = \left\langle \frac{\vecl{Q}\cdot\vecl{q}}{m_p^2}
h_{AP}(q^2) (\vecs{\sigma}_a\cdot\vecs{\sigma}_b) \right\rangle 
\quad\,\,\approx \mathcal{M}_{GT}^{'AP}$ 
& $\tilde{h}_{AP}(q^2)$                                         \\
$\tilde{\mathcal{M}}^{AP'}_{T} = \left\langle \frac{\vecl{Q}\cdot\vecl{q}}{m_p^2} 
h_{AP}(q^2) S_{ab} \right\rangle
\qquad\quad\,\,\,\,\approx \mathcal{M}_{T}^{'AP}$	
& $\tilde{h}_{AP}(q^2)$                                         \\
$\mathcal{M}_{GT}^{\prime q_0 PP'} = \left\langle \frac{q_0 \vecl{q}^2}{m_p^3} 
h_{PP}(q^2) (\vecs{\sigma}_a\cdot\vecs{\sigma}_b) \right\rangle 
\approx 10^{-2}\mathcal{M}_{GT}^{\prime P'P'}$ 
& $\tilde{h}_{PP}(q^2)$                                         \\
$\mathcal{M}_T^{\prime q_0 PP'} = \left\langle \frac{q_0\vecl{q}^2}{m_p^3}  
h_{PP}(q^2) S_{ab} \right\rangle
\quad\quad\,\,\approx 10^{-2} \mathcal{M}_T^{\prime P'P'}$ 
& $\tilde{h}_{PP}(q^2)$                                         \\
\hline
\end{tabular}
\caption{\label{nme_ff} Double beta decay Fermi ($\mathcal{M}_F$), Gamow-Teller ($\mathcal{M}_{GT}$) and tensor ($\mathcal{M}_T$) NMEs appearing in Eqs.~\eqref{eq:nme1}-\eqref{eq:nme5}, with the associated reduced form factor product $\tilde{h}(q^2)$. The NMEs are calculated using the functions $h_\circ(q^2) = v(q^2)\tilde h_\circ(q^2)$ enhanced by the neutrino potential Eq.~\eqref{eq:neutrinopotential} for short-range mechanisms and standard light neutrino exchange, Eq.~\eqref{eq:neutrinopotentialLR}. The subscript $X$ stands for $X = V, W, T_1$ for which the same form factor shape parameter $m_V$ applies. The Pauli matrices in the space of the spins of the individual nucleons $a$, $b$ are represented as $\vecs{\sigma}_{a,b}$ and the tensor NMEs are calculated over $S_{ab} = 3(\vecs{\sigma}_a \cdot \vecl{q}) (\vecs{\sigma}_b \cdot \vecl{q}) - (\vecs{\sigma}_a \cdot \vecs{\sigma}_b)$.}
\end{table}

In addition to the product of the reduced nucleon form factors, the NMEs listed in Table~\ref{nme_ff} also contain the so called neutrino potential describing the $q$ dependence of the underlying particle physics mediator of $0\nu\beta\beta$ decay. Here we follow the formulation of \cite{Simkovic:1999re} and \cite{Barea:2013bz} where the two-body transition operator is constructed in momentum space as the product of the neutrino potential $v(q)$ times the product of the reduced form factors $\tilde{h}(q^2)$. In the case of the short-range mechanisms we consider here, the neutrino potential is especially simple; as point-like operators, they are described by a Dirac delta function in configuration space, $\delta(\vecl{r}_a - \vecl{r}_b)$, hence in momentum space it is a $q$-independent constant. Following the usual normalization the short-range neutrino potential is \cite{Barea:2013bz, Simkovic:1999re}
\begin{align}
\label{eq:neutrinopotential}
	v(q^2) = \frac{2}{\pi}\frac{1}{m_e m_p}.
\end{align}

We also consider the standard light neutrino exchange mechanism with the NME
\begin{align}
\label{eq:nme_nu}
	M_\nu &= g_V^2 \nme_F - g_A^2 \nme_{GT}^{AA} 
	       + \frac{g_A g_{P}}{6} 
	         \left(\nme^{\prime AP}_{GT} + \nme^{\prime AP}_T\right) \nonumber\\
          &+ \frac{(g_V+g_W)^2}{12}
             \left(-2\nme^{\prime WW}_{GT} + \nme^{\prime WW}_T\right) 
           - \frac{g_{P}^2}{48} 
             \left(\nme^{\prime\prime PP}_{GT} 
                 + \nme^{\prime\prime PP}_T
             \right).
\end{align}
Note that this is fully analogous to $\nme_3$ in Eq.~\eqref{eq:nme3} in the case where the quark currents have the same chirality, but the crucial difference is that the NMEs in Eq.~\eqref{eq:nme_nu} are calculated with the appropriate neutrino potential in momentum space \cite{Barea:2013bz},
\begin{align}
\label{eq:neutrinopotentialLR}
	v(q) = \frac{2}{\pi}\frac{1}{q(q + \tilde A)}.
\end{align}
Here, the neutrino mass has been neglected in comparison with the neutrino momentum $q\sim 100$~MeV, and $\tilde{A}$ is the closure energy, taken from Ref.~\cite{Haxton:1985am} or estimated by the systematics, $\tilde{A} = 1.12\sqrt{A}$~MeV. This describes the long-range exchange of an essentially massless neutrino mediating $0\nu\beta\beta$ decay in this case. As noted earlier, the relative sign between the GT and T terms in Eq.~\eqref{eq:nme_nu} is different than in our previous papers \cite{Barea:2009zza, barea12, Barea:2013bz, Barea:2015kwa} and other literature.

Our derivation of the NMEs performed within the phenomenological framework of the nucleon form factors can be compared with an alternative way which has been developed in the literature over recent years. It is based on chiral effective field theory \cite{Weinberg:1991um}, i.e. the effective theory describing interactions at low energy in terms of baryons, mesons, photons and leptons \cite{Cirigliano:2017djv, Cirigliano:2018hja, Cirigliano:2018yza, Cirigliano:2019vdj}. In this approach the process of hadronization is replaced by a perturbative expansion in terms of $q/\Lambda_\chi$ reflecting the approximate chiral symmetry of QCD, where $\Lambda_\chi \simeq m_p \approx 1$~GeV is the chiral symmetry breaking scale. The chiral Lagrangian on which the corresponding calculation is based should then incorporate all possible terms invariant under the chiral symmetry $SU(2)_L\times SU(2)_R$ in the same way as the corresponding quark-level operators. Each term then comes with a so called low energy constant (LEC) parametrizing the non-perturbative nature of QCD. Thus, the LECs play a role similar to that of the nuclear form factors arising in hadronization and their reliable determination, e.g. using lattice QCD input, is necessary to calculate the $\ovbb$ decay rate in the chiral EFT framework. The benefit of this approach is that one can avoid the factorization of the nucleon currents, which is a necessary approximation in the hadronization procedure.

\subsection{Determination of NMEs in the IBM-2}
\label{sec:nme_numeric}
In order to evaluate the NMEs we make use of the microscopic interacting boson model (IBM-2) \cite{ARIMA1977205, iac87} which has the advantage that it can be used to all nuclei of interest. The interacting boson model  has been one of the most successful models in reproducing collective features of the low-lying levels of medium as well as heavy nuclei, and  is one of the few models that can be used consistently to all nuclei of interest. We have already studied different mechanisms systematically using the microscopic interacting boson model (IBM-2) \cite{Barea:2009zza, barea12, Barea:2013bz, barea13b, kotila14, Barea:2015kwa, Barea:2015zfa} and this study adds the short-range non-standard mechanisms of double beta decay to the list. 

The method of evaluation is discussed in detail in \cite{Barea:2009zza, Barea:2015kwa}. We use the interacting boson model with isospin restoration \cite{ Barea:2015kwa} in which isospin is restored by enforcing $M_F^{2\nu} = 0$ as in the Quasi-particle Random Phase Approximation (QRPA) calculations of \cite{Simkovic:2013qiy, Fang:2015zha}. Here we briefly mention the logic of the method, which  is a mapping of the fermion operator $H$ onto a boson space and its evaluation with bosonic wave functions. The mapping \cite{OTSUKA19781} can be done to leading order (LO), next to leading order (NLO), etc.. In Ref.~\cite{Barea:2009zza} it was shown, by explicit calculations, that NLO terms give, in general, negligible contributions, $\leq1\%$. The matrix elements of the mapped operators are then evaluated with realistic wave functions, taken either from the literature, when available, or obtained from a fit to the observed energies and other properties ($B(E2)$ values, quadrupole moments, $B(M1)$ values, magnetic moments, etc.). The values of parameters used in the current calculations are given in Appendix~A.

The single-particle and single-hole energies and strengths of interaction were evaluated and discussed in detail in Ref.~\cite{PhysRevC.94.034320} where the occupancies of the single particle levels were calculated in order to satisfy a twofold goal: to asses the goodness of the single particle energies and check the reliability of the used wave functions. Both tests are particularly important in the case of nuclei involved in double beta decay, as they affect the evaluation of the NMEs and then their reliability \cite{eng15}. The energies of the single particle levels constitute a very important input for the calculation of the occupancies in the method used in Ref.~\cite{PhysRevC.94.034320}. In principle those energies can be considered as input parameters that can be fitted to reproduce the experimental occupancies. Instead of fitting, the single particle energies were extracted from experimental data on nuclei with a particle more or one particle less than a shell closure. These single particle energy sets were then used to calculate the occupancies of several nuclei of interest in double beta decay. Finally, the results were compared with other theoretical calculations and experimental occupancies, when available, and good correspondence was obtained. As part of the calculation single particle energies for several major shells were updated to values given in Appendix~B. 

Finally, an additional improvement is the introduction of short-range correlations in the nuclear structure calculation. These are of crucial importance for short-range non-standard mechanisms and they can be taken into account by multiplying the potential $v(r)$ in coordinate space by a correlation function $f(r)$ squared. The most commonly used correlation function is the Jastrow function,
\begin{align}
	f_J(r) = 1 - ce^{-ar^2}(1 - br^2),
\end{align}
with $a = 1.1\,\text{fm}^{-2}$, $b = 0.68\,\text{fm}^{-2}$ and $c = 1$ for the phenomenological Miller-Spencer parametrization \cite{MILLER1976562}, and $a = 1.59\,\text{fm}^{-2}$, $b = 1.45\,\text{fm}^{-2}$ and $c = 0.92$ for the Argonne parametrization \cite{Simkovic:2009pp}. Since our formulation is in momentum space, we take short-range correlations into account by using the Fourier-Bessel transform of $f_J(r)$.

\subsubsection{Numerical Values of the NMEs}
\begin{table}[t!]
	\centering
	\scalebox{0.73}{
		\begin{tabular}{r|rrrrrrrrrrrrrrrr}
			\hline
			Isotope & $\nme_F$ & $\nme^{AA}_{GT}$ &
			$\nme^{AT_1}_{GT}$ & $\nme^{T_1T_1}_{GT}$ & $\nme^{'WW}_{GT}$ & $\nme^{'WW}_T$ &
			$\nme^{'AP}_{GT}$ & $\nme^{'AP}_T$ & $\nme^{'PT_1}_{GT}$ & $\nme^{'PT_1}_{T}$ &
			$\nme^{'P'P'}_{GT}$ & $\nme^{'P'P'}_{T}$ & $\nme^{''PP}_{GT}$ &
			$\nme^{''PP}_{T}$ \\
			\hline
			${}^{76}$Ge	& $-48.89$	& $170.0$	& $174.3$	& $173.5$	& $-2.945$	& $-6.541$	& $2.110$	& $-1.310$	& $2.255$	& $-1.183$	& $0.798$	& $-0.271$	& $0.028$	& $-0.022$	\\
			${}^{82}$Se	& $-41.22$	& $140.7$	& $144.3$	& $143.6$	& $-2.456$	& $-6.206$	& $1.758$	& $-1.249$	& $1.878$	& $-1.183$	& $0.660$	& $-0.259$	& $0.024$	& $-0.021$	\\
			${}^{96}$Zr	& $-35.31$	& $124.3$	& $128.5$	& $128.8$	& $-3.116$	& $5.436$	& $1.523$	& $1.090$	& $1.652$	& $0.984$	& $0.613$	& $0.228$	& $0.020$	& $0.019$	\\
			${}^{100}$Mo	& $-51.96$	& $181.9$	& $188.1$	& $188.6$	& $-4.590$	& $8.055$	& $2.273$	& $1.590$	& $2.464$	& $1.128$	& $0.910$	& $0.317$	& $0.029$	& $0.027$	\\
			${}^{110}$Pd	& $-43.52$	& $151.2$	& $156.5$	& $157.0$	& $-3.945$	& $6.816$	& $1.892$	& $1.356$	& $2.055$	& $1.223$	& $0.762$	& $0.271$	& $0.024$	& $0.023$	\\
			${}^{116}$Cd	& $-32.45$	& $110.5$	& $114.6$	& $115.2$	& $-3.069$	& $4.222$	& $1.374$	& $0.843$	& $1.497$	& $0.760$	& $0.565$	& $0.169$	& $0.017$	& $0.015$	\\
			${}^{124}$Sn	& $-33.19$	& $104.2$	& $106.7$	& $106.1$	& $-1.701$	& $-3.655$	& $1.321$	& $-0.723$	& $1.407$	& $-0.651$	& $0.489$	& $-0.146$	& $0.018$	& $-0.012$	\\
			${}^{128}$Te	& $-41.82$	& $131.7$	& $134.9$	& $134.1$	& $-2.439$	& $-4.519$	& $1.667$	& $-0.890$	& $1.776$	& $-1.433$	& $0.617$	& $-0.178$	& $0.023$	& $-0.015$	\\
			${}^{130}$Te	& $-38.05$	& $119.7$	& $122.6$	& $121.9$	& $-1.951$	& $-4.105$	& $1.514$	& $-0.807$	& $1.613$	& $-0.726$	& $0.561$	& $-0.160$	& $0.021$	& $-0.014$	\\
			${}^{134}$Xe	& $-39.45$	& $124.7$	& $127.8$	& $127.2$	& $-2.111$	& $-4.191$	& $1.564$	& $-0.823$	& $1.669$	& $-0.741$	& $0.585$	& $-0.163$	& $0.021$	& $-0.014$	\\
			${}^{136}$Xe	& $-29.83$	& $94.18$	& $96.56$	& $96.09$	& $-1.625$	& $-3.158$	& $1.177$	& $-0.620$	& $1.257$	& $-0.558$	& $0.442$	& $-0.123$	& $0.016$	& $-0.011$	\\
			${}^{148}$Nd	& $-31.71$	& $103.0$	& $106.0$	& $105.8$	& $-2.145$	& $2.557$	& $1.346$	& $0.510$	& $1.445$	& $0.460$	& $0.508$	& $0.104$	& $0.018$	& $0.009$	\\
			${}^{150}$Nd	& $-30.18$	& $100.0$	& $103.2$	& $103.1$	& $-2.230$	& $2.955$	& $1.292$	& $0.581$	& $1.392$	& $0.523$	& $0.497$	& $0.116$	& $0.017$	& $0.010$	\\
			${}^{154}$Sm	& $-31.83$	& $107.1$	& $110.7$	& $110.9$	& $-2.618$	& $3.397$	& $1.356$	& $0.668$	& $1.467$	& $0.601$	& $0.536$	& $0.135$	& $0.018$	& $0.012$	\\
			${}^{160}$Gd	& $-41.43$	& $142.9$	& $148.0$	& $148.6$	& $-3.808$	& $5.231$	& $1.776$	& $1.023$	& $1.931$	& $0.920$	& $0.722$	& $0.205$	& $0.023$	& $0.018$	\\
			${}^{198}$Pt	& $-31.87$	& $104.4$	& $108.4$	& $109.0$	& $-2.992$	& $3.172$	& $1.334$	& $0.626$	& $1.454$	& $0.564$	& $0.546$	& $0.119$	& $0.017$	& $0.011$	\\
			${}^{232}$Th	& $-44.04$	& $154.2$	& $159.7$	& $160.3$	& $-4.116$	& $6.146$	& $1.900$	& $1.185$	& $2.067$	& $1.063$	& $0.783$	& $0.230$	& $0.024$	& $0.021$	\\
			${}^{238}$U	& $-52.48$	& $183.1$	& $189.7$	& $190.5$	& $-4.981$	& $7.206$	& $2.255$	& $1.393$	& $2.456$	& $1.251$	& $0.932$	& $0.272$	& $0.029$	& $0.024$	\\
			\hline
		\end{tabular}
	}
	\caption{NMEs for short-range $0\nu\beta\beta$ decay mechanisms evaluated in the IBM-2 as described in the text and to be used in Eqs.~\eqref{eq:nme1} - \eqref{eq:nme5}. The values of the last four NMEs in Table~\ref{nme_ff} are not listed as they are derived from other NMEs as indicated therein.}
	\label{tab:NMEs1}
\end{table}
We present the numerical values all NMEs necessary to evaluate Eqs.~\eqref{eq:nme1} - \eqref{eq:nme5} for the short-range mechanisms and for all relevant $0\nu\beta\beta$ decaying isotopes in Table~\ref{tab:NMEs1}. They have been calculated within the IBM-2 as discussed above. This represents the first complete calculation of the NMEs needed for the description of short-range mechanisms of neutrinoless double beta decay. Note that the last four NMEs in Table~\ref{nme_ff} are not listed as they are derived from other NMEs as indicated therein. Likewise, Table~\ref{tab:lightNMEs} contains the NMEs for the standard light neutrino exchange mechanism, cf. Eq.~\eqref{eq:nme_nu}. We remind the reader that in our convention where we calculate the NMEs using the reduced nucleon form factors $F_X(q^2)/g_X$, the NMEs in Tabs.~\ref{tab:NMEs1} and \ref{tab:lightNMEs} do not contain the form factor charges. They instead explicitly appear as coefficients in the expressions for $\mathcal{M}_1$ to $\mathcal{M}_5$ and for $\mathcal{M}_\nu$.

\begin{table}[t!]
	\centering
	\begin{tabular}{r|rrrrrrrr}
		\hline
		Isotope & $\nme_F$ & $\nme^{AA}_{GT}$ & $\nme^{'AP}_{GT}$ &
		$\nme^{'AP}_T$ & $\nme^{'WW}_{GT}$ & $\nme^{'WW}_T$ &
		$\nme^{''PP}_{GT}$ & $\nme^{''PP}_T$ \\
		\hline
		${}^{76}$Ge  & $-0.780$ & $6.062$ & $0.036$ & $-0.010$ & $0.089$ & $-0.035$ & $3.4\times10^{-4}$ & $-1.4\times10^{-4} $ \\
		${}^{82}$Se  & $-0.667$ & $4.928$ & $0.030$ & $-0.010$ & $0.073$ & $-0.034$ & $4.1\times10^{-4}$ & $-1.3\times10^{-4} $ \\
		${}^{96}$Zr  & $-0.361$ & $4.317$ & $0.027$ & $0.009$ & $0.065$ & $0.032$ & $3.1\times10^{-4}$ & $1.2\times10^{-4} $ \\
		${}^{100}$Mo & $-0.511$ & $5.553$ & $0.038$ & $0.012$ & $0.096$ & $0.041$ & $4.7\times10^{-4}$ & $1.6\times10^{-4} $ \\
		${}^{110}$Pd & $-0.425$ & $4.432$ & $0.032$ & $0.009$ & $0.080$ & $0.036$ & $3.9\times10^{-4}$ & $1.4\times10^{-4} $ \\
		${}^{116}$Cd & $-0.335$ & $3.173$ & $0.023$ & $0.005$ & $0.058$ & $0.023$ & $2.9\times10^{-4}$ & $8.7\times10^{-5} $ \\
		${}^{124}$Sn & $-0.572$ & $3.370$ & $0.021$ & $-0.005$ & $0.053$ & $-0.018$ & $2.5\times10^{-4}$ & $-7.5\times10^{-5}$ \\
		${}^{128}$Te & $-0.718$ & $4.321$ & $0.027$ & $-0.005$ & $0.067$ & $-0.023$ & $3.1\times10^{-4}$ & $-9.1\times10^{-5} $ \\
		${}^{130}$Te & $-0.651$ & $3.894$ & $0.024$ & $-0.006$ & $0.061$ & $-0.021$ & $2.8\times10^{-4}$ & $-8.3\times10^{-5}$ \\
		${}^{134}$Xe & $-0.686$ & $4.211$ & $0.026$ & $-0.005$ & $0.064$ & $-0.023$ & $3.0\times10^{-4}$ & $-8.3\times10^{-5}$ \\
		${}^{136}$Xe & $-0.522$ & $3.203$ & $0.019$ & $-0.005$ & $0.048$ & $-0.016$ & $2.2\times10^{-4}$ & $-6.3\times10^{-5}$ \\
		${}^{148}$Nd & $-0.363$ & $2.517$ & $0.020$ & $0.005$ & $0.053$ & $0.014$ & $2.6\times10^{-4}$ & $5.3\times10^{-5} $ \\
		${}^{150}$Nd & $-0.507$ & $3.753$ & $0.032$ & $0.005$ & $0.083$ & $0.027$ & $4.1\times10^{-4}$ & $9.7\times10^{-5} $ \\
		${}^{154}$Sm & $-0.340$ & $2.984$ & $0.022$ & $0.005$ & $0.056$ & $0.018$ & $2.7\times10^{-4}$ & $6.9\times10^{-5} $ \\
		${}^{160}$Gd & $-0.415$ & $4.224$ & $0.030$ & $0.009$ & $0.074$ & $0.027$ & $3.6\times10^{-4}$ & $1.1\times10^{-4} $ \\
		${}^{198}$Pt & $-0.329$ & $2.270$ & $0.021$ & $0.005$ & $0.054$ & $0.014$ & $2.7\times10^{-4}$ & $6.1\times10^{-5} $ \\
		${}^{232}$Th & $-0.444$ & $4.169$ & $0.032$ & $0.009$ & $0.079$ & $0.032$ & $3.9\times10^{-4}$ & $1.2\times10^{-4} $ \\
		${}^{238}$U  & $-0.525$ & $4.962$ & $0.038$ & $0.009$ & $0.093$ & $0.036$ & $4.6\times10^{-4}$ & $1.4\times10^{-4} $ \\
		\hline
	\end{tabular}
	\caption{NMEs for the standard light neutrino exchange $0\nu\beta\beta$ decay mechanism evaluated in the IBM-2 as described in the text and to be used in Eq.~\eqref{eq:nme_nu}.}
	\label{tab:lightNMEs}
\end{table}
By specifically separating the value of $g_A$ we allow for the possibility of a quenching of the axial-vector coupling. Even though quenching of $g_A$ goes beyond the topic of this study, we would like to remind that it is well known from single beta decay and electron capture that $g_A$ is renormalized in models of nuclei. Quenching of $g_A$ in $2\nu\beta\beta$-decay, consistent with single-beta decay, has also been observed~\cite{Barea:2013bz, Barea:2015kwa} (for a review see \cite{10.3389/fphy.2017.00055}). However, the question of whether or not $g_A$ in $0\nu\beta\beta$ decay is renormalized as much as in $2\nu\beta\beta$ is of much debate. This problem is currently being addressed both experimentally, by employing single and double charge exchange reactions \cite{PhysRevC.86.044603, Cappuzzello2018}, and theoretically, by using effective field theories to estimate the effect of non-nucleonic degrees of freedom \cite{PhysRevLett.107.062501}. Quenching of $g_A$ arises from the omission of non-nucleonic degrees of freedom and from the limited model space in which the calculations are done. The former effect is not expected to be present in $\ovbb$ decay since the average neutrino momentum is $\sim 100$ MeV, while in $2\nu\beta\beta$ decay is of the order of $1-2$ MeV. The latter effect instead appears both in $\ovbb$ and $2\nu\beta\beta$ decays. This consideration suggests to use an effective value of $g_A^\text{eff} = 1.0$, in between the free value $g_A = 1.269$ and the value observed in $2\nu\beta\beta$ decay, $g_A\sim 0.6$. We henceforth use this value.

\subsubsection{Comparison with Earlier Results}
From the NMEs in Tables~\ref{tab:NMEs1} and \ref{tab:lightNMEs} one can calculate the NMEs for the standard mass mechanism, $\nme_\nu$ and heavy neutrino exchange $\nme_{\nu_h} = \nme_3^{LL}$ to compare with earlier calculations. To this end, it is convenient to introduce the quantities
\begin{align}
	\nme_{GT} &= \nme_{GT}^{AA} - \frac{g_P}{6g_A}\nme_{GT}^{'AP} 
	           + \frac{(g_V+g_W)^2}{6g_A^2}\nme_{GT}^{'WW} 
	           + \frac{g_P^2}{48g_A^2}\nme_{GT}^{''PP} \\
	\nme_T    &= \frac{g_P}{6g_A}\nme_{T}^{'AP} 
	           + \frac{(g_V+g_W)^2}{12g_A^2}\nme_{T}^{'WW} 
	           - \frac{g_P^2}{48g_A^2}\nme_{T}^{''PP} 
\end{align}
and write $\nme_\nu$ as
\begin{align}
	\nme_\nu = g_A^2 \left[\left(\frac{g_V}{g_A}\right)^2\nme_F -\nme_{GT} + \nme_T \right],
	\label{eq:lightnunme}
\end{align}
and similarly for $\nme_{\nu_h}$. 

\begin{table}[t!]
	\centering
	\begin{tabular}{@{\extracolsep{4pt}}r|rrrr|r|rrrr@{}}
		\hline
		Isotope 		& $\nme_F^{\text{old}}$ & $\nme_{GT}^{\text{old}}$ & $\nme_T^{\text{old}}$ & $\nme_\nu^{\text{old}}$ & $\tilde{\nme}_\nu^{\text{old}}$ & $\nme_F$ & $\nme_{GT}$ & $\nme_T$ & $\nme_\nu$ \\ \hline
		${}^{76}$Ge		& $-0.68$	& $4.49$	& $-0.23$	& $-4.94$	& $-5.40$	& $-0.78$	& $5.58$	& $-0.28$	& $-6.64$	\\
		${}^{82}$Se		& $-0.6$	& $3.59$	& $-0.23$	& $-3.96$	& $-4.42$	& $-0.67$	& $4.52$	& $-0.27$	& $-5.46$	\\
		${}^{96}$Zr		& $-0.33$	& $2.51$	& $0.11$	& $-2.95$	& $-2.73$	& $-0.36$	& $3.95$	& $0.25$	& $-4.07$	\\
		${}^{100}$Mo	& $-0.48$	& $3.73$	& $0.19$	& $-4.40$	& $-4.02$	& $-0.51$	& $5.08$	& $0.32$	& $-5.27$	\\
		${}^{110}$Pd	& $-0.40$	& $3.59$	& $0.21$	& $-4.20$	& $-3.78$	& $-0.43$	& $4.03$	& $0.24$	& $-4.21$	\\
		${}^{116}$Cd	& $-0.33$	& $2.76$	& $0.14$	& $-3.23$	& $-2.95$	& $-0.34$	& $2.89$	& $0.12$	& $-3.11$	\\
		${}^{124}$Sn	& $-0.57$	& $2.96$	& $-0.12$	& $-3.41$	& $-3.65$	& $-0.57$	& $3.10$	& $-0.12$	& $-3.79$	\\
		${}^{128}$Te	& $-0.72$	& $3.80$	& $-0.15$	& $-4.37$	& $-4.67$	& $-0.72$	& $3.97$	& $-0.12$	& $-4.80$	\\
		${}^{130}$Te	& $-0.65$	& $3.43$	& $-0.13$	& $-3.95$	& $-4.21$	& $-0.65$	& $3.59$	& $-0.16$	& $-4.40$	\\
		${}^{134}$Xe	& $-0.68$	& $3.77$	& $-0.15$	& $-4.30$	& $-4.60$	& $-0.69$	& $3.86$	& $-0.12$	& $-4.67$	\\
		${}^{136}$Xe	& $-0.52$	& $2.83$	& $-0.10$	& $-3.25$	& $-3.45$	& $-0.52$	& $2.96$	& $-0.12$	& $-3.60$	\\
		${}^{148}$Nd	& $-0.38$	& $2.00$	& $0.07$	& $-2.45$	& $-2.31$	& $-0.36$	& $2.28$	& $0.12$	& $-2.52$	\\
		${}^{150}$Nd	& $-0.39$	& $2.33$	& $0.10$	& $-2.82$	& $-2.62$	& $-0.51$	& $3.37$	& $0.12$	& $-3.76$	\\
		${}^{154}$Sm	& $-0.36$	& $2.49$	& $0.11$	& $-2.96$	& $-2.74$	& $-0.34$	& $2.71$	& $0.12$	& $-2.93$	\\
		${}^{160}$Gd	& $-0.45$	& $3.64$	& $0.17$	& $-4.26$	& $-3.92$	& $-0.42$	& $3.84$	& $0.25$	& $-4.00$	\\
		${}^{198}$Pt	& $-0.33$	& $1.90$	& $0.09$	& $-2.32$	& $-2.14$	& $-0.33$	& $2.02$	& $0.12$	& $-2.23$	\\
		${}^{232}$Th	& $-0.44$	& $3.58$	& $0.18$	& $-4.20$	& $-3.84$	& $-0.44$	& $3.76$	& $0.25$	& $-3.95$	\\
		${}^{238}$U		& $-0.53$	& $4.27$	& $0.21$	& $-5.01$	& $-4.59$	& $-0.53$	& $4.47$	& $0.24$	& $-4.75$	\\
		\hline
	\end{tabular}
	\caption{Comparison between the light neutrino exchange NMEs calculated in this work and those calculated in~\cite{Barea:2015kwa} using the quenched value $g_A = 1.0$ and the convention that $\nme_\nu < 0$. The ``old" $F$, $GT$ and $T$ NMEs of Table~I in \cite{Barea:2015kwa} are combined in the NMEs $\nme_\nu^\text{old}$ and $\tilde{\nme}_\nu^\text{old}$ using a negative and positive sign of the tensor NME relative to that of the $GT$ NME, respectively.}
	\label{tab:nmecomplight}
\end{table}
\begin{table}[t!]
	\centering
	\begin{tabular}{@{\extracolsep{3pt}}r|rrrr|r|rrrr@{}}
		\hline
		Isotope & $\nme_{\nu_h,\,F}^{\text{old}}$ & $\nme_{\nu_h,\,GT}^{\text{old}}$ & $\nme_{\nu_h,\,T}^{\text{old}}$ & $\nme_{\nu_h}^{\text{old}}$ & $\tilde{\nme}_{\nu_h}^{\text{old}}$ & $\nme_{\nu_h,\,F}$ & $\nme_{\nu_h,\,GT}$ & $\nme_{\nu_h,\,T}$ & $\nme_{\nu_h}$ \\ \hline
		${}^{76}$Ge		& $-42.8$	& $104$	& $-26.9$	& $-120$	& $-174$	& $-48.9$	& $115$	& $-36.3$	& $-200$	\\
		${}^{82}$Se		& $-37.1$	& $87.2$	& $-27.3$	& $-97.0$	& $-152$	& $-41.2$	& $94.7$	& $-34.5$	& $-171$	\\
		${}^{96}$Zr		& $-29.2$	& $67.9$	& $12.7$	& $-110$	& $-84.4$	& $-35.3$	& $80.2$	& $30.2$	& $-85.4$	\\
		${}^{100}$Mo	& $-46.8$	& $111$	& $24.2$	& $-182$	& $-134$	& $-52.0$	& $116$	& $44.1$	& $-124$	\\
		${}^{110}$Pd	& $-41.4$	& $100$	& $27.7$	& $-169$	& $-114$	& $-43.5$	& $96.2$	& $37.5$	& $-102$	\\
		${}^{116}$Cd	& $-31.2$	& $73.9$	& $16.9$	& $-122$	& $-88.2$	& $-32.5$	& $69.6$	& $23.3$	& $-78.8$	\\
		${}^{124}$Sn	& $-33.1$	& $73.7$	& $-14.9$	& $-91.9$	& $-122$	& $-33.2$	& $70.3$	& $-20.0$	& $-124$	\\
		${}^{128}$Te	& $-41.7$	& $93.4$	& $-18.3$	& $-117$	& $-153$	& $-41.8$	& $87.9$	& $-24.7$	& $-154$	\\
		${}^{130}$Te	& $-37.9$	& $84.8$	& $-16.6$	& $-106$	& $-139$	& $-38.1$	& $80.8$	& $-22.4$	& $-141$	\\
		${}^{134}$Xe	& $-39.3$	& $86.6$	& $-19.8$	& $-106$	& $-146$	& $-39.5$	& $84.1$	& $-22.8$	& $-146$	\\
		${}^{136}$Xe	& $-29.7$	& $66.8$	& $-12.7$	& $-83.8$	& $-109$	& $-29.8$	& $63.5$	& $-17.2$	& $-111$	\\
		${}^{148}$Nd	& $-32.7$	& $72.8$	& $9.60$	& $-115$	& $-95.9$	& $-31.7$	& $66.8$	& $14.1$	& $-84.4$	\\
		${}^{150}$Nd	& $-35.6$	& $81.1$	& $13.2$	& $-130$	& $-104$	& $-30.2$	& $64.5$	& $16.1$	& $-78.6$	\\
		${}^{154}$Sm	& $-33.7$	& $78.1$	& $13.8$	& $-126$	& $-98.0$	& $-31.8$	& $68.6$	& $18.6$	& $-81.9$	\\
		${}^{160}$Gd	& $-44.6$	& $106$	& $21.5$	& $-172$	& $-129$	& $-41.4$	& $90.8$	& $28.5$	& $-104$	\\
		${}^{198}$Pt	& $-31.9$	& $71.4$	& $12.8$	& $-116$	& $-90.5$	& $-31.9$	& $64.7$	& $17.3$	& $-79.3$	\\
		${}^{232}$Th	& $-44.0$	& $107$	& $24.4$	& $-175$	& $-127$	& $-44.0$	& $98.1$	& $33.0$	& $-109$	\\
		${}^{238}$U		& $-52.5$	& $127$	& $28.7$	& $-208$	& $-151$	& $-52.5$	& $116$	& $38.8$	& $-130$	\\
		\hline
	\end{tabular}
	\caption{As Tab.~\ref{tab:nmecomplight}, but comparing the heavy neutrino exchange NMEs calculated in this work and those given in Table~IV of \cite{Barea:2015kwa}.}
	\label{tab:nmecompheavy}
\end{table}
The values of the NMEs in the present work are compared with those in \cite{Barea:2015kwa} in Table~\ref{tab:nmecomplight} for light neutrino exchange and in Table~\ref{tab:nmecompheavy} for heavy neutrino exchange. Comparing the old and new values of the $F$, $GT$ and $T$ matrix elements one can see that the effect of improved single particle energies is sizeable in $^{76}$Ge, $^{82}$Se, $^{96}$Zr, $^{150}$Nd and small otherwise. The main difference between the calculation reported in~\cite{Barea:2015kwa} and the present one is in the sign of the tensor matrix element in Eq.~\eqref{eq:lightnunme}. The present derivation gives a sign of the $\nme_T$ term relative to that of $\nme_F$ which is opposite to the one employed in Ref.~\cite{Barea:2015kwa}. This correction has little effect on the standard mass mechanism, for which $\nme_T$ is small, but has considerable effect on the short-range mechanisms. Additionally, one can see that the matrix elements $\nme_F$, $\nme_{GT}$, $\nme_T$ for both light and heavy neutrino exchange are of the same order of magnitude in all elements with $GT$ being the dominant term. This is due to the fact that the individual contributions given in Tables~\ref{tab:nmecomplight} and \ref{tab:nmecompheavy} are all of the same order of magnitude and that the dominant term in $\nme_3$ is $\nme_{GT}^{AA}$. The only difference comes from the sign of the tensor terms, $\nme_T^{\prime AP}$, $\nme_T^{\prime WW}$, $\nme_T^{\prime\prime PP}$, which is different for the p-p and h-h case from the p-h and h-p case.

\begin{figure}[t!]
	\centering
	\includegraphics[clip,width=0.65\textwidth]{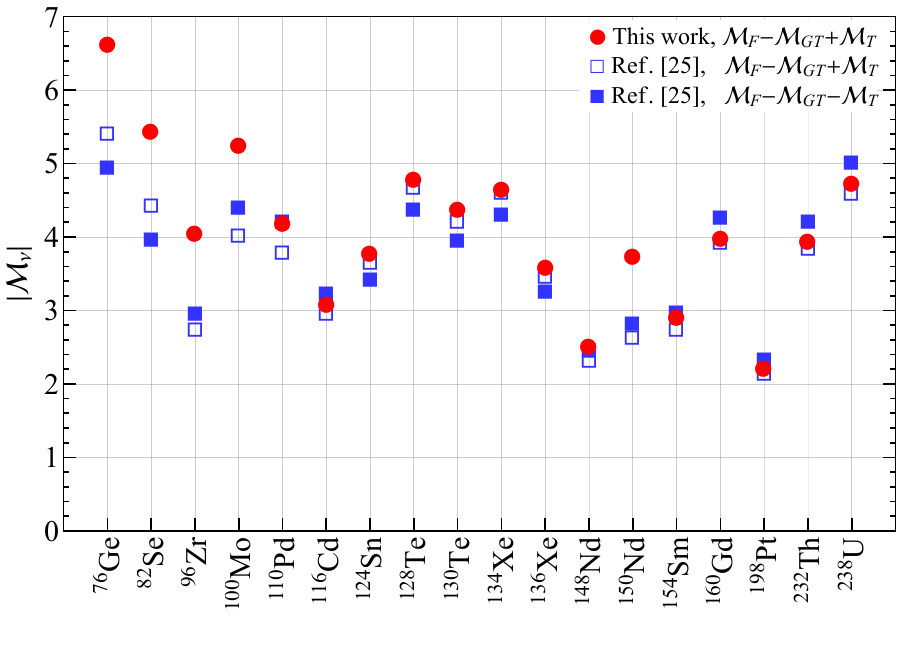}
	\caption{Comparison between the light neutrino exchange IBM-2 NMEs $\nme_\nu$ calculated in this work (red circles) and the ones calculated in~\cite{Barea:2015kwa} (solid blue squares), assuming the quenched value $g_A = 1.0$. We show also the old total NME $\tilde{\nme}^\text{old}_\nu$ incorporating the (old) tensor part but with the correct sign (empty blue squares).}
	\label{fig:nmescomplight}
\end{figure}
\begin{figure}[t!]
	\centering
	\includegraphics[clip,width=0.65\textwidth]{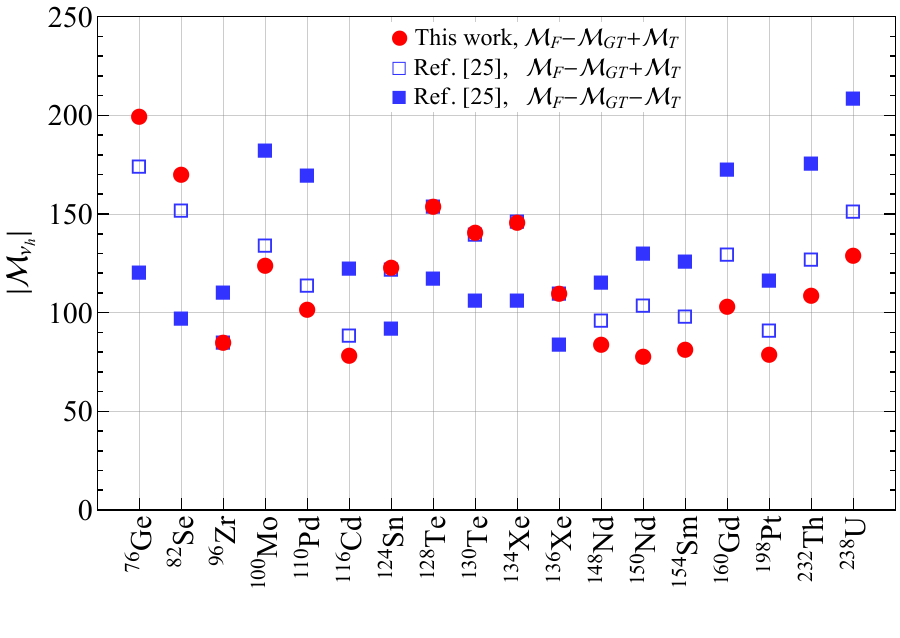}
	\caption{As Fig.~\ref{fig:nmescomplight}, but showing the comparison for the heavy-neutrino-exchange NME $\mathcal{M}_{\nu_h}$.}
	\label{fig:nmescompheavy}
\end{figure}
In Figs.~\ref{fig:nmescomplight} and \ref{fig:nmescompheavy} we, respectively, compare the compound NMEs $\nme_\nu$ and $\nme_{\nu_h}$ in the different calculations: present work (red circles), Ref.~\cite{Barea:2015kwa} (filled blue squares) and Ref.~\cite{Barea:2015kwa} with the correct sign for the tensor term (empty blue squares). This allows disentangling the effect of the new single particle energies from that induced by the sign of the tensor NME. As already mentioned, for light neutrino exchange (Fig.~\ref{fig:nmescomplight}), the sign of the tensor term has relatively little impact, whereas the single particle energies lead to a sizeable increase of $\nme_\nu$ in lighter isotopes. On the other hand, Fig.~\ref{fig:nmescompheavy} demonstrates the strong effect of the sign of the tensor term in essentially all isotopes.

In comparison with calculations other than IBM-2 we note that our improved results for the standard mass mechanism are very similar to those in QRPA in all isotopes~\cite{Simkovic:2007vu}, but still differ from those of the Shell Model~\cite{Menendez:2011zza}. For the short-range mechanisms the obtained numbers are again similar to the QRPA in the case in which both neutron and proton are particle-like (p-p) or hole-like (h-h), while different in the case in which neutrons are hole-like and protons are particle-like or vice versa (p-h and h-p). We also note that, although not discussed here, the main source of uncertainty for the matrix elements $\nme_{\nu_h}$ is the parametrization of the short-range correlations. For example, for $^{76}$Ge, QRPA reports~\cite{Faessler:2011rv} $\nme_{\nu_h} = 32.6$ for Miller-Spencer, $\nme_{\nu_h} = 233$ for Argonne and $\nme_{\nu_h} = 352$ for CD-Bonn parametrization, a factor of ten difference. In the present paper we use the Argonne parametrization and obtain $\nme_{\nu_h} = -200$ (see Table~\ref{tab:nmecompheavy}) in reasonable agreement with the QRPA result, except for the overall sign in Eq.~\eqref{eq:lightnunme}, which, as indicated above, is opposite to that of QRPA.

\subsubsection{Compound NMEs for Short-range Mechanisms}
\begin{table}[t!]
	\centering
	\begin{tabular}{@{\extracolsep{3pt}}r|rrrrrrrr@{}}
		\hline
		Isotope & $\nme_1^{XX}$ & $\nme_1^{XY}$ & $\nme_2^{XX}$ & $\nme_3^{XX}$ & $\nme_3^{XY}$ & $\nme_4^{XX}$ & $\nme_5^{XX}$ & $\nme_5^{XY}$ \\ \hline
		${}^{76}$Ge	& $5300$	& $-5400$	& $-174$	& $-200$	& $99.8$	& $-158$	& $202$	& $-301$ \\
		${}^{82}$Se	& $4030$	& $-4110$	& $-144$	& $-171$	& $83.3$	& $-134$	& $114$	& $-199$ \\
		${}^{96}$Zr	& $8500$	& $-8570$	& $-129$	& $-85.4$	& $57.7$	& $-88.6$	& $832$	& $-904$ \\
		${}^{100}$Mo	& $12400$	& $-12500$	& $-189$	& $-124$	& $83.9$	& $-134$	& $1230$	& $-1340$ \\
		${}^{110}$Pd	& $10400$	& $-10500$	& $-157$	& $-102$	& $69.3$	& $-107$	& $1030$	& $-1120$ \\
		${}^{116}$Cd	& $7420$	& $-7480$	& $-115$	& $-78.8$	& $52.0$	& $-80.4$	& $702$	& $-768$ \\
		${}^{124}$Sn	& $3450$	& $-3520$	& $-106$	& $-124$	& $56.2$	& $-95.3$	& $157$	& $-224$ \\
		${}^{128}$Te	& $4410$	& $-4500$	& $-134$	& $-154$	& $72.0$	& $-130$	& $205$	& $-291$ \\
		${}^{130}$Te	& $4030$	& $-4110$	& $-122$	& $-141$	& $64.4$	& $-109$	& $187$	& $-264$ \\
		${}^{134}$Xe	& $4240$	& $-4320$	& $-127$	& $-146$	& $67.6$	& $-114$	& $196$	& $-277$ \\
		${}^{136}$Xe	& $3210$	& $-3270$	& $-96.1$	& $-111$	& $51.2$	& $-86.0$	& $147$	& $-208$ \\
		${}^{148}$Nd	& $6180$	& $-6240$	& $-106$	& $-84.4$	& $46.2$	& $-77.2$	& $583$	& $-648$ \\
		${}^{150}$Nd	& $6190$	& $-6250$	& $-103$	& $-78.6$	& $45.5$	& $-74.2$	& $591$	& $-652$ \\
		${}^{154}$Sm	& $6780$	& $-6840$	& $-111$	& $-81.9$	& $50.0$	& $-79.4$	& $638$	& $-703$ \\
		${}^{160}$Gd	& $9370$	& $-9450$	& $-149$	& $-104$	& $68.2$	& $-105$	& $886$	& $-970$ \\
		${}^{198}$Pt	& $6720$	& $-6780$	& $-109$	& $-79.3$	& $49.3$	& $-77.9$	& $616$	& $-681$ \\
		${}^{232}$Th	& $10200$	& $-10300$	& $-160$	& $-109$	& $74.0$	& $-112$	& $978$	& $-1070$ \\
		${}^{238}$U	& $12200$	& $-12300$	& $-191$	& $-130$	& $88.1$	& $-134$	& $1160$	& $-1260$ \\
		\hline
	\end{tabular}
	\caption{Compound NMEs $\nme_1$ to $\nme_5$ for all distinct quark current chirality combinations, calculated using the quenched value $g_A = 1.0$.}
	\label{tab:nme-compound}
\end{table}
Concluding our calculation of NMEs, in Tab.~\ref{tab:nme-compound} we summarize for clarity the numeric values of the compound NMEs $\nme_1$ to $\nme_5$ relevant for short-range $0\nu\beta\beta$ contributions, as defined in Eqs.~\eqref{eq:nme1} - \eqref{eq:nme5}. They are listed for all distinct combinations of quark chiralities and they are calculated using the quenched value $g_A = 1.0$. The values of light neutrino exchange NMEs $\nme_\nu$ calculated in our approach are shown in the last column of Table~\ref{tab:nmecomplight}.

We note that the NMEs $\mathcal{M}_1$ and $\mathcal{M}_5$ are generally enhanced due to the large pseudo-scalar charges $g_P$ and $g_{P'}$, though in $\mathcal{M}_5$ this is often compensated by the suppressed component NMEs. The enhancement is especially strong in isotopes with the same sign for $\mathcal{M'}^{P'P'}_{GT}$ and $\mathcal{M'}^{P'P'}_T$ which arises from particle-particle versus particle-hole configurations of the nucleons. This, along with the large PSFs discussed below, makes $^{100}$Mo an ideal isotope to probe the corresponding mechanisms from a theoretical point of view.

\section{Leptonic Phase Space and Decay Rate}
\label{sec:PSFs}

\subsection{Leptonic Matrix Elements}

Besides the NMEs, the calculation of the $0\nu\beta\beta$ decay rate requires the calculation of the so called leptonic phase space factors. Here, we follow the numerical approach of Ref.~\cite{Kotila:2012zza}. Because of Pauli-blocking of the inner states, the nucleons are expected to decay largely at the surface of the nucleus, which means that the electron wave function can be approximated by its value at the nuclear radius $r = R_A$.

Since we are interested only in $0^+ \to 0^+$ $0\nu\beta\beta$ transitions, nucleon operators of a certain parity must be combined with partial leptonic wave functions of the same parity. Specifically, the parity-even operators will be accompanied by $S_{1/2} - S_{1/2}$ and $P_{1/2} - P_{1/2}$ electron wave functions, while parity-odd ones will go together with the $S_{1/2} - P_{1/2}$ combination of wave functions. In this study we restrict ourselves only to the $S_{1/2} - S_{1/2}$ approximation, which allows us to drop the parity-odd nucleon operators from our calculation. The leptonic squared matrix elements for $S_{1/2} - S_{1/2}$ wave functions, summed over the electron spins $s_1$ and $s_2$, then read \cite{Graf:2018ozy}
\begin{align} 
\label{eq:PSFs1}
	\sum_{s_1,s_2}(\bar{e}_1(1+\gamma_5)e_2^c)
	(\bar{e}_1(1\pm\gamma_5)e_2^c)^\dagger
	(1-P_{e_1 e_2})/2 &= 
	f_{11\pm}^{(0)} + f_{11\pm}^{(1)}\,\hat{\vecl{p}}_1\cdot\hat{\vecl{p}}_2, \\
	\sum_{s_1,s_2}(\bar{e}_1\gamma_\mu \gamma_5 e_2^c)
	(\bar{e}_1\gamma_\nu \gamma_5 e_2^c)^\dagger 
	(1-P_{e_1e_2})/2 &= 
	\frac{1}{16}\left(f^{(0)}_{66} 
	+ f^{(1)}_{66}\,\vecl{\op p}_1\cdot\vecl{\op p}_2 \right), 
	\, (\mu, \nu = 0), \\
\label{eq:PSFs3}
	\sum_{s_1,s_2}(\bar{e}_1\gamma_\mu\gamma_5 e_2^c)
	(\bar{e}_1(1\pm\gamma_5)e_2^c)^\dagger
	(1-P_{e_1e_2})/2 &= 
	\mp \frac{1}{4}f_{16}^{(0)}, \quad\qquad\qquad\qquad\,\, (\mu = 0),
\end{align}
where the scalar product between the asymptotic electron momentum vectors is parametrized as $\vecl{\op p}_1\cdot\vecl{\op p}_2 = \cos\theta$ with the opening angle $0 \leq \theta \leq \pi$. The term $(1-P_{e_1 e_2})$ indicates that the matrix element is anti-symmetrized over the electrons. The result for Eq.~\eqref{eq:PSFs1} when both currents are left-handed is the same as the one shown when both currents are right-handed. Since we are interested only in $0^+ \to 0^+$ transitions in the $S_{1/2} - S_{1/2}$ approximation, we omit phase space factors corresponding to $\mu = j$ or $\nu = j$. Further, in Eqs.~\eqref{eq:PSFs1} - \eqref{eq:PSFs3} we have used the quantities $f^{(0,1)}_{ij} \equiv f^{(0,1)}_{ij}(E_1, E_2)$ defined as
\begin{align}
\label{eq:PSFscal}
	f_{11\pm}^{(0)} &=  \pm|f^{-1-1}|^2\pm|f_{11}|^2+|f{^{-1}}_1|^2+|{f_1}^{-1}|^2,               &
	f_{11\pm}^{(1)} &= -2\left({f^{-1}}_1{f_1}^{-1}\pm f^{-1-1}f_{11}\right), \\
\label{eq:PSFcal2}
	f_{66}^{(0)}    &= 16 \left(|f^{-1-1}|^2+|f_{11}|^2\right),               &
	f_{66}^{(1)}    &= 32 f^{-1-1}f_{11},                                     \\
\label{eq:PSFcal3}
	f_{16}^{(0)}    &=  4 \left(|f_{11}|^2-|f^{-1-1}|^2\right),               &
	f_{16}^{(1)}    &=  0.
\end{align}
Here, the definitions in terms of electron wave functions $g_{-1}(E)$ and $f_1(E)$ evaluated at the nuclear surface apply, $f^{-1-1} = g_{-1}(E_1)g_{-1}(E_2)$, $f_{11} = f_1(E_1)f_1(E_2)$, ${f^{-1}}_1 = g_{-1}(E_1)f_1(E_2)$, ${f_1}^{-1} = f_1(E_1)g_{-1}(E_2)$. When compared to Refs.~\cite{Pas:2000vn} and \cite{Tomoda:1990rs} our results agree but we also introduce additional factors $f_{11-}^{(0,1)}$ which appear as a result of the interference between the left- and right-handed scalar electron currents. In fact, these terms are not independent of the others as they can be expressed as $f_{11-}^{(0,1)} = f_{11+}^{(0,1)} - \frac{1}{8}f_{66\pm}^{(0,1)}$. 

In determining the squared leptonic matrix elements, we numerically calculate the electron wave functions according to \cite{Kotila:2012zza}, taking into account the finite nuclear size and electron cloud screening corrections. 

\subsection{Differential Decay Distributions}
\label{sec:diff-distributions}

The NMEs presented in the previous section and the squared leptonic matrix elements shown in Eqs.~\eqref{eq:PSFs1} - \eqref{eq:PSFs3} can now be combined to calculate the rate of $0^+ \to 0^+$ $0\nu\beta\beta$ decay. The fully differential rate is expressed as~\cite{Doi:1981, Doi:1983, Tomoda:1990rs}
\begin{align}
\label{eq:differentialrate}
	\frac{d^2\Gamma}{dE_1 d\!\cos\theta} = 
	C\, w(E_1) \left(a(E_1) + b(E_1)\cos\theta\right),
\end{align}
with
\begin{align}
	C      = \frac{G_F^4 \cos^4\theta_C m_e^2}{16\pi^5},\qquad
	w(E_1) = E_1 E_2 p_1 p_2, 
\end{align}
and where $E_2$, $p_1 = \sqrt{E_1^2 - m_e^2}$ and $p_2 = \sqrt{E_2^2 - m_e^2}$ are understood to be functions of $E_1$ due to overall energy conservation, $E_2 = Q_{\beta\beta} + 2m_e - E_1$. Here, $Q_{\beta\beta}$ is the so called double beta decay $Q$ value of the given isotope, i.e. the kinetic energy release of the electrons.

The coefficients $a(E_1)$ and $b(E_1)$ in Eq.~\eqref{eq:differentialrate} are, respectively, given by
\begin{align}
\label{eq:aterm}
	a(E_1) &=  f_{11+}^{(0)}
	   \left|\sum_{I=1}^3\epsilon_I^L\nme_I+\epsilon_{\nu}\nme_{\nu}\right|^2 
	 + f_{11+}^{(0)}\left|\sum_{I=1}^3\epsilon_I^R\nme_I\right|^2 
	 + \frac{1}{16} f_{66}^{(0)} 
	   \left|\sum_{I=4}^5\epsilon_I\nme_I\right|^2                  \nonumber\\
	&+\phantom{\frac{1}{4}} f_{11-}^{(0)}\times 2\,
	   \text{Re}\left[\left(\sum_{I=1}^3\epsilon_I^L\nme_I 
	   + \epsilon_\nu\nme_\nu\right)
	   \left(\sum_{I=1}^3\epsilon_I^R\nme_I\right)^*\right]         \nonumber\\
	& + \frac{1}{4} f_{16\phantom{-}}^{(0)}\times 2\,
	    \text{Re}\left[\left(\sum_{I=1}^3\epsilon_I^L\nme_I 
	   - \sum_{I=1}^3\epsilon_I^R\nme_I +\epsilon_\nu\nme_\nu\right)
       \left(\sum_{I=4}^5\epsilon_I\nme_I\right)^*\right],
\end{align}
and
\begin{align}
\label{eq:bterm}
	b(E_1) &= f_{11+}^{(1)}
	     \left|\sum_{I=1}^3\epsilon_I^L\nme_I + \epsilon_\nu\nme_\nu\right|^2 
	 + f_{11+}^{(1)}\left|\sum_{I=1}^3\epsilon_I^R\nme_I\right|^2  
	+ \frac{1}{16} f_{66}^{(1)}\left|\sum_{I=4}^5\epsilon_I\nme_I\right|^2.
\end{align}
These expressions are valid under the presence of any combination of short-range mechanisms, with associated particle coefficients $\epsilon_I$, and the standard light neutrino exchange where the coefficient $\epsilon_\nu$ is defined by $\epsilon_\nu = m_{\beta\beta} / m_e$. Here, $m_{\beta\beta}$ is the usual effective $0\nu\beta\beta$ mass given in Eq.~\eqref{eq:bbmass}. The NMEs $M_I$ and $M_\nu$ are defined in Eqs.~\eqref{eq:nme1} - \eqref{eq:nme5} and \eqref{eq:nme_nu}, respectively, where the summations are over the different short-range current types $i = 1,\dots,5$ including their different chiralities, $I = (i,XYZ)$ with $X,Y,Z \in \{L,R\}$. A distinction is made between short-range mechanisms of type $i = 1,2,3$ for which the scalar current is left-handed or right-handed. This is indicated by $\epsilon_I^L$ and $\epsilon_I^R$, respectively, where the sum is only over the corresponding terms. This distinction represents the interference behaviour between terms of different electron chiralities. For example, the first term on the right-hand side of Eq.~\eqref{eq:aterm} describes the contributions of and the interference among the $i = 1,2,3$ short-range mechanisms $\epsilon_i^{XYL}$ with left-handed electron chiralities (but including all quark current chiralities) and that of the standard light neutrino exchange. Likewise the second term describes the contributions of $i = 1,2,3$ short-range mechanisms $\epsilon_I^R$ with right-handed electron chiralities including their cross interference, whereas the third term contains the interference between these two classes, $(\epsilon_{1,2,3}^L, \epsilon_\nu)$ with $\epsilon_{1,2,3}^R$. The other terms appearing in Eqs.~\eqref{eq:aterm} and \eqref{eq:bterm} can be understood in a similar way where the electron-energy dependent factors $f_{ij}^{(0,1)} \equiv f_{ij}^{(0,1)}(E_1)$ describe the correctly associated squared lepton matrix elements as defined in Eqs.~\eqref{eq:PSFscal}-\eqref{eq:PSFcal3}. Note that the interference term between short-range operators of type $i = 1,2,3$ and $i = 4,5$ vanishes in $b(E_1)$ due to $f_{16}^{(1)} = 0$ in Eq.~\eqref{eq:PSFcal3}.

The fully differential decay rate Eq.~\eqref{eq:differentialrate} contains the complete kinematic information and integrating over the whole electron phase space will yield the total rate. Of experimental interest are the distribution over the single electron energy and the angular correlation. The single electron energy distribution is simply given by
\begin{align}
\label{eq:energydistro}
	\frac{d\Gamma}{dE_1} = 2C w(E_1)a(E_1),
\end{align}
and the energy-dependent angular correlation is introduced as $\alpha(E_1) = b(E_1) / a(E_1).$ The latter has the property $- 1 < \alpha(E_1) < +1$ and as it appears in front of the $\cos\theta$ term, it describes the likelihood for the electrons to be emitted back-to-back ($\alpha(E_1) \gtrsim -1$), collinearly ($\alpha(E_1) \lesssim +1$) or isotropically ($\alpha(E_1) \approx 0$). Defining
\begin{gather} 
	A = \int_{m_e}^{Q_{\beta\beta} + m_e} dE_1 w(E_1) a(E_1), \qquad
	B = \int_{m_e}^{Q_{\beta\beta} + m_e} dE_1 w(E_1) b(E_1),
\end{gather}
and their ratio $K = B/A$, the angular distribution reads
\begin{align}
	\frac{d\Gamma}{d\!\cos\theta} = \frac{\Gamma}{2}\left(1 + K\cos\theta\right).
\end{align}

With the given information we determine the single electron distribution $d\Gamma/dE_1$ and the angular correlation $\alpha(E_1)$ for the three relevant phase space factors that occur for short-range operators: $f_{11+}^{(0,1)}$ (for mechanisms $i=1,2,3$ with a scalar electron current), $f_{66}^{(0,1)}$ (for mechanisms $i=4,5$ with an axial-vector electron current), $f_{16}^{(0)}$ (for interference between the two classes) and $f_{11-}^{(0)}$ (for interference between $i = 1,2,3$ of different lepton chirality). As already noted, $f_{16}^{(1)}$ vanishes, as does $f_{11-}^{(1)}$. The electron phase space distributions $f_{11+}^{(0,1)}$ also apply for the standard mass mechanism, calculated in the closure approximation.

\begin{figure}[t!]
	\centering
	\includegraphics[width=0.48\textwidth]{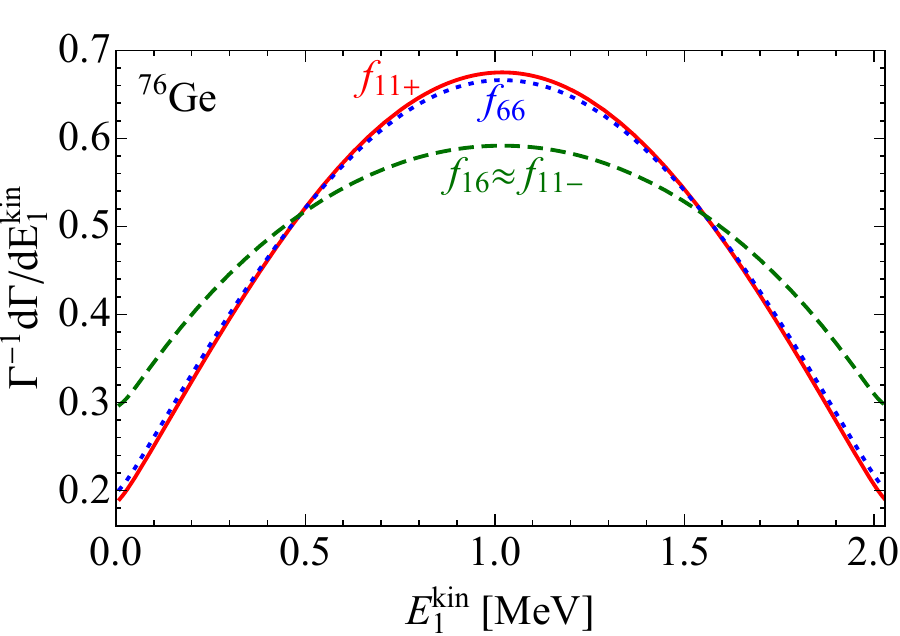}
	\includegraphics[width=0.495\textwidth]{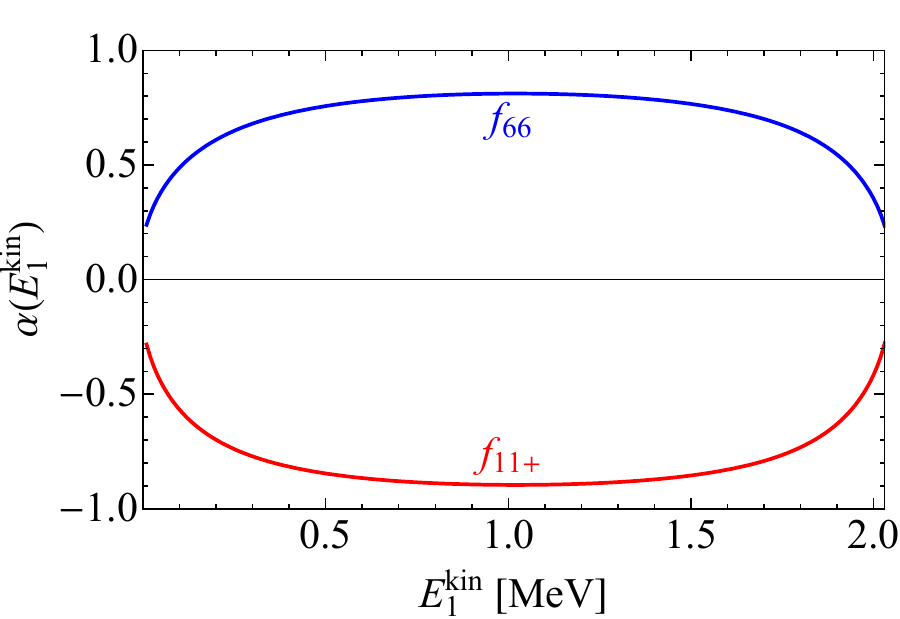}
	\caption{Normalized single electron energy distributions $\Gamma^{-1}d\Gamma/dE^\text{kin}_1$ (left) and angular correlation $\alpha(E^\text{kin}_1)$ (right) for $^{76}$Ge as a function of the kinetic energy $E^\text{kin}_1 = E_1 - m_e$. Shown are the phase space factors for Eq.~\eqref{eq:aterm} in the former and for Eq.~\eqref{eq:bterm} in the latter.}
	\label{fig:distros}
\end{figure}

The resulting single energy distribution and angular correlation were already presented in Ref.~\cite{Graf:2018ozy} for several isotopes, but in Fig.~\ref{fig:distros}~(left) we illustrate the normalized single energy distributions for $^{76}$Ge as functions of the kinetic energy $E_1^\text{kin} = E_1 - m_e$ of one of the electrons, i.e. the range is from zero up to $Q_{\beta\beta}$ value. As can be seen, the term $f^{(0)}_{11-}$ produces an energy distribution virtually indistinguishable from that of $f^{(0)}_{16}$. All mechanisms produce a hill-like shaped energy distribution and observing the single energy spectrum is not expected to help distinguish between the standard mass mechanism (corresponding to $f^{(0)}_{11+}$) and any of the short-range mechanisms. The angular correlation $\alpha(E^\text{kin}_1)$, shown in Fig.~\ref{fig:distros}~(right), can distinguish between different mechanisms, namely short-range mechanisms of type $i = 4,5$ produce electrons that are emitted collinearly whereas for $i = 1,2,3$ and the standard mass mechanism, they are dominantly back-to-back. As mentioned, the factors $f_{16}^{(1)}$ and $f_{11-}^{(1)}$ vanish. There is therefore no change of the angular correlation due to interference and the angular correlation is an incoherent sum over contributions.

\subsection{Total Decay Rate}
\label{sec:total-rate}

Finally, we can integrate over the whole electron phase space to determine the total decay rate $\Gamma$ and the decay half life $T_{1/2}$,
\begin{align}
\label{eq:totalrate}
	\Gamma = \frac{\ln 2}{T_{1/2}} 
	       = 2 C\int_{m_e}^{Q_{\beta\beta} + m_e} dE_1 w(E_1) a(E_1).
\end{align}
To facilitate calculation of the total rate under the presence of one or more mechanisms, we define the integrated PSFs \cite{Kotila:2012zza}
\begin{align}
\label{eq:Gs}
	G^{(0,1)}_{ij} = \frac{2C}{\ln 2}\frac{g^{(0,1)}_{ij}}{4 R_A^2}
				   \int^{Q_{\beta\beta} + m_e}_{m_e} dE_1 
				   w(E_1) f^{(0,1)}_{ij}(E_1,Q_{\beta\beta} + 2m_e - E_1),
\end{align}
with $g^{(0,1)}_{11\pm} = 1$, $g^{(0,1)}_{66} = 1/16$, $g^{(0)}_{16} = 1/4$, $g^{(1)}_{16} = 0$. The factor $1/R_A^2$ has been introduced to conform with our convention where the NMEs are made dimensionless by multiplying with the nuclear radius $R_A$. The numerical values of the PSFs $G_{ij}^{(0,1)}$ are given in Table~\ref{tab:psfs}, in units of $10^{-15}\,\text{yr}^{-1}$. As mentioned earlier, both $G_{16}^{(1)}$ and $G_{11-}^{(1)}$ vanish, corresponding to the absence of interference in the angular part $b(E_1)$.
\begin{table}
	\centering
	\begin{tabular}{r|rrrrrr}
		\hline
		\multirow{2}{*}{Isotope} & $G_{11+}^{(0)}$ & $G_{11-}^{(0)}$ &
		$G_{66}^{(0)}$ & $G_{16}^{(0)}$ & $G_{11}^{(1)}$ & $G_{66}^{(1)}$ \\
		& \multicolumn{6}{c}{[$10^{-15}$~yr$^{-1}$]} \\
		\hline
		${}^{76}$Ge	& $2.360$	& $-0.280$	& $1.320$	& $0.870$	& $-1.954$	& $0.977$ \\
		${}^{82}$Se	& $10.19$	& $-0.712$	& $5.450$	& $2.925$	& $-9.079$	& $4.539$ \\
		${}^{96}$Zr	& $20.58$	& $-1.190$	& $10.88$	& $5.403$	& $-21.62$	& $9.335$ \\
		${}^{100}$Mo	& $15.91$	& $-1.053$	& $8.482$	& $4.456$	& $-14.25$	& $7.125$ \\
		${}^{110}$Pd	& $4.807$	& $-0.541$	& $2.674$	& $1.730$	& $-4.014$	& $2.007$ \\
		${}^{116}$Cd	& $16.69$	& $-1.187$	& $8.938$	& $4.843$	& $-19.37$	& $7.414$ \\
		${}^{124}$Sn	& $9.028$	& $-0.843$	& $4.935$	& $2.976$	& $-7.760$	& $3.880$ \\
		${}^{128}$Te	& $0.585$	& $-0.156$	& $0.371$	& $0.313$	& $-0.390$	& $0.195$ \\
		${}^{130}$Te	& $14.20$	& $-1.142$	& $7.672$	& $4.367$	& $-12.45$	& $6.223$ \\
		${}^{134}$Xe	& $0.597$	& $-0.164$	& $0.380$	& $0.323$	& $-0.394$	& $0.197$ \\
		${}^{136}$Xe	& $14.56$	& $-1.197$	& $7.876$	& $4.524$	& $-12.72$	& $6.361$ \\
		${}^{148}$Nd	& $10.07$	& $-1.084$	& $5.579$	& $3.548$	& $-14.19$	& $4.246$ \\
		${}^{150}$Nd	& $62.98$	& $-3.125$	& $33.05$	& $15.44$	& $-57.83$	& $28.91$ \\
		${}^{154}$Sm	& $3.005$	& $-0.539$	& $1.772$	& $1.338$	& $-2.291$	& $1.145$ \\
		${}^{160}$Gd	& $9.526$	& $-1.129$	& $5.321$	& $3.506$	& $-7.917$	& $3.958$ \\
		${}^{198}$Pt	& $7.513$	& $-1.305$	& $4.409$	& $3.278$	& $-5.844$	& $2.922$ \\
		${}^{232}$Th	& $13.87$	& $-2.419$	& $8.144$	& $6.019$	& $-10.92$	& $5.457$ \\
		${}^{238}$U	& $33.45$	& $-4.176$	& $18.81$	& $12.46$	& $-28.02$	& $14.01$ \\
		\hline
	\end{tabular}
	\caption{PSFs in units of $10^{-15}~\text{yr}^{-1}$ used in the calculation of the total decay rate for the standard light neutrino exchange and short-range mechanisms. The PSFs corresponding to $f_{11-}^{(1)}$ and $f_{16}^{(1)}$ vanish.}
	\label{tab:psfs}
\end{table}

With the above PSFs, the inverse $0\nu\beta\beta$ decay half-life can be written
\begin{align}
\label{eq:totalhalflife}
	T_{1/2}^{-1} &= 
	   G_{11+}^{(0)}
	   \left|\sum_{I=1}^3\epsilon_I^L\nme_I + \epsilon_\nu\nme_\nu\right|^2 
	 + G_{11+}^{(0)}\left|\sum_{I=1}^3\epsilon_I^R\nme_I\right|^2 
	 + G_{66}^{(0)}\left|\sum_{I=4}^5\epsilon_I\nme_I\right|^2   \nonumber\\
	&+ G_{11-}^{(0)}\times 2\,
	   \text{Re}\left[\left(\sum_{I=1}^3\epsilon_I^L\nme_I 
	 + \epsilon_\nu\nme_\nu\right)
	   \left(\sum_{I=1}^3\epsilon_I^R\nme_I\right)^*\right]         \nonumber\\
	&+ G_{16\phantom{-}}^{(0)} \times 2\,
	   \text{Re}\left[\left(\sum_{I=1}^3\epsilon_I^L\nme_I 
	 - \sum_{I=1}^3\epsilon_I^R\nme_I + \epsilon_\nu\nme_\nu\right)
	   \left(\sum_{I=4}^5\epsilon_I\nme_I\right)^*\right].
\end{align}
Expressed in this way, the inverse half life now only depends on the NMEs in Tables~\ref{tab:nme-compound} and \ref{tab:nmecomplight}~(last column), the PSFs in Table~\ref{tab:psfs} and the coefficients $\epsilon_I$, $\epsilon_\nu = m_{\beta\beta}/m_e$ encapsulating the particle physics aspects.

\section{Results}
\label{sec:numresults}

\subsection{Bounds on the Effective Neutrino Mass}
\begin{table}[t!]
	\centering
	\begin{tabular}{rcc|c|rrrrrrrr}
		\hline
	    \multirow{2}{*}{Isotope}& \multirow{2}{*}{$T_{1/2}^\text{exp}$~[yr]}& &
		$|m_{\beta\beta}|$ &
		$|\epsilon_1^{XX}|$ & $|\epsilon_1^{XY}|$  & 
		$|\epsilon_2^{XX}|$                        & 
		$|\epsilon_3^{XX}|$ & $|\epsilon_3^{XY}|$  & 
		$|\,\epsilon_4\,|$               & 
		$|\epsilon_5^{XX}|$ & $|\epsilon_5^{XY}|$  \\
		& & & [meV] & \multicolumn{8}{c}{$[10^{-10}]$} \\
		\hline
		$\prescript{76}{}{\text{Ge}}$  & $1.8\times 10^{26}$ & \cite{Agostini:2020xta} & 
		$118$	& $2.90$	& $2.84$	& $88.4$	& $77.1$	& $154$	& $130$	& $102$	& $68.1$	\\
		$\prescript{82}{}{\text{Se}}$  & $2.4\times 10^{24}$ & \cite{PhysRevLett.120.232502} 	& $599$	& $15.9$	& $15.5$	& $445$	& $375$	& $768$	& $654$	& $764$	& $440$	\\
		$\prescript{96}{}{\text{Zr}}$  & $9.2\times 10^{21}$ & \cite{ARGYRIADES2010168} 		& $9130$	& $85.5$	& $84.8$	& $5640$	& $8510$	& $12600$	& $11300$	& $1200$	& $1110$	\\
		$\prescript{100}{}{\text{Mo}}$ & $1.1\times 10^{24}$ & \cite{PhysRevD.92.072011} 		& $733$	& $6.10$	& $6.04$	& $401$	& $608$	& $901$	& $774$	& $84.1$	& $77.5$	\\
		$\prescript{116}{}{\text{Cd}}$ & $2.2\times 10^{23}$ & \cite{PhysRevD.98.092007} 		& $2720$	& $22.3$	& $22.1$	& $1430$	& $2090$	& $3170$	& $2800$	& $321$	& $294$	\\
		$\prescript{128}{}{\text{Te}}$ & $1.1\times 10^{23}$ & \cite{Arnaboldi:2002te}			& $13300$	& $283$	& $277$	& $9300$	& $8080$	& $17300$	& $12100$	& $7630$	& $5390$	\\
		$\prescript{130}{}{\text{Te}}$ & $3.2\times 10^{25}$ & \cite{collaboration2019improved}	& $252$	& $5.38$	& $5.27$	& $178$	& $153$	& $336$	& $270$	& $158$	& $112$	\\
		$\prescript{136}{}{\text{Xe}}$ & $1.1\times 10^{26}$ & \cite{PhysRevLett.117.082503} 	& $114$	& $2.50$	& $2.45$	& $83.4$	& $72.5$	& $157$	& $127$	& $74$	& $52.4$	\\
		$\prescript{150}{}{\text{Nd}}$ & $2.0\times 10^{22}$ & \cite{PhysRevD.94.072003} 		& $3830$	& $45.5$	& $45.1$	& $2730$	& $3590$	& $6190$	& $5240$	& $659$	& $596$	\\
		\hline
	\end{tabular}
	\caption{Upper limits on the effective $0\nu\beta\beta$ mass $|m_{\beta\beta}|$ and the short-range $\epsilon_I$ couplings in units of $10^{-10}$ from current experimental bounds $T_{1/2}^\text{exp}$ at 90\%~CL, assuming a single contribution at a time and $g_A = 1.0$. The chiralities of the involved quark currents are specified: The label $XX$ stands for the case when both chiralities are the same, $XX = RR, LL$ and $XY$ applies if the chiralities are different, $XY = RL,LR$. The limit on $\epsilon_4$ applies for all chirality combinations.}
	\label{tab:bounds}
\end{table}
With $\epsilon_\nu = m_{\beta\beta}/m_e$ and the other short-range $\epsilon_I$ set to zero, Eq.~\eqref{eq:totalhalflife} simplifies to the well know formula for light neutrino exchange,
\begin{align}
	T_{1/2}^{-1} = \frac{|m_{\beta\beta}|^2}{m_e^2} G_{11+}^{(0)} |\mathcal{M}_\nu|^2.
\end{align}
Using the updated NME values for the light neutrino exchange mechanism shown in Tab.~\ref{tab:nmecomplight}~(last column) we can set new limits on the effective $0\nu\beta\beta$ mass $|m_{\beta\beta}|$. For isotopes with existing experimental bounds on the $0\nu\beta\beta$ decay rate, the resulting limits at 90\% CL are summarized in Table~\ref{tab:bounds}. As mentioned, the axial coupling is set to $g_A = 1.0$. Generally, the limits have improved compared to the previous analysis \cite{Barea:2015kwa}. This is a consequence of the better experimental limits for $^{76}$Ge, $^{82}$Se, $^{130}$Te and $^{136}$Xe as well as of the updated single particle energies in the NMEs for $^{76}$Ge, $^{82}$Se, $^{96}$Zr and $^{150}$Nd.

\begin{figure}[t!]
	\centering
	\includegraphics[width=0.6\textwidth]{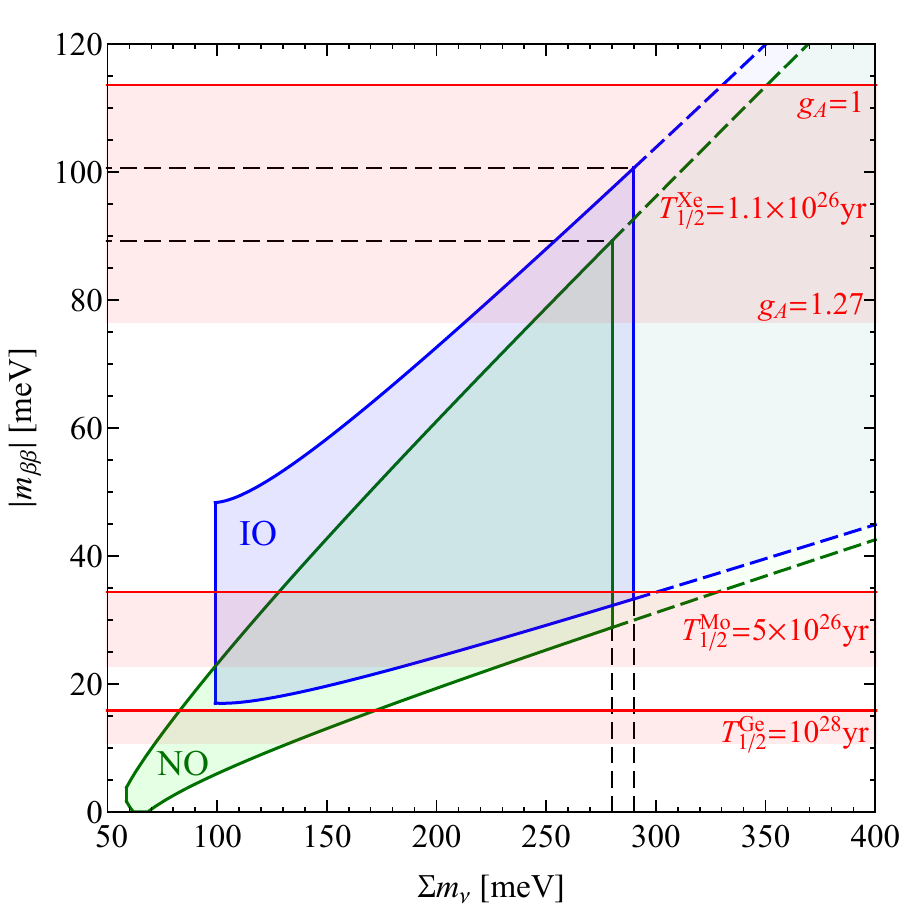}
	\caption{Relation between the $0\nu\beta\beta$ mass $|m_{\beta\beta}|$ and the sum of neutrino masses $\Sigma m_\nu$ for normally (NO) and inversely ordered (IO) neutrinos with the oscillation parameters fixed to the current best fit values. The dark shaded regions denote the parameter space allowed by the limits on $\Sigma m_\nu$ at 95\% CL from cosmological searches. The horizontal bars indicate the current upper limit on $|m_{\beta\beta}|$ and future sensitivities of $0\nu\beta\beta$ decay searches with an unquenched ($g_A = 1.27$, bottom edge) and quenched ($g_A = 1.0$, top edge) value of the axial coupling.}
	\label{fig:massplot}
\end{figure}
In Fig.~\ref{fig:massplot}, we compare the existing limit and future sensitivities in a plot correlating the $0\nu\beta\beta$ mass $|m_{\beta\beta}|$ with the sum of neutrino masses $\Sigma m_\nu = m_{\nu_1} + m_{\nu_2} + m_{\nu_3}$ for the standard picture of three active neutrinos. The shaded regions indicate, as usual, the allowed parameter space for normally (NO) and inversely (IO) ordered neutrino spectra by varying over the Majorana $CP$ phases, where we take the best fit values of the oscillation angles and mass-squared differences as given in \cite{deSalas:2020pgw}. Using our NMEs, the currently best limit is set by the KamLAND-Zen collaboration $T_{1/2}(^{136}\text{Xe}) > 1.1\times 10^{26}$~yr \cite{PhysRevLett.117.082503} resulting in $|m_{\beta\beta}| < 114$~meV at 90\%~CL for $g_A = 1.0$. The recent final result from GERDA with $T_{1/2}(^{76}\text{Ge}) > 1.8\times 10^{26}$~yr \cite{Agostini:2020xta} corresponds to an essentially equal limit of $|m_{\beta\beta}| < 118$~meV at 90\%~CL. In Fig.~\ref{fig:massplot} we also illustrate the corresponding limit assuming no quenching with $g_A = 1.27$, giving $|m_{\beta\beta}| < 76$~meV. In addition to the current limit we also show two examples of prospective sensitivities $T_{1/2}(^{100}\text{Mo}) = 5\times 10^{26}$~yr expected at AMoRE-II \cite{Alenkov:2019jis} and $T_{1/2}(^{76}\text{Ge}) = 10^{28}$~yr for LEGEND-1000 \cite{Zsigmond:2020bfx}. The latter will probe the full IO regime and a large chunk of the NO regime.

Neutrino masses are also probed by the cosmological effect of the relic neutrino background on the cosmic microwave background and the structure of the universe. Current observations are compatible with no effect arising from neutrino masses setting stringent limits on $\Sigma m_\nu$ down to $\Sigma m_\nu < 150$~meV at 90\%~CL \cite{RoyChoudhury:2019hls}. The limit generally depends on the neutrino ordering due to different priors in the statistical analysis and it is affected by the choice of the astrophysical data. It can also be weakened if an underlying cosmological model other than the standard minimal $\Lambda$CDM is used. In Fig.~\ref{fig:massplot} we show the most conservative limits arising from a choice of cosmological models surveyed in Ref.~\cite{RoyChoudhury:2019hls}. Namely, $\Sigma m_\nu < 280$~meV (NO) and $\Sigma m_\nu < 290$~meV (IO) at 95\%~CL arise in the $\Lambda$CDM with non-zero neutrino masses and a free scaling of the so-called weak lensing amplitude $A_\text{lens}$ ($\Lambda$CDM + $\Sigma m_\nu$ + $A_\text{lens}$). These limits correspond to $|m_{\beta\beta}| < 89$~meV (NO) and $|m_{\beta\beta}| < 101$~meV.

\subsection{Bounds on Effective Short-Range Mechanisms}
\label{sec:bounds-eff-couplings}
We can likewise assume that only a single short-range contribution is present by setting all other coefficients to zero and specifically assuming that the standard light neutrino contribution is negligible. Equation~\eqref{eq:totalhalflife} then reduces to 
\begin{align}
	T_{1/2}^{-1} = |\epsilon_I|^2 G_I |\mathcal{M}_I|^2,
\end{align}
with the PSF $G_I$ and NME $\mathcal{M}_I$ depending on the type of contribution. From the current non-observation of $0\nu\beta\beta$ decay we can then set upper limits on the effective $\epsilon_I$ couplings. These are also shown in Table~\ref{tab:bounds}, using our calculated PSFs and NMEs with $g_A = 1.0$. Different chiralities of the quark currents in the operators lead to different bounds as indicated, where $\epsilon_i^{XX}$ denotes the case where the chiralities of the two quark currents are equal, $XX=LL,RR$, whereas $\epsilon_i^{XY}$ indicates that they are different, $XY=RL,LR$. For $\epsilon_2$, the quark currents are required to be equal, cf. Eq.~\eqref{eq:tensor-current}, and for $\epsilon_4$, the bounds do not depend on the choice of the quark chiralities. Considering that a single $\epsilon_I$ contributes at a time, the limits do not depend on the lepton chirality as the corresponding PSFs are independent of it.

Numerically, the best limits for all $\epsilon_I$ are currently derived from the KamLAND-Zen constraint $T_{1/2}(^{136}\text{Xe}) > 1.1\times 10^{26}$~yr, except for $\epsilon_3^{XY}$ where the GERDA constraint is slightly better. In any case, the KamLAND-Zen and GERDA bounds result in essentially equally stringent limits in most of the cases, and they are of the order $10^{-10}$ to $10^{-8}$. For $\epsilon_1$ and $\epsilon_5$, in addition to the improved experimental bounds, the limits are the most stringent due to enhanced values of the nucleon current charges, specifically the large value of the intrinsic pseudoscalar charge $g_{P'}$, see Eq.~\eqref{eq:fpprime}. In case of $\epsilon_3$ the sign of the tensor nuclear matrix elements also plays an important role.

\begin{table}[t!]
	\centering
	\begin{tabular}{rcc|rrrrrrrrr}
		\hline
		\multirow{2}{*}{Isotope}& \multirow{2}{*}{$T_{1/2}^\text{exp}$~[yr]}& &
		$|c_1^{XX}|$ & $|c_1^{XY}|$ & 
		$|c_2^{XX}|$                & 
		$|c_3^{XX}|$ & $|c_3^{XY}|$ & 
		$|c_4^{XX}|$ & $|c_4^{XY}|$ & 
		$|c_5^{XX}|$ & $|c_5^{XY}|$ \\
		& & & \multicolumn{9}{c}{$[10^{-10}]$} \\
		\hline
		$\prescript{76}{}{\text{Ge}}$  & $1.8\times 10^{26}$ & \cite{Agostini:2020xta} 					& 1.42 & 0.948 & 611 & 101 & 177 & 286 & 185 & 50.3 & 22.9 \\
		$\prescript{82}{}{\text{Se}}$  & $2.4\times 10^{24}$ & \cite{PhysRevLett.120.232502}	 	& 7.74 & 5.19 & 2630 & 494 & 882 & 1450 & 934 & 361 & 148 \\
		$\prescript{96}{}{\text{Zr}}$  & $9.2\times 10^{21}$ & \cite{ARGYRIADES2010168} 			& 42.9 & 28.5 & 26900 & 11200 & 14500 & 17300 & 16100 & 616 & 372 \\
		$\prescript{100}{}{\text{Mo}}$ & $1.1\times 10^{24}$ & \cite{PhysRevD.92.072011} 			& 3.06 & 2.03 & 1930 & 800 & 1040 & 1200 & 1110 & 43.1 & 26.1 \\
		$\prescript{116}{}{\text{Cd}}$ & $2.2\times 10^{23}$ & \cite{PhysRevD.98.092007} 			& 11.2 & 7.40 & 7390 & 2760 & 3650 & 4470 & 4000 & 165 & 98.9 \\
		$\prescript{128}{}{\text{Te}}$ & $1.1\times 10^{23}$ & \cite{Arnaboldi:2002te} 				& 139 & 92.6 & 76800 & 10600 & 19900 & 26400 & 17300 & 3820 & 1810 \\
		$\prescript{130}{}{\text{Te}}$ & $3.2\times 10^{25}$ & \cite{collaboration2019improved}		& 2.64 & 1.76 & 1490 & 202 & 387 & 589 & 386 & 79.2 & 37.5 \\
		$\prescript{136}{}{\text{Xe}}$ & $1.1\times 10^{26}$ & \cite{PhysRevLett.117.082503} 		& 1.23 & 0.819 & 717 & 95.4 & 180 & 277 & 181 & 37.2 & 17.6 \\
		$\prescript{150}{}{\text{Nd}}$ & $2.0\times 10^{22}$ & \cite{PhysRevD.94.072003} 			& 22.8 & 15.1 & 18200 & 4720 & 7120 & 8720 & 7490 & 337 & 201 \\
		\hline
	\end{tabular}
	\caption{As Table~\ref{tab:bounds}, but for the short-range couplings $c_I = \epsilon_I(m_W)$ in units of $10^{-10}$, defined at the scale $m_W = 80.4$~GeV and omitting $|m_{\beta\beta}|$. Compared to Table~\ref{tab:bounds}, the limits on $c_4$ depend on whether the quark currents have the same ($XX$) or opposite ($XY$) chirality.}
	\label{tab:boundsQCD}
\end{table}
The limits in Table~\ref{tab:bounds} on the effective couplings apply at the QCD scale $\Lambda_\text{QCD}\approx 1$~GeV. As described in \cite{Graf:2018ozy} following \cite{Mahajan:2013ixa, Gonzalez:2015ady} one can instead define the couplings at the electroweak scale $m_W = 80.4$~GeV and evolve them to $\Lambda_\text{QCD}$, where the appropriate bound can be set employing the experimental limit on the $0\nu\beta\beta$ decay half life. Because different operators mix radiatively, a single contribution at $m_W$ may generally induce several contributions at $\Lambda_\text{QCD}$. The limits obtained in this way can be compared more directly with constraints derived from collider experiments. The resulting bounds on the couplings $c_I = \epsilon_I(m_W)$ at $m_W$, including QCD running effects, are displayed in Table~\ref{tab:boundsQCD}. Note that the limit on $|\epsilon_4|$ splits into two different values $|c_4^{XX}|$ and $|c_4^{XY}|$, since the different quark current chiralities affect the running. Numerically, the limits can be weaker or stronger than those at $\Lambda_\text{QCD}$ due to the overlapping effect of the QCD corrections and the mixing of operators. The already stringent limits on $\epsilon_1^{XX}$ and $\epsilon_1^{XY}$ improve further at $m_W$ and $c_1^{XY}$ is the most strongly constrained coupling by KamLAND-Zen. On the other hand, the limit on $c_2^{XX}$ is relatively much weaker than that on $\epsilon_2^{XX}$. This is an effect of the renormalization group mixing with $c_1^{XX}$ and partial cancellation with this induced term.

\begin{figure}[t!]
	\centering
	\includegraphics[width=0.8\textwidth]{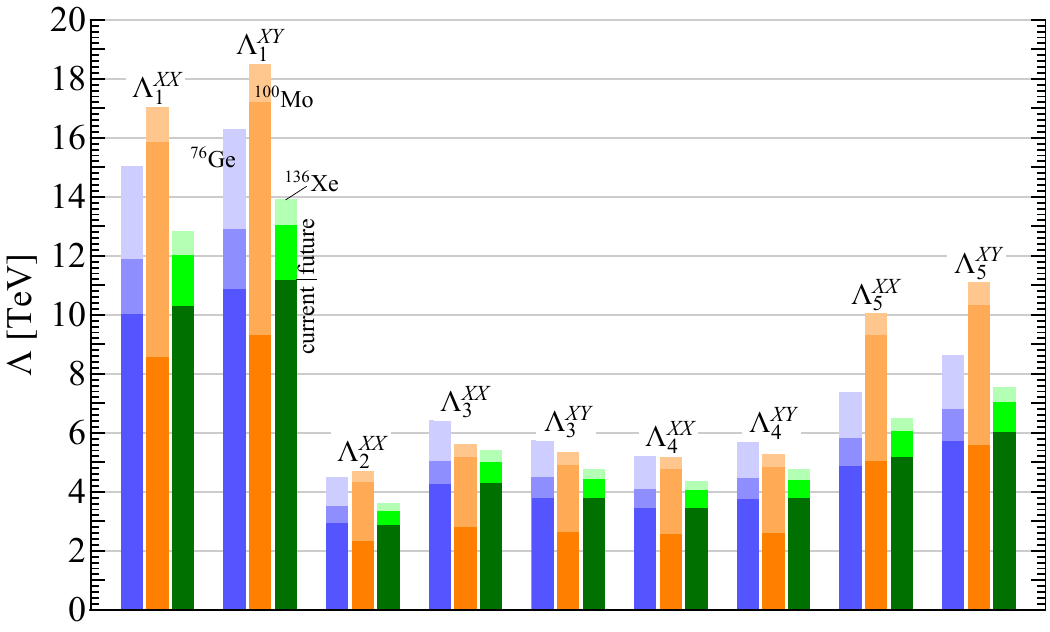}
	\caption{Lower limits on the effective short-range operator scales $\Lambda_I$ defined at $m_W$ and assuming all other contributions are zero. The limits are from the current bounds (dark shade) and two future sensitivities (lighter shades) in $^{76}$Ge at $(1.8, 10, 100)\times 10^{26}$~yr (left, blue), $^{100}$Mo at $(0.011, 5, 10)\times 10^{26}$~yr (middle, orange) and $^{136}$Xe at $(1.1, 5, 9.2)\times 10^{26}$~yr (right, green).}
	\label{fig:operatorscales}
\end{figure}
The effective short-range operator couplings can be interpreted in terms of effective New Physics operator scales $\Lambda_I$ where we simply match
\begin{align}
	\frac{1}{\Lambda_I^5} = \frac{G_F^2\cos^2\theta_C}{2m_p}c_I,
\end{align}
using the couplings $c_I$ defined at the electroweak scale. In Fig.~\ref{fig:operatorscales} we illustrate the current bounds and expected future sensitivities in $^{76}$Ge (blue), $^{100}$Mo (orange) and $^{136}$Xe (green). The coloured bars indicate the lower bound on the given operator scale where the darkest shade corresponds to the current limit and the two increasingly lighter shades represent expected future sensitivities. For the three isotopes, the setups are: (i) $T_{1/2}(^{76}\text{Ge})/(10^{26}$~yr) = 1.8 (GERDA \cite{Agostini:2020xta}, current), 10 (LEGEND-200 \cite{Zsigmond:2020bfx}), 100 (LEGEND-1000 \cite{Zsigmond:2020bfx}); (ii) $T_{1/2}(^{100}\text{Mo})/(10^{26}$~yr) = 0.011 (NEMO-3 \cite{PhysRevD.92.072011}, current), 5 (AMoRE-II \cite{Alenkov:2019jis}), 10 (CUPID \cite{Pavan:2020ipz}); (iii) $T_{1/2}(^{136}\text{Xe})/(10^{26}$~yr) = 1.1 (KamLAND-Zen-400 \cite{PhysRevLett.117.082503}, current), 5 (KamLAND-Zen-800 \cite{Gando:2020cxo}), 9.2 (nEXO \cite{Pocar:2020zqz}). As before, we assume only one short-range contribution to be present at $m_W$ and we neglect any contribution from light neutrino exchange. As can be seen, the strong limits on $c_1^{XX}$ and $c_1^{XY}$ probe operator scales up to 18~TeV. The weakest limits, applying to $c_{2,3,4}$, still probe scales of order 4-6~TeV.

\subsection{Interference between Light Neutrino Exchange and Short-Range Mechanisms}

So far we have only considered one mechanism (operator) to be present at a given time, either the light neutrino exchange or one of the short-range operators. We now discuss the effect of two or more mechanisms operating at the same time. A large number of combinations are of course possible but at least the standard light neutrino contribution is expected to be present at some level in any case. This is because any New Physics scenario that  generates a $\Delta L = 2$ short-range operator is also expected to generate Majorana neutrino masses at a level to explain neutrino oscillations. Therefore, it is reasonable to look into the interference of one of the non-standard short-range mechanisms with the standard light neutrino exchange. We here discuss a few illustrative scenarios.

We first consider the interference with the operator associated with $\epsilon_3^{LLL}$. As we have seen in Sec.~\ref{sec:light-heavy-neutrinos}, it is triggered by heavy sterile neutrinos. Under the presence of $\epsilon_\nu$ and $\epsilon_3^{LLL}$, Eq.~\eqref{eq:totalhalflife} simplifies to
\begin{align}
\label{eq:mnu-3LLL}
	T_{1/2}^{-1} = 
	G_{11+}^{(0)}
	\left|\frac{m_{\beta\beta}}{m_e}\nme_\nu + \epsilon_3^{LLL}\nme_3^{LL}\right|^2.
\end{align}
Because light neutrino exchange and the operator associated with $\epsilon_3^{LLL}$ have the same leptonic structure, the two contributions add coherently. The same behaviour occurs for all operators of type $\epsilon_{1,2,3}$ with a left-handed leptonic current. Depending on the complex phases of the NMEs and the particle physics parameters $m_{\beta\beta}$, $\epsilon_3^{LLL}$, the interference can be constructive or destructive. The NMEs are conventionally defined to be real with values given in Sec.~\ref{sec:nme}. In the given scenario, both NMEs are negative. We can choose $m_{\beta\beta}$ to be real and positive and the interference is described by the relative phase of $\epsilon_3^{LLL}$. The largest effect then arises when $\epsilon_3^{LLL}$ is real and positive (constructive) or negative (destructive). Specifically, if $\epsilon_3^{LLL} = - |m_{\beta\beta}|/m_e (\mathcal{M}_\nu / \mathcal{M}_3^{LL})$, both contributions cancel each other.

\begin{figure}[t!]
	\centering
	\includegraphics[width=0.482\textwidth]{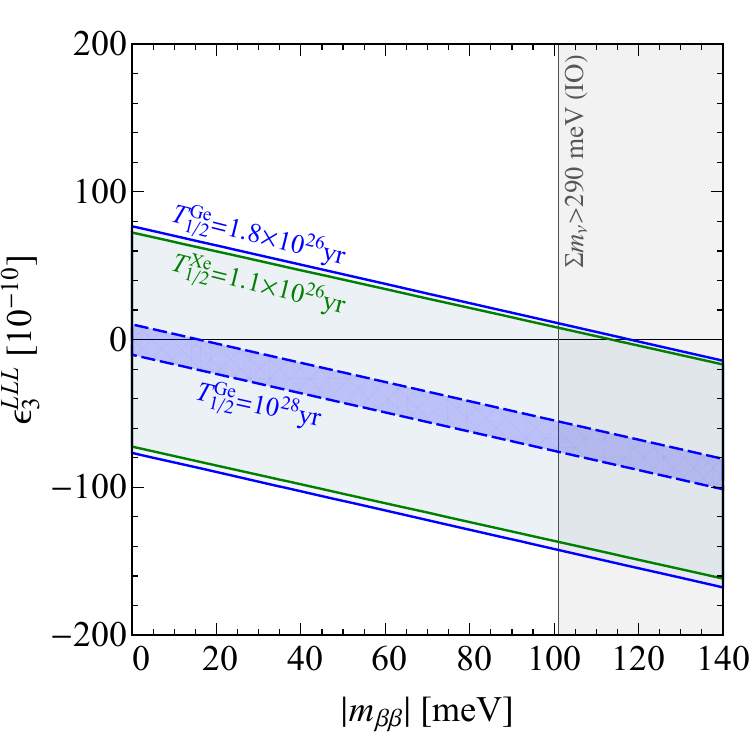}
	\includegraphics[width=0.47\textwidth]{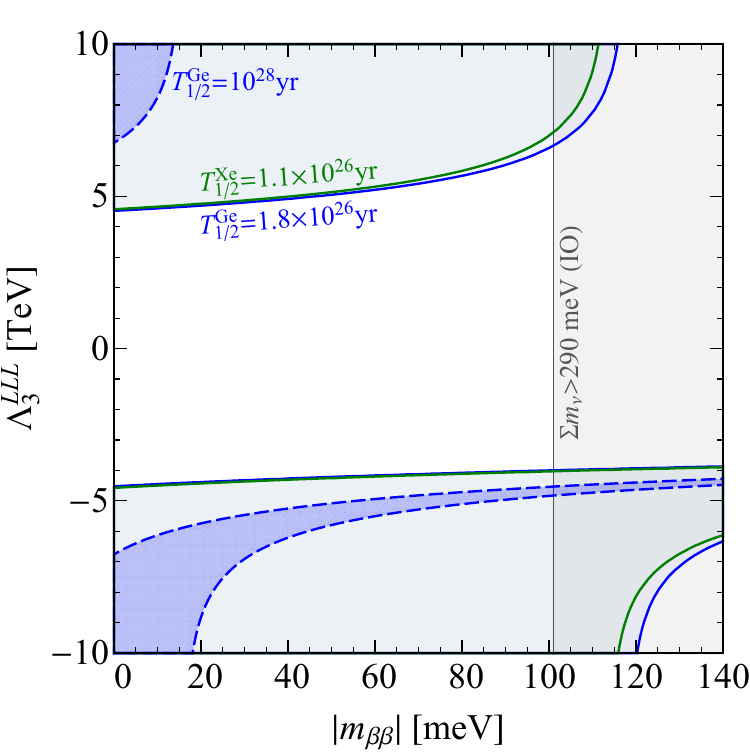}
	\caption{Constraints on the effective $0\nu\beta\beta$ decay mass $|m_{\beta\beta}|$ and the short-range operator coupling $\epsilon_3^{LLL}$ (left) as well as the associated operator scale $\Lambda_3^{LLL}$ (right). All other effective couplings are set to zero. The highlighted regions denote the allowed parameter space from the current limits $T_{1/2}(^{76}\text{Ge}) > 1.8\times 10^{26}$~yr (light blue) and $T_{1/2}(^{136}\text{Xe}) > 1.1\times 10^{26}$~yr (light green), as well as the future sensitivity $T_{1/2}(^{76}\text{Ge}) = 10^{28}$~yr (dashed blue). The grey shaded area on the right is excluded assuming the bound on the sum of neutrino masses of $\Sigma m_\nu > 290$~meV from cosmological observations for an inverse neutrino mass ordering.}
	\label{fig:mnu-eps3LLL}
\end{figure}
The general constraints on the $(|m_{\beta\beta}|, \epsilon_3^{LLL})$ parameter space are depicted in Fig.~\ref{fig:mnu-eps3LLL}~(left). As discussed, we take both $m_{\beta\beta}$ and $\epsilon_3^{LLL}$ to be relatively real with $m_{\beta\beta} > 0$ by convention. The light shaded areas are allowed given the current limits from $0\nu\beta\beta$ decays searches in $^{76}$Ge and $^{136}$Xe, whereas the dark shaded area denotes the sensitivity from future searches at $T_{1/2}(^{76}\text{Ge}) = 10^{28}$~yr. The combination of contributions in Eq.~\eqref{eq:mnu-3LLL} leads to the linear relation between the variables and no independent limits can be set on them. From cosmological observations we can infer the upper limit $|m_{\beta\beta}| < 101$~meV, see Fig.~\ref{fig:massplot}, and neither $|m_{\beta\beta}|$ nor $|\epsilon_3^{LLL}|$ can be arbitrarily large given this additional constraint. Thus allowing a contribution $0 \leq |m_{\beta\beta}| < 101$~meV from light neutrino exchange, $\epsilon_3^{LLL}$ is currently constrained to the interval $-137\times 10^{-10} < \epsilon_3^{LLL} < 72.5\times 10^{-10}$, compared to $|\epsilon_3^{LLL}| < 72.5\times 10^{-10}$ in the case it is the only contribution. In Fig.~\ref{fig:mnu-eps3LLL}~(right), we show the equivalent plot in the $(|m_{\beta\beta}|, \Lambda_3^{LLL})$ parameter plane, where the effective operator scale is defined through $1/(\Lambda_3^{LLL})^5 = G_F^2 \cos^2\theta_C \epsilon_3^{LLL}/(2m_p)$. The current experimental constraints give $|\Lambda_3^{LLL}| \gtrsim 4.5$~TeV. If $m_{\beta\beta}$ is not restricted further independently, e.g. by inference from an improved measurement of $\Sigma m_\nu$, the future constraint $T_{1/2}(^{76}\text{Ge}) = 10^{28}$~yr will still allow $\Lambda_3^{LLL} \approx - 4.8$~TeV due to destructive interference.

\begin{figure}[t!]
	\centering
	\includegraphics[width=0.482\textwidth]{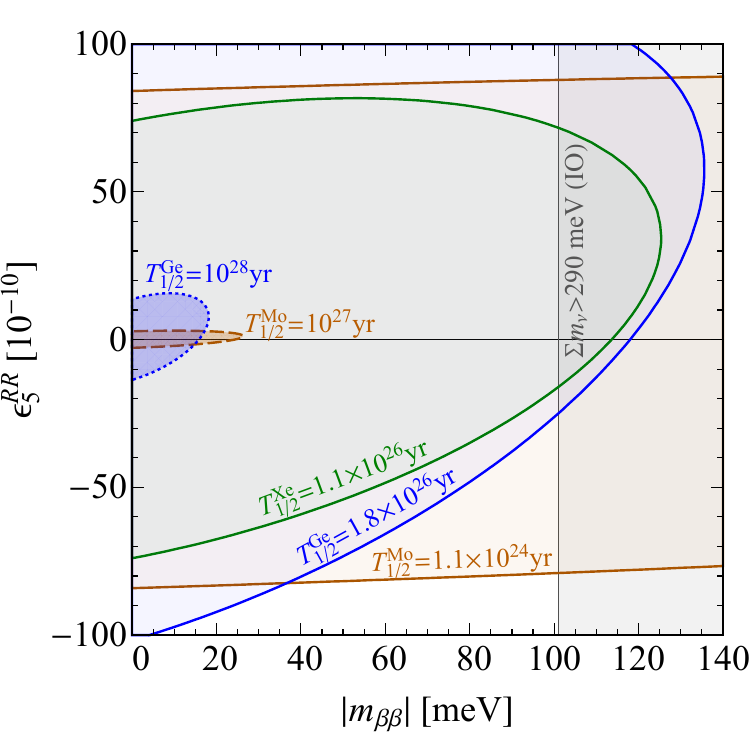}
	\includegraphics[width=0.47\textwidth]{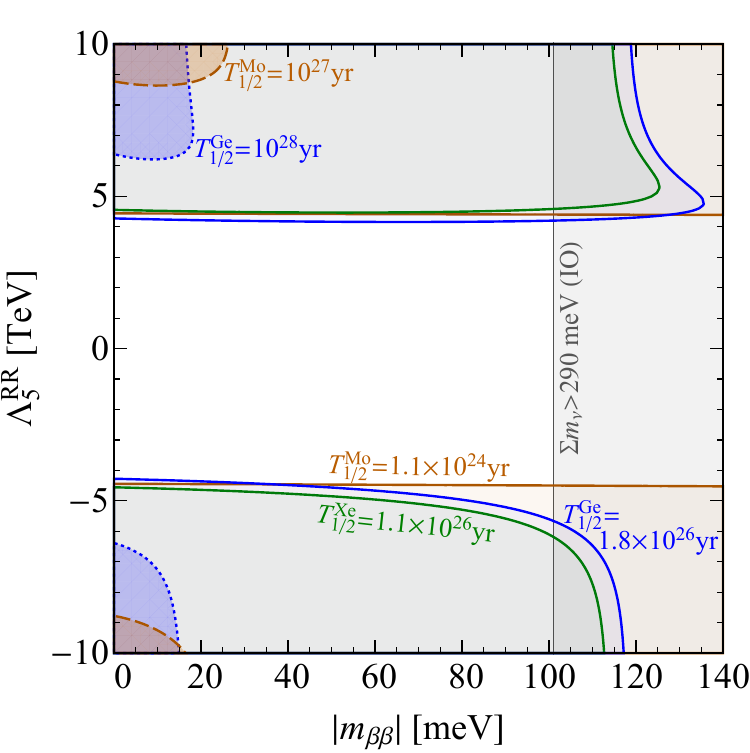}
	\caption{As Fig.~\ref{fig:mnu-eps3LLL}, but for the short-range operator coupling $\epsilon_5^{RR}$ and associated scale $\Lambda_5^{RR}$, also showing constraints from the current limit and future sensitivity in $^{100}$Mo.}
	\label{fig:mnu-eps5RR}
\end{figure}
In the case of the interference between the standard light neutrino contribution with one operator of the type $\epsilon_{1,2,3}$ with a right-handed lepton current or of the type $\epsilon_{4,5}$, the overlap is suppressed by the interference between the different lepton currents involved. We here discuss the example $\epsilon_5^{RR}$ in which case Eq.~\eqref{eq:totalhalflife} simplifies to
\begin{align}
\label{eq:nu5RR}
	T_{1/2}^{-1} &= 
	  G_{11+}^{(0)}|\nme_\nu|^2 \frac{|m_{\beta\beta}|^2}{m_e^2}
	+ G_{66}^{(0)}|\mathcal{M}_5^{RR}|^2 |\epsilon_5^{RR}|^2
	+ 2 G_{16}^{(0)}(\nme_\nu\nme_5^{RR})
	\text{Re}\left[\frac{m_{\beta\beta}}{m_e}\epsilon_5^{RR*}\right] \nonumber\\
	&= A |m_{\beta\beta}|^2 + B |\epsilon_5^{RR}|^2 
	 - 2 C |m_{\beta\beta}||\epsilon_5^{RR}|\cos(\alpha-\beta).
\end{align}
Here, $A = G_{11+}^{(0)}|\nme_\nu|^2/m_e^2$, $B = G_{66}^{(0)}|\mathcal{M}_5^{RR}|^2$, $C = G_{16}^{(0)}|\nme_\nu||\nme_5^{RR}|/m_e$ are positive coefficients (we consider the NMEs to be real with $\nme_\nu$, $\nme_5^{RR}$ having opposite signs, cf. Tab.~\ref{tab:nme-compound}), and $\alpha$, $\beta$ are the complex phases of $m_{\beta\beta}$, $\epsilon_5^{RR}$, respectively. We again consider that the relative phase between $m_{\beta\beta}$ and $\epsilon_5^{RR}$ is $\alpha - \beta = 0, \pi$ in which case Eq.~\eqref{eq:nu5RR} is a quadratic function in $|m_{\beta\beta}|$ and $\epsilon_5^{RR}$ and for a given value of $T_{1/2}^{-1}$ represent an ellipse. This is shown in Fig.~\ref{fig:mnu-eps5RR}~(left) where the tilting is determined by the size of the PSF $G_{16}^{(0)}$ relative to $G_{11+}^{(0)}$ and $G_{66}^{(0)}$. The currently most stringent constraint is set in $^{136}$Xe but the limit on $\epsilon_5^{RR}$ from $^{100}$Mo is competitive despite the much weaker half life limit. This is a consequence of enhanced NME $\nme_5^{RR}$ in $^{100}$Mo, see Table~\ref{tab:nme-compound}. Fig.~\ref{fig:mnu-eps5RR}~(right) shows the equivalent plot for the effective operator scale $\Lambda_5^{RR}$. As can be seen in Table~\ref{tab:psfs}, the PSFs $G_{16}^{(0)}$ applicable to all contributions of type $\epsilon_{4,5}$ are generally quite sizeable resulting in a comparatively strong interference. On the other hand, the PSF $G_{11-}^{(0)}$ regulates the interference with operators of type $\epsilon_{1,2,3}^R$ with right-handed lepton currents, see Eq.~\eqref{eq:totalhalflife}, which is suppressed by the electron mass compared to the beta decay $Q_{\beta\beta}$ value.

\subsection{Constraints on New Physics Scenarios}

The above constraints on the effective neutrino mass and short-range operator couplings can be interpreted in terms of the New Physics scenarios introduced in Sec.~\ref{sec:newphysics}.

\subsubsection{Light and Heavy Sterile Neutrinos}

In the sterile neutrino case discussed in Sec.~\ref{sec:light-heavy-neutrinos}, we consider the simplified scenario where a single sterile neutrino of mass $m_N$ with mixing $V_{eN}$ to the electron neutrino contributes to $0\nu\beta\beta$ decay. The limiting cases where the sterile neutrino is much lighter and heavier than 100~MeV were discussed in Sec.~\ref{sec:light-heavy-neutrinos}. Currently, the most stringent limit in Table~\ref{tab:bounds} on $0\nu\beta\beta$ decay contributions of heavy sterile neutrinos is set in $^{136}$Xe,
\begin{align}
	\epsilon_3^{LLL} < 72.5\times 10^{-10} \quad \Rightarrow \quad 
	\left(\sum_{i=1}^{n_N} \frac{V_{eN_i}^2}{m_{N_i}}\right)^{-1} > 1.3\times 10^8~\text{GeV},
\end{align}
assuming that the contributions from the light SM neutrinos are negligible.

\begin{figure}[t!]
	\centering
	\includegraphics[width=0.90\textwidth]{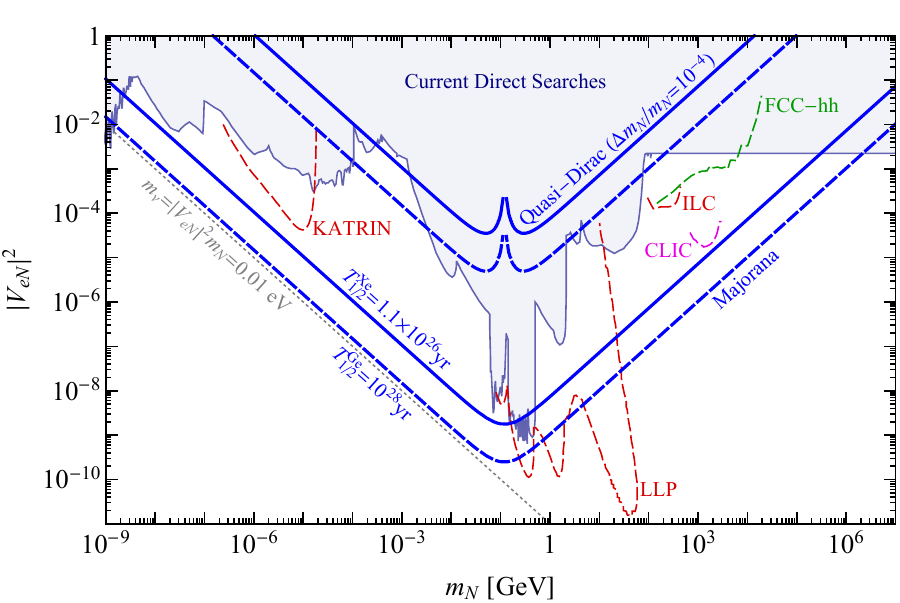}
	\caption{Upper limit on the active-sterile neutrino mixing strength $|V_{eN}|^2$ as a function of the sterile neutrino mass $m_N$ from current $0\nu\beta\beta$ decay searches (solid curves) and at future sensitivities with $T_{1/2} = 10^{28}$~yr (dashed curves). The sterile neutrino is assumed to be of Majorana or Quasi-Dirac nature as indicated and contributions from light neutrinos are neglected. The blue shaded area is excluded by current data from neutrino oscillations, beta decays, meson decays, colliders and electroweak precision measurements. The dashed contours indicate the estimated future sensitivity in Tritium decays (KATRIN), long-lived particle searches (LLP) and at colliders (FCC-hh, ILC, CLIC). The diagonal line gives the seesaw relation of light neutrino mass generation, $m_\nu = |V_{eN}|^2 m_N = 0.01$~eV.}
	\label{fig:sterileneutrino}
\end{figure}
To approximately incorporate the intermediate range $m_N \approx 100$~MeV as well, we use the interpolation \cite{Faessler:2014kka, Barea:2015zfa}
\begin{align}
\label{eq:sterile-100mev}
	T_{1/2}^{-1} = G^{(0)}_{11+} 
	|\mathcal{M}_3^{LL}|^2
	\left(\frac{m_p m_N}{\langle {\mathbf q}^2 \rangle + m_N^2}\right)^2 |V_{eN}|^4,
\end{align}
with the average momentum transfer $\langle {\mathbf q}^2 \rangle = m_p m_e |\mathcal{M}_3^{LL} / \mathcal{M}_\nu|$. In Fig.~\ref{fig:sterileneutrino}, we show the current limit and future sensitivity in the $(m_N, |V_{eN}|^2)$ parameter space. The region above the $0\nu\beta\beta$ bottom-most contours indicated are ruled out by the corresponding observation, assuming that the sterile neutrino is of a Majorana nature. We compare the $0\nu\beta\beta$ decay constraints with other searches for sterile neutrinos which are being pursued in neutrino oscillations, single beta decays, meson decays, at colliders and in electroweak precision measurements. The most recent searches are generally summarized in Ref.~\cite{Bolton:2019pcu} and collider signatures are reviewed in Refs.~\cite{Cai:2017mow, Das:2018hph}. The shaded area is excluded by current data and the dashed lines give examples of sensitivities in future searches. This includes the Tritium decay experiment KATRIN \cite{Mertens:2018vuu}, searches for long-lived particles (LLP, the shape is mainly determined by the planned DUNE \cite{Krasnov:2019kdc}, SHiP \cite{SHiP:2018xqw} and FCC-ee collider \cite{Blondel:2014bra}) and high energy colliders FCC-hh \cite{Pascoli:2018heg}, ILC \cite{Banerjee:2015gca} and CLIC \cite{Chakraborty:2018khw, Das:2018usr}. As can be seen, future $0\nu\beta\beta$ decay searches at a level of $T_{1/2} \approx 10^{28}$~yr will be able to probe mixing strengths expected for light neutrino neutrino mass generation via the Seesaw mechanism, $m_\nu = |V_{eN}|^2 m_N ~ 0.01$~eV for $m_N \lesssim 100$~MeV. Likewise, $0\nu\beta\beta$ decay searches probe very heavy Majorana neutrinos with masses up to $m_N \approx 10^6$~GeV where electroweak precision measurements can otherwise set comparatively weak limits of order $|V_{eN}|^2 \lesssim 10^{-3}$.

We stress that this strong sensitivity of $0\nu\beta\beta$ decay searches applies to purely Majorana neutrinos, which are difficult to reconcile with the lightness of active neutrinos for $m_N \gtrsim 1$~GeV. For sterile neutrinos with such masses it is more natural that they form Quasi-Dirac states where LNV is suppressed by a small mass splitting. In Fig.~\ref{fig:sterileneutrino}, we also show the sensitivity towards such a scenario where two Majorana neutrinos with a relative mass splitting of $\Delta m_N / m_N = 10^{-4}$ form a Quasi-Dirac pair, partially cancelling their contributions to $0\nu\beta\beta$ decay. While the sensitivity is strongly reduced, $0\nu\beta\beta$ decay searches are still competitive at this level for $m_N \approx 1$~MeV and $m_N \approx 100$~GeV.

\subsubsection{Left-Right Symmetry}
\begin{figure}[t!]
	\centering
	\includegraphics[width=0.60\textwidth]{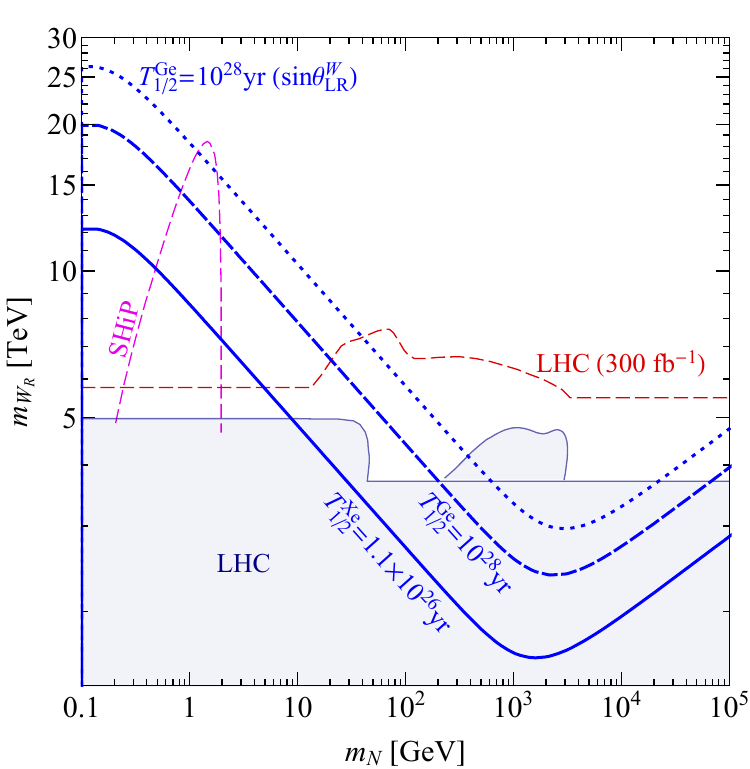}
	\caption{Lower limit on the right-handed $W_R$ boson mass $m_{W_R}$ as a function of the right-handed neutrino mass $m_N$ from current $0\nu\beta\beta$ decay searches (solid curve) and the corresponding future sensitivities with $T_{1/2}(^{76}\text{Ge}) = 10^{28}$~yr (dashed curve) in the LRSM with negligible $W$ boson mixing. The dotted curve indicates the future sensitivity on the scenario where the $W$ boson mixing is $\sin\theta_{LR}^W = m_W^2 / m_{W_R}^2$. The blue shaded area is excluded by current data from the LHC and the dashed contours indicate the estimated future sensitivity at the LHC with 300~fb$^{-1}$ and at SHiP.}
	\label{fig:lr-rpv}
\end{figure}
In Fig.~\ref{fig:lr-rpv}, we show the limits from $0\nu\beta\beta$ decay searches on the right-handed $W_R$ boson mass $m_{W_R}$ and the heavy neutrino mass $m_N$ in the LRSM introduced in Sec.~\ref{sec:lrsm}. Here, we consider a simplified scenario with one lepton generation, i.e. a single heavy neutrino $N$ and $U^R_{e1} = 1$. We also choose the so-called manifest left-right symmetric case with $g_R = g$, $\cos\theta_C^R = \cos\theta_C$ and take $m_{\Delta_R^{--}} = m_{W_R}$ for the mass of the doubly charged triplet Higgs. The solid and dashed $0\nu\beta\beta$ curves give the lower limit on $m_{W_R}$ where we additionally neglect the $W$ boson mixing, $\sin\theta_\text{LR}^W = 0$, thus $\epsilon_3^{RRR}$ is the only contribution. The rise of the $0\nu\beta\beta$ curves to the right of $m_N \approx 10^3$~GeV in Fig.~\ref{fig:lr-rpv} results from the doubly charged Higgs contribution in Eq.~\eqref{eq:doubly-charged-higgs} increasing linearly with $m_N$. Note, though, that too large values of $m_N$, compared to $m_{W_R}$, are not natural as they would require non-perturbative Yukawa couplings with the triplet Higgs.

The $0\nu\beta\beta$ decay limits are compared with the direct limits from the LHC and the future SHiP experiment. The current LHC limits arise from dijet, $e + E_\text{miss}$ \cite{Nemevsek:2018bbt} and $eejj$ signatures \cite{Aaboud:2018spl}. The future LHC limits are estimated for 300~fb$^{-1}$ of luminosity and are taken from \cite{Nemevsek:2018bbt}. The dijet and $e + E_\text{miss}$ signatures are largely independent of the heavy neutrino mass in the applicable kinematic regimes and are sensitive to $m_{W_R} \approx 4 - 7$~TeV. On the other hand, the SHiP experiment would probe heavy neutrinos produced mainly in $D$ meson decays and the strong sensitivity around $m_N \approx 1$~GeV shown is taken from Ref.~\cite{Mandal:2017tab}. As can be seen, $0\nu\beta\beta$ decay searches are especially sensitive for $m_N \lesssim 20$~GeV. Note that we only consider heavy neutrino masses as light as $m_N = 100$~MeV where the short-range contribution assumption is reasonable. For $m_N \approx 100$~MeV we incorporate the approximation in Eq.~\eqref{eq:sterile-100mev}; masses around and below this scale can be incorporated using an analysis of the relevant dim-7 operators \cite{Ali:2006iu, Ali:2007ec, Deppisch:2012nb, Cirigliano:2017djv} and by including the mass dependence of the neutrino potential~\cite{Faessler:2014kka, Barea:2015zfa, Dekens:2020ttz}.

Both the LHC and SHiP limits were derived assuming negligible $W$ boson mixing; those based on the lifetime of the heavy neutrino will be affected and need to be re-assessed. We nevertheless also include the sensitivity of future $0\nu\beta\beta$ decay searches for $\sin\theta_\text{LR}^W = m_W^2 / m_{W_R}^2$, i.e. the generically maximal value expected, where all three operators $\epsilon_3^{RRR}$, $\epsilon_3^{LLR}$, $\epsilon_3^{LRR}$ contribute. Future searches are then expected to be sensitive up to $m_{W_R} \approx 26$~TeV.

\subsubsection{$R$-Parity Violating Supersymmetry}

Assuming gluino dominance, $R$-parity violating supersymmetry will induce the contributions in Eq.~\eqref{eq:rpv-epsilons}. Neglecting any other contributions, including those from light neutrinos, Eq.~\eqref{eq:totalhalflife} simplifies to
\begin{align}
	T_{1/2}^{-1} = 
	G_{11+}^{(0)}
	\left(1.95\nme_1^{RR} - 2.88\nme_2^{RR}\right)^2
	\left(\frac{8\pi\alpha_s\lambda'^{2}_{111}}{9\cos^2\theta_C}\frac{G_F^{-2}}{m^4_{\tilde q}}\frac{m_p}{m_{\tilde g}}\right)^2,
\end{align}
where the numerical factors in front of the NMEs take into account the effect of QCD running, i.e. we here interpret the coupling strength $\lambda'_{111}$ at $m_W$. Using the current KamLAND-Zen bound $T_{1/2}(^{136}\text{Xe}) > 1.1\times 10^{26}$~yr, this can be translated into an upper limit on $\lambda'_{111}$,
\begin{align}
\label{eq:rpv-lambdalimit}
	\lambda'_{111} < 7.0\times 10^{-3} 
	                 \left(\frac{m_{\tilde q}}{1\,\text{TeV}}\right)^2
	                 \left(\frac{m_{\tilde g}}{1\,\text{TeV}}\right)^{1/2}.
\end{align}
This compares to the limit $\lambda'_{111} < 7.2\times 10^{-3}$ in \cite{Arbelaez:2016uto}\footnote{Reference \cite{Arbelaez:2016uto} contains updated experimental constraints and includes the effects of QCD running to the scale 1~TeV compared to \cite{Hirsch:1995ek}.} for the same squark and gluino masses and the above KamLAND-Zen bound. Somewhat surprisingly, the limits are thus of a very similar size; whereas in our case, the strong sensitivity is predominantly due to the enhanced value of the NME $\nme_1^{RR}$ resulting from the large pseudoscalar form factor in Eq.~\eqref{eq:fpprime}, in Ref.~\cite{Arbelaez:2016uto} it is an effect of the QCD running and operator mixing.

If $0\nu\beta\beta$ decay is not observed in future experiments with a sensitivity approaching $T_{1/2}(^{100}\text{Mo}) = 10^{27}$~yr, the limit will improve to 
\begin{align}
	\lambda'_{111} < 2.0\times 10^{-3} 
	\left(\frac{m_{\tilde q}}{1\,\text{TeV}}\right)^2
	\left(\frac{m_{\tilde g}}{1\,\text{TeV}}\right)^{1/2}.
\end{align}
This is mainly a result of the strong sensitivity to $\epsilon_1^{RR}$ especially in $^{100}$Mo, see Sec.~\ref{sec:bounds-eff-couplings}. As mentioned, the derived limit is based on the assumption of gluino dominance. It will be important to re-evaluate the impact of $0\nu\beta\beta$ decay searches on the $R$-parity violating supersymmetry in light of the new results and the current constraints from direct searches for supersymmetric particles.

\section{Summary and Conclusion}
\label{sec:summary}

Signatures of total lepton number violation are crucial if we want to understand the origin of neutrino masses, which constitute a key open issue in particle physics. Neutrinoless double beta decay has so far been the only practical means to probe light Majorana neutrino masses at scales indicated by neutrino oscillations. In addition it is sensitive to New Physics contributions from exotic particles and interactions coupling to first-generation quarks and electrons. Within an Effective Field Theory (EFT) framework, $0\nu\beta\beta$ decay searches strongly constrain contributions of that form. In this work we have concentrated on short-range contributions which result from integrating out exotic particles much heavier than the energy scale $m_F \approx 100$~MeV of double beta decay, leading to effective dimension-9 operators of the form $\Lambda^{-5}\bar u\bar u d d\bar e\bar e$. In addition, we update calculations for the standard light neutrino exchange mechanism to analyze the interplay with short-range contributions.

We have presented a first complete numerical evaluation of the NMEs needed for the description of short-range non-standard mechanisms of $0\nu\beta\beta$ decay. The calculation is performed within the framework of IBM-2 with restoration of the isospin properties of the Fermi transition operator. We also use updated single particle energies extracted from experimental data on nuclei with one nucleon removed or added from shell closure. We include additional NMEs that become important when the latest values of the nucleon form factors are taken into account. However, the main difference to previous calculations is in the sign of the tensor NMEs; the present derivation gives a sign of the tensor term $\mathcal{M}_T$ which is opposite to that in e.g. \cite{Barea:2015kwa}. This change has little effect on the standard mechanism, for which $\mathcal{M}_T$ is small $\approx 1$\%, but it is sizeable for short-range mechanisms.

As noted, we have performed our calculation in the phenomenological framework of the interacting boson model, using nucleon currents in the impulse approximation including higher-order terms in the nucleon momentum transfer determined in \cite{Graf:2018ozy}. We model pion-mediated modes via enhanced pseudo-scalar nucleon form factors informed by PCAC and lattice QCD calculations. In our numerical results we consider a possible quenching of the axial-vector coupling by choosing $g_A = 1.0$ compared to the unquenched value $g_A = 1.27$. We follow this classical approach in contrast to \emph{ab initio} methods based on chiral EFT interactions \cite{Weinberg:1991um}. Such formulations promise the determination of NMEs with controllable errors, e.g. may address part of the quenching problem \cite{Gysbers:2019uyb}. Calculations of the standard light neutrino exchange NME $\mathcal{M}_\nu$ following this approach have become possible for the lightest double beta decay isotopes $^{48}$Ca \cite{Yao:2019rck, Belley:2020ejd, Novario:2020dmr}, $^{76}$Ge and $^{82}$Se \cite{Belley:2020ejd}, indicating noticeably smaller values than those from phenomenological models such as IBM-2, see \cite{Yao:2020azz} for a recent review. If confirmed, this will require an understanding for such a deviation as well further studies to apply ab initio methods to heavier nuclei. NMEs should ideally be verified experimentally by employing single and double charge exchange reactions \cite{PhysRevC.86.044603, Cappuzzello2018}. Chiral EFT techniques have been used to reveal a potentially sizeable short-range contribution in standard light neutrino exchange \cite{Cirigliano:2018hja} and to calculate exotic contributions \cite{Cirigliano:2017djv,Cirigliano:2018yza}.

In addition to the NMEs calculated in our approach we also present the full set of leptonic PSFs for all relevant isotopes, determined numerically including effects from the finite nuclear size and electron cloud screening corrections. This allows us to set updated limits on the effective couplings of all possible short-range operators contributing to $0\nu\beta\beta$ decay. Considering one operator at a time, the current limits correspond to operator scales ranging between 3 TeV to 10 TeV, where the strongest sensitivity is achieved for operators enhanced by pion-mediated corrections, in agreement with previous analyses \cite{faessler:1997ph, Prezeau:2003xn, Peng:2015haa, Cirigliano:2018yza}, in our case arising from enhanced pseudo-scalar form factors. We further illustrate the interplay between different contributions by considering the interference between the standard light neutrino exchange with one short-range contribution $\epsilon_I$ thus setting constraints on the combined parameter space $(m_{\beta\beta}, \epsilon_I)$. Finally, we apply the effective operator framework to three example New Physics scenarios, namely the SM with sterile neutrinos, Left-Right symmetry and $R$-parity violating supersymmetry. Here, we set updated constraints on simplified parameter spaces and compare them with limits coming from other searches.

Searches for lepton number violating signatures, with $0\nu\beta\beta$ decay as the most prominent example, are crucial for our understanding of neutrinos and physics beyond the SM in general. Given that no clear sign of New Physics has been seen so far, short-range operators as those considered in this work provide a model-agnostic means to probe the presence of lepton number violating physics. Due to the strong suppression --- the $0\nu\beta\beta$ decay rate scales as $\propto \Lambda^{-10}$ --- future experimental advances increasing the sensitivity by up to two orders of magnitude to half lives $T_{1/2}^{0\nu\beta\beta} \approx 10^{27-28}$~yr will only result in modest improvements in constraining $\Lambda$, see Fig.~\ref{fig:operatorscales}. Detailed analyses such as our work and \cite{Cirigliano:2018yza} are still important as these operator scales $\Lambda \approx 4 - 18$~TeV are in a regime relevant for the LHC and potential future colliders. If an exotic short-range contribution were to be observed, it would indicate that light neutrino masses have their origin around the TeV scale. It would also have profound consequences on possible explanations of the matter-antimatter asymmetry of the universe, with the observation of non-standard $\ovbb$ decay contributions disfavouring baryogenesis mechanisms operating above the electroweak scale \cite{Deppisch:2015yqa, Deppisch:2017ecm}.

\section*{Acknowledgements}
The authors would like to thank Jose Barea for providing the code to calculate standard mechanisms of double beta decay in IBM-2 and Patrick Bolton for sharing the direct sterile neutrino search limits and sensitivities. The authors would also like to thank Martin Hirsch for a careful reading of the manuscript and useful discussions. This work was supported in part by the U.S. Department of Energy (Grant No. DE-FG-02-91ER-40608) and the UK Royal Society International Exchange program. The work of J. K. was supported by the  Academy of Finland (Grant Nos. 314733, 320062). L. G. and F. F. D. acknowledge support from the UK Science and Technology Facilities Council (STFC) via a Consolidated Grant (Reference ST/P00072X/1).

\section*{Appendix A: Parameters of the IBM-2 Hamiltonian}
A detailed description of the IBM-2 Hamiltonian is given in \cite{iac87} and \cite{otsukacode}. For most nuclei, the Hamiltonian parameters are taken from the literature \cite{Duval83, Kaup83, Kaup79, Shlomo92, Isacker80, Kim96, Giannatiempo91, Sambataro82, Iachello96, Puddu80, ScholtenPhD, Bijker80, Barfield83, kot12b}. The new calculations are done using the program NPBOS \cite{otsukacode}. They include energies, B(E2) values, quadrupole moments, B(M1) values, magnetic moments, etc.. For the semi-magic nuclei $^{124}$Sn and $^{136}$Xe, we have obtained the parameters by a fit to the energy of the low lying states using the same procedure as in Ref.~\cite{Iachello96} for $^{116}$Sn. A compilation of the used parameters is given in Table~\ref{tab:ibm2parameters}.
\begin{table*}[htb!]
\begin{centering}
\scalebox{0.68}{
\begin{tabular}{cccccccccccccccccccc}
\toprule
Nucleus  & $\epsilon_{d_{\nu}}$  & $\epsilon_{d_{\pi}}$  & $\kappa$  & $\chi_{\nu}$  & $\chi_{\pi}$  & $\xi_{1}$  & $\xi_{2}$  & $\xi_{3}$  & $c_{\nu}^{(0)}$  & $c_{\nu}^{(2)}$  & $c_{\nu}^{(4)}$  & $c_{\pi}^{(0)}$  & $c_{\pi}^{(2)}$  & $c_{\pi}^{(4)}$  & $\omega_{\nu\nu}$  & $\omega_{\pi\pi}$  & $\omega_{\nu\pi}$  & $w_{\nu}$  & $y_{\nu}$\tabularnewline
\colrule
$^{76}\mbox{Ge}$ \cite{Duval83}  & 1.20  & 1.20  & -0.21  & 1.00  & -1.20  & -0.05  & 0.10  & -0.05  &  &  &  &  &  &  &  &  &  &  & \tabularnewline
$^{76}\mbox{Se}$ \cite{Kaup83}  & 0.96  & 0.96  & -0.16  & 0.50  & -0.90  &  &  & -0.10  &  &  &  &  &  &  &  &  &  &  & \tabularnewline
$^{82}\mbox{Se}$ \cite{Kaup83}  & 1.00  & 1.00  & -0.28  & 1.14  & -0.90  &  &  & -0.10  &  &  &  &  &  &  &  &  &  &  & \tabularnewline
$^{82}\mbox{Kr}$ \cite{Kaup79}  & 1.15  & 1.15  & -0.19  & 0.93  & -1.13  & -0.10  &  & -0.10  &  &  &  &  &  &  &  &  &  &  & \tabularnewline
$^{96}\mbox{Zr}$\footnotemark[1]  & 1.00  & 1.00  & -0.20  & -2.20  & 0.65  &  &  &  &  &  &  &  &  &  & 0.17  & 0.17  & 0.33  &  & \tabularnewline
$^{96}\mbox{Mo}$ \cite{Shlomo92}  & 0.73  & 1.10  & -0.09  & -1.20  & 0.40  & -0.10  & 0.10  & -0.10  & -0.50  & 0.10  &  &  &  &  &  &  &  &  & \tabularnewline
$^{100}\mbox{Mo}$ \cite{Shlomo92}  & 0.55  & 1.00  & -0.06  & -1.20  & 0.40  & -0.10  & 0.10  & -0.10  & -0.60  & 0.20  & 0.10  &  &  &  &  &  &  &  & \tabularnewline
$^{100}\mbox{Ru}$ \cite{Isacker80}  & 0.89  & 0.89  & -0.18  & -1.00  & 0.40  &  &  &  & 0.60  & 0.09  & -0.13  &  &  &  &  &  &  &  & \tabularnewline
$^{110}\mbox{Pd}$ \cite{Kim96}  & 0.78  & 0.60  & -0.13  & 0.00  & -0.30  & 0.20  & 0.04  & 0.00  & -0.26  & -0.29  & -0.30  & -0.26  & -0.29  & -0.03  &  &  &  &  & \tabularnewline
$^{110}\mbox{Cd}$ \cite{Giannatiempo91}  & 0.92  & 0.92  & -0.15  & -1.10  & -0.80  & 1.10  & 0.109  & 1.10  & 0.07  & -0.17  & 0.16  &  &  &  &  &  &  &  & \tabularnewline
$^{116}\mbox{Cd}$ \cite{Sambataro82}  & 0.85  & 0.85  & -0.27  & -0.58  & 0.00  & -0.18  & 0.24  & -0.18  & -0.15  & -0.06  &  &  &  &  &  &  &  &  & \tabularnewline
$^{116}\mbox{Sn}$ \cite{Iachello96}  & 1.32  &  &  &  &  &  &  &  & -0.50  & -0.22  & -0.07  &  &  &  &  &  &  & -0.06  & 0.04\tabularnewline
$^{124}\mbox{Sn}$\footnotemark[2]  & 1.10  &  &  &  &  &  &  &  & -0.30  & -0.16  & -0.20  &  &  &  &  &  &  & 0.30  & 0.02\tabularnewline
$^{124}\mbox{Te}$ \cite{Sambataro82}  & 0.82  & 0.82  & -0.15  & 0.00  & -1.20  & -0.18  & 0.24  & -0.18  & 0.10  &  &  &  &  &  &  &  &  &  & \tabularnewline
$^{128}\mbox{Te}$ \cite{Sambataro82}  & 0.93  & 0.93  & -0.17  & 0.50  & -1.20  & -0.18  & 0.24  & -0.18  & 0.30  & 0.22  &  &  &  &  &  &  &  &  & \tabularnewline
$^{128}\mbox{Xe}$ \cite{Puddu80}  & 0.70  & 0.70  & -0.17  & 0.33  & -0.80  & -0.18  & 0.24  & -0.18  & 0.30  &  &  &  &  &  &  &  &  &  & \tabularnewline
$^{130}\mbox{Te}$ \cite{Sambataro82}  & 1.05  & 1.05  & -0.20  & 0.90  & -1.20  & -0.18  & 0.24  & -0.18  & 0.30  & 0.22  &  &  &  &  &  &  &  &  & \tabularnewline
$^{130}\mbox{Xe}$ \cite{Puddu80}  & 0.76  & 0.76  & -0.19  & 0.50  & -0.80  & -0.18  & 0.24  & -0.18  & 0.30  & 0.22  &  &  &  &  &  &  &  &  & \tabularnewline
$^{136}\mbox{Xe}$\footnotemark[2]  &  & 1.31  &  &  &  &  &  &  &  &  &  & -0.04  & 0.01  & -0.02  &  &  &  &  & \tabularnewline
$^{136}\mbox{Ba}$ \cite{Puddu80}  & 1.03  & 1.03  & -0.23  & 1.00  & -0.90  & -0.18  & 0.24  & -0.18  & 0.30  & 0.10  &  &  &  &  &  &  &  &  & \tabularnewline
$^{148}\mbox{Nd}$ \cite{ScholtenPhD}  & 0.70  & 0.70  & -0.10  & -0.80  & -1.20  & -0.12  & 0.24  & 0.90  &  &  &  & 0.40  & 0.20  &  &  &  &  &  & \tabularnewline
$^{148}\mbox{Sm}$ \cite{ScholtenPhD}  & 0.95  & 0.95  & -0.12  & 0.00  & -1.30  & -0.12  & 0.24  & 0.90  &  &  &  &  & 0.05  &  &  &  &  &  & \tabularnewline
$^{150}\mbox{Nd}$ \cite{ScholtenPhD}  & 0.47  & 0.47  & -0.07  & -1.00  & -1.20  & -0.12  & 0.24  & 0.90  &  &  &  & 0.40  & 0.20  &  &  &  &  &  & \tabularnewline
$^{150}\mbox{Sm}$ \cite{ScholtenPhD}  & 0.70  & 0.70  & -0.08  & -0.80  & -1.30  & -0.12  & 0.24  & 0.90  &  &  &  &  & 0.05  &  &  &  &  &  & \tabularnewline
$^{154}\mbox{Sm}$ \cite{ScholtenPhD}  & 0.43  & 0.43  & -0.08  & -1.10  & -1.30  & -0.12  & 0.24  & 0.90  &  &  &  &  & 0.05  &  &  &  &  &  & \tabularnewline
$^{154}\mbox{Gd}$ \cite{ScholtenPhD}  & 0.55  & 0.55  & -0.08  & -1.00  & -1.00  & -0.12  & 0.24  & 0.90  &  &  &  & -0.20  & -0.10  &  &  &  &  &  & \tabularnewline
$^{160}\mbox{Gd}$ \cite{kot12b}  & 0.42  & 0.42  & -0.05  & -0.80  & -1.00  & 0.08  & 0.08  & 0.08  &  &  &  & -0.20  & -0.10  &  &  &  &  &  & \tabularnewline
$^{160}\mbox{Dy}$ \cite{kot12b}  & 0.44  & 0.44  & -0.06  & -0.80  & -0.90  & 0.08  & 0.08  & 0.08  &  &  &  & -0.05  & -0.15  &  &  &  &  &  & \tabularnewline
$^{198}\mbox{Pt}$ \cite{Bijker80}  & 0.58  & 0.58  & -0.18  & 1.05  & -0.80  & -0.10  & 0.08  & -0.10  & 0.00  & 0.02  & 0.00  &  &  &  &  &  &  &  & \tabularnewline
$^{198}\mbox{Hg}$ \cite{Barfield83}  & 0.55  & 0.55  & -0.21  & 1.00  & -0.40  &  & 0.08  &  & 0.37  & 0.25  & 0.16  &  &  &  &  &  &  &  & \tabularnewline
$^{232}\mbox{Th}$\footnotemark[1]  & 0.26  & 0.26  & -0.05  & -0.80  & -1.45  &0.20  & 0.20  &0.20  &   &   &   &  &  &  &  &  &  &  & \tabularnewline
$^{232}\mbox{U}$\footnotemark[1]  & 0.28  & 0.28  & -0.05  & -1.00  & -1.30  &0.12  & 0.12  &0.12  &   &   &   &0.20  &0.10  &  &  &  &  &  & \tabularnewline
$^{238}\mbox{U}$\footnotemark[1]  & 0.22  & 0.22  & -0.05  & -0.40  & -1.30  &0.12  & 0.12  &0.12  &   &   &   &0.20  &0.10  &  &  &  &  &  & \tabularnewline
$^{238}\mbox{Pu}$\footnotemark[1]  & 0.24  & 0.24  & -0.05  & -0.60  & -0.05  &0.12  & 0.12  &0.12  &   &0.02   &   &  &0.05  &-0.09  &  &  &  &  & \tabularnewline

\botrule
\end{tabular}}
\footnotetext[1]{Parameters fitted to reproduce the spectroscopic
data of the low lying energy states.}
\footnotetext[2]{GS parameters fitted to reproduce the spectroscopic data of the low lying energy states.}
\par\end{centering}
\caption{Hamiltonian parameters employed in the IBM-2 calculation of the wave functions along with their references.}
\label{tab:ibm2parameters}
\end{table*}

\section*{Appendix B: Surface Delta Interaction Strength Values A1 and Single-Particle and Hole Energies}
The reliability of single-particle and -hole energies as well as the interaction strengths in connection with IBM-2 wave functions was studied in \cite{PhysRevC.94.034320} by comparing recently measured occupation probabilities of initial and final states of interest in double beta decay. The pair structure constants were generated as usual by diagonalizing the surface delta interaction (SDI) in the two identical particle states, pp, nn, where the strength of the (isovector) interaction, $A_{1}$,  is obtained by fitting the 2$^{+}$-0$^{+}$ energy difference in nuclei with either two protons (proton holes) or two neutrons (neutron holes). The used single particle energies and $A_1$ values are given in Tables~\ref{tab:shell28-50} - \ref{tab:spl82-126}.
\begin{table}[h]
\begin{tabular}{cccc}
\toprule
Orbital & %
\begin{tabular}{c}
Protons\tabularnewline
(particles)\tabularnewline
$A\sim76$\tabularnewline
$A_{1}=0.299$\tabularnewline
\end{tabular} & %
\begin{tabular}{c}
Protons\tabularnewline
(holes)\tabularnewline
$A\sim100$\tabularnewline
$A_{1}=0.239$\tabularnewline
\end{tabular} & %
\begin{tabular}{c}
Neutrons\tabularnewline
(holes)\tabularnewline
$A\sim76$\tabularnewline
$A_{1}=0.237$\tabularnewline
\end{tabular}\tabularnewline
\colrule
$2p_{1/2}$ & 1.179 & 0.678 & 0.588\tabularnewline
$2p_{3/2}$ & 0.000 & 1.107 & 1.095\tabularnewline
$1f_{5/2}$ & 0.340 & 1.518 & 1.451\tabularnewline
$1g_{9/2}$ & 2.640 & 0.000 & 0.000\tabularnewline
\botrule
\end{tabular}
\caption{Single particle energies and isovector SDI strength parameters $A_{1}$ in MeV used for the 28-50 shell \cite{PhysRevC.94.034320}.}
\label{tab:shell28-50}
\end{table}
\begin{table}[h]
\begin{tabular}{cccccc}
\toprule
Orbital & %
\begin{tabular}{c}
Protons \tabularnewline
(particles)\tabularnewline
$A\sim130$\tabularnewline
$A_{1}=0.222$\tabularnewline
\end{tabular} & %
\begin{tabular}{c}
Protons \tabularnewline
(particles)\tabularnewline
$A\sim150$\tabularnewline
$A_{1}=0.223$\tabularnewline
\end{tabular} & %
\begin{tabular}{c}
Protons\tabularnewline
(holes)\tabularnewline
$A\sim198$\tabularnewline
$A_{1}=0.200$\tabularnewline
\end{tabular} & %
\begin{tabular}{c}
Neutrons \tabularnewline
(particles)\tabularnewline
$A\sim100$\tabularnewline
$A_{1}=0.242$\tabularnewline
\end{tabular} & %
\begin{tabular}{c}
Neutrons \tabularnewline
(holes)\tabularnewline
$A\sim130$\tabularnewline
$A_{1}=0.163$\tabularnewline
\end{tabular}\tabularnewline
\colrule
$3s_{1/2}$ & 2.990 & 0.719 &0.000	& 0.775 & 0.332\tabularnewline
$2d_{3/2}$ & 2.440 & 0.466 &0.350	& 1.142 & 0.000\tabularnewline
$2d_{5/2}$ & 0.962 & 0.365 &1.670	& 0.000 & 1.654\tabularnewline
$1g_{7/2}$ & 0.000 & 0.000 &2.700	& 0.172 & 2.434\tabularnewline
$1h_{11/2}$ & 2.792 & 0.668 &1.340	& 2.868 & 0.069\tabularnewline
\botrule
\end{tabular}
\caption{Single particle energies and isovector SDI strength parameters $A_{1}$ in MeV used for the 50-82 shell~\cite{PhysRevC.94.034320}.}
\label{tab:shell50-82}
\end{table}
\begin{table}[h]
\begin{tabular}{cccc}
\toprule
Orbital & %
\begin{tabular}{c}
Protons \tabularnewline
(particles)\tabularnewline
$A\sim232$\tabularnewline
$A_{1}=0.147$\tabularnewline
\end{tabular} & %
\begin{tabular}{c}
Neutrons \tabularnewline
(particles)\tabularnewline
$A\sim150$\tabularnewline
$A_{1}=0.133$\tabularnewline
\end{tabular} & %
\begin{tabular}{c}
Neutrons\tabularnewline
(holes)\tabularnewline
$A\sim198$\tabularnewline
$A_{1}=0.150$\tabularnewline
\end{tabular}\tabularnewline
\colrule
$3p_{1/2}$ &3.633	& 1.363  &0.000	\tabularnewline
$3p_{3/2}$ &3.119	& 0.854  &0.900	\tabularnewline
$2f_{5/2}$ &2.826	& 2.005  &0.570	\tabularnewline
$2f_{7/2}$ &0.896	& 0.000  &2.340	\tabularnewline
$1h_{9/2}$ &0.000	& 1.561  &3.410	\tabularnewline
$1i_{13/2}$ &1.608	& 3.700  &1.630	\tabularnewline
\botrule
\end{tabular}
\caption{Single particle energies and isovector SDI strength parameters $A_{1}$ in MeV used for the 82-126 shell~\cite{PhysRevC.94.034320}.}
\label{tab:shell82-126}
\end{table}
\begin{table}[h]
\begin{tabular}{cc}
\toprule
Orbital  & %
\begin{tabular}{c}
Neutrons\tabularnewline
(particles)\tabularnewline
$A\sim232$\tabularnewline
$A_{1}=$0.089\tabularnewline
\end{tabular} \\ %
\colrule
$4s_{1/2}$  & 2.032  \tabularnewline
$3d_{3/2}$  & 2.538  \tabularnewline
$3d_{5/2}$  & 1.567  \tabularnewline
$2g_{7/2}$  & 2.491  \tabularnewline
$2g_{9/2}$  & 0.000  \tabularnewline
$1i_{11/2}$  & 0.779  \tabularnewline
$1j_{15/2}$  & 1.423  \tabularnewline
\botrule
\end{tabular}
\caption{SDI strength values $A_{1}$ and single particle energies (in MeV) in the $N=126-184$ shell.}
\label{tab:spl82-126}
\end{table}

\clearpage

\bibliography{literature}

\begin{thebibliography}{100}

\bibitem{Otten:2008zz}
E.~W. Otten and C.~Weinheimer,
\newblock Rept. Prog. Phys. {\bf 71}, 086201 (2008), [0909.2104].

\bibitem{Aker:2019uuj}
KATRIN, M.~Aker {\em et~al.},
\newblock Phys. Rev. Lett. {\bf 123}, 221802 (2019), [1909.06048].

\bibitem{Ade:2015xua}
Planck Collaboration, P.~A.~R. Ade {\em et~al.},
\newblock Astron. Astrophys. {\bf 594}, A13 (2016), [1502.01589].

\bibitem{Minkowski:1977sc}
P.~Minkowski,
\newblock Phys.Lett. {\bf B67}, 421 (1977).

\bibitem{mohapatra:1979ia}
R.~N. Mohapatra and G.~Senjanovic,
\newblock Phys. Rev. Lett. {\bf 44}, 912 (1980).

\bibitem{Yanagida:1979as}
T.~Yanagida,
\newblock Conf.Proc. {\bf C7902131}, 95 (1979).

\bibitem{seesaw:1979}
M.~Gell-Mann, P.~Ramond and R.~Slansky,
\newblock Conf.Proc. {\bf C790927}, 315 (1979), [1306.4669].

\bibitem{Schechter:1980gr}
J.~Schechter and J.~W.~F. Valle,
\newblock Phys. Rev. {\bf D22}, 2227 (1980).

\bibitem{Agostini:2020xta}
GERDA, M.~Agostini {\em et~al.},
\newblock 2009.06079.

\bibitem{Pas:1999fc}
H.~P{\"a}s, M.~Hirsch, H.~Klapdor-Kleingrothaus and S.~Kovalenko,
\newblock Phys.Lett. {\bf B453}, 194 (1999).

\bibitem{Pas:2000vn}
H.~Pas, M.~Hirsch, H.~V. Klapdor-Kleingrothaus and S.~G. Kovalenko,
\newblock Phys. Lett. {\bf B498}, 35 (2001), [hep-ph/0008182].

\bibitem{delAguila:2011gr}
F.~del Aguila, A.~Aparici, S.~Bhattacharya, A.~Santamaria and J.~Wudka,
\newblock JHEP {\bf 05}, 133 (2012), [1111.6960].

\bibitem{delAguila:2012nu}
F.~del Aguila, A.~Aparici, S.~Bhattacharya, A.~Santamaria and J.~Wudka,
\newblock JHEP {\bf 06}, 146 (2012), [1204.5986].

\bibitem{Deppisch:2017ecm}
F.~F. Deppisch, L.~Graf, J.~Harz and W.-C. Huang,
\newblock Phys. Rev. {\bf D98}, 055029 (2018), [1711.10432].

\bibitem{Kim:2020vjv}
Y.-H. Kim,
\newblock 2004.02510.

\bibitem{Doi:1981}
M.~Doi {\em et~al.},
\newblock Phys. Theor. Phys. {\bf 66}, 1739 (1983).

\bibitem{Doi:1983}
M.~Doi {\em et~al.},
\newblock Phys. Theor. Phys. {\bf 69}, 602 (1983).

\bibitem{Tomoda:1990rs}
T.~Tomoda,
\newblock Rept. Prog. Phys. {\bf 54}, 53 (1991).

\bibitem{Ali:2006iu}
A.~Ali, A.~Borisov and D.~Zhuridov,
\newblock hep-ph/0606072.

\bibitem{Ali:2007ec}
A.~Ali, A.~V. Borisov and D.~V. Zhuridov,
\newblock Phys. Rev. {\bf D76}, 093009 (2007), [0706.4165].

\bibitem{Cirigliano:2017djv}
V.~Cirigliano, W.~Dekens, J.~de~Vries, M.~Graesser and E.~Mereghetti,
\newblock JHEP {\bf 12}, 082 (2017), [1708.09390].

\bibitem{Helo:2016vsi}
J.~C. Helo, M.~Hirsch and T.~Ota,
\newblock JHEP {\bf 06}, 006 (2016), [1602.03362].

\bibitem{Deppisch:2014zta}
F.~F. Deppisch, T.~E. Gonzalo, S.~Patra, N.~Sahu and U.~Sarkar,
\newblock Phys.Rev. {\bf D91}, 015018 (2015), [1410.6427].

\bibitem{Deppisch:2017vne}
F.~F. Deppisch, C.~Hati, S.~Patra, P.~Pritimita and U.~Sarkar,
\newblock Phys. Rev. D {\bf 97}, 035005 (2018), [1701.02107].

\bibitem{Cirigliano:2018yza}
V.~Cirigliano, W.~Dekens, J.~de~Vries, M.~Graesser and E.~Mereghetti,
\newblock JHEP {\bf 12}, 097 (2018), [1806.02780].

\bibitem{Li:2020flq}
G.~Li, M.~Ramsey-Musolf and J.~C. Vasquez,
\newblock 2009.01257.

\bibitem{Graf:2018ozy}
L.~Graf, F.~F. Deppisch, F.~Iachello and J.~Kotila,
\newblock Phys. Rev. D {\bf 98}, 095023 (2018).

\bibitem{Barea:2009zza}
J.~Barea and F.~Iachello,
\newblock Phys. Rev. {\bf C79}, 044301 (2009).

\bibitem{Barea:2013bz}
J.~Barea, J.~Kotila and F.~Iachello,
\newblock Phys. Rev. {\bf C87}, 014315 (2013), [1301.4203].

\bibitem{Barea:2015kwa}
J.~Barea, J.~Kotila and F.~Iachello,
\newblock Phys. Rev. {\bf C91}, 034304 (2015), [1506.08530].

\bibitem{Simkovic:2007vu}
F.~Simkovic, A.~Faessler, V.~Rodin, P.~Vogel and J.~Engel,
\newblock Phys. Rev. {\bf C77}, 045503 (2008), [0710.2055].

\bibitem{Simkovic:2013qiy}
F.~Simkovic, V.~Rodin, A.~Faessler and P.~Vogel,
\newblock Phys. Rev. {\bf C87}, 045501 (2013), [1302.1509].

\bibitem{Suhonen:1991sk}
J.~Suhonen,
\newblock J. Phys. {\bf G19}, 139 (1993).

\bibitem{Suhonen:2012wd}
J.~Suhonen,
\newblock AIP Conf. Proc. {\bf 1488}, 326 (2012).

\bibitem{Caurier:2007xz}
E.~Caurier, F.~Nowacki and A.~Poves,
\newblock Int. J. Mod. Phys. {\bf E16}, 552 (2007).

\bibitem{Menendez:2008jp}
J.~Menendez, A.~Poves, E.~Caurier and F.~Nowacki,
\newblock Nucl. Phys. {\bf A818}, 139 (2009), [0801.3760].

\bibitem{Rodriguez:2010mn}
T.~R. Rodriguez and G.~Martinez-Pinedo,
\newblock Phys. Rev. Lett. {\bf 105}, 252503 (2010), [1008.5260].

\bibitem{Babu:2001ex}
K.~S. Babu and C.~N. Leung,
\newblock Nucl. Phys. {\bf B619}, 667 (2001), [hep-ph/0106054].

\bibitem{deGouvea:2007xp}
A.~de~Gouvea and J.~Jenkins,
\newblock Phys.Rev. {\bf D77}, 013008 (2008), [0708.1344].

\bibitem{Pati:1974yy}
J.~C. Pati and A.~Salam,
\newblock Phys. Rev. {\bf D10}, 275 (1974).

\bibitem{Mohapatra:1974gc}
R.~N. Mohapatra and J.~C. Pati,
\newblock Phys. Rev. {\bf D11}, 2558 (1975).

\bibitem{Senjanovic:1975rk}
G.~Senjanovic and R.~N. Mohapatra,
\newblock Phys.Rev. {\bf D12}, 1502 (1975).

\bibitem{Dimopoulos:1988jw}
S.~Dimopoulos and L.~J. Hall,
\newblock Phys. Lett. {\bf B207}, 210 (1988).

\bibitem{Hall:1983id}
L.~J. Hall and M.~Suzuki,
\newblock Nucl.Phys. {\bf B231}, 419 (1984).

\bibitem{Mohapatra:1986su}
R.~N. Mohapatra,
\newblock Phys. Rev. {\bf D34}, 3457 (1986).

\bibitem{Hirsch:1995ek}
M.~Hirsch, H.~Klapdor-Kleingrothaus and S.~Kovalenko,
\newblock Phys.Rev. {\bf D53}, 1329 (1996), [hep-ph/9502385].

\bibitem{Adler:1975he}
S.~L. Adler {\em et~al.},
\newblock Phys. Rev. {\bf D11}, 3309 (1975),
\newblock [,507(1975)].

\bibitem{PhysRev.112.1375}
S.~Weinberg,
\newblock Phys. Rev. {\bf 112}, 1375 (1958).

\bibitem{Gonzalez-Alonso:2018omy}
M.~Gonz\'alez-Alonso, O.~Naviliat-Cuncic and N.~Severijns,
\newblock Prog. Part. Nucl. Phys. {\bf 104}, 165 (2019), [1803.08732].

\bibitem{Simkovic:1999re}
F.~Simkovic, G.~Pantis, J.~D. Vergados and A.~Faessler,
\newblock Phys. Rev. {\bf C60}, 055502 (1999), [hep-ph/9905509].

\bibitem{Schindler:2006jq}
M.~R. Schindler and S.~Scherer,
\newblock Eur. Phys. J. {\bf A32}, 429 (2007), [hep-ph/0608325],
\newblock [,59(2006)].

\bibitem{Bernard:2001rs}
V.~Bernard, L.~Elouadrhiri and U.-G. Meissner,
\newblock J. Phys. {\bf G28}, R1 (2002), [hep-ph/0107088].

\bibitem{Andreev:2012fj}
MuCap, V.~A. Andreev {\em et~al.},
\newblock Phys. Rev. Lett. {\bf 110}, 012504 (2013), [1210.6545].

\bibitem{barea12}
J.~Barea, J.~Kotila and F.~Iachello,
\newblock Phys. Rev. Lett. {\bf 109}, 042501 (2012).

\bibitem{Haxton:1985am}
W.~Haxton and G.~Stephenson,
\newblock Prog.Part.Nucl.Phys. {\bf 12}, 409 (1984).

\bibitem{Weinberg:1991um}
S.~Weinberg,
\newblock Nucl. Phys. B {\bf 363}, 3 (1991).

\bibitem{Cirigliano:2018hja}
V.~Cirigliano {\em et~al.},
\newblock Phys. Rev. Lett. {\bf 120}, 202001 (2018), [1802.10097].

\bibitem{Cirigliano:2019vdj}
V.~Cirigliano {\em et~al.},
\newblock Phys. Rev. C {\bf 100}, 055504 (2019), [1907.11254].

\bibitem{ARIMA1977205}
A.~Arima, T.~Ohtsuka, F.~Iachello and I.~Talmi,
\newblock Phys. Lett. B {\bf 66}, 205  (1977).

\bibitem{iac87}
F.~Iachello and A.~Arima,
\newblock {\em The Interacting Boson Model} (Cambridge University Press, 1987).

\bibitem{barea13b}
J.~Barea, J.~Kotila and F.~Iachello,
\newblock Phys. Rev. C {\bf 87}, 057301 (2013).

\bibitem{kotila14}
J.~Kotila, J.~Barea and F.~Iachello,
\newblock Phys. Rev. C {\bf 89}, 064319 (2014).

\bibitem{Barea:2015zfa}
J.~Barea, J.~Kotila and F.~Iachello,
\newblock Phys. Rev. {\bf D92}, 093001 (2015), [1509.01925].

\bibitem{Fang:2015zha}
D.-L. Fang, A.~Faessler and F.~Simkovic,
\newblock Phys. Rev. C {\bf 92}, 044301 (2015), [1508.02097].

\bibitem{OTSUKA19781}
T.~Otsuka, A.~Arima and F.~Iachello,
\newblock Nucl. Phys. A {\bf 309}, 1  (1978).

\bibitem{PhysRevC.94.034320}
J.~Kotila and J.~Barea,
\newblock Phys. Rev. C {\bf 94}, 034320 (2016).

\bibitem{eng15}
J.~Engel,
\newblock Journal of Physics G: Nuclear and Particle Physics {\bf 42}, 034017
  (2015).

\bibitem{MILLER1976562}
G.~A. Miller and J.~E. Spencer,
\newblock Ann. Phys. (NY) {\bf 100}, 562  (1976).

\bibitem{Simkovic:2009pp}
F.~Simkovic, A.~Faessler, H.~Muther, V.~Rodin and M.~Stauf,
\newblock Phys. Rev. {\bf C79}, 055501 (2009), [0902.0331].

\bibitem{10.3389/fphy.2017.00055}
J.~T. Suhonen,
\newblock Frontiers in Physics {\bf 5}, 55 (2017).

\bibitem{PhysRevC.86.044603}
P.~Puppe {\em et~al.},
\newblock Phys. Rev. C {\bf 86}, 044603 (2012).

\bibitem{Cappuzzello2018}
F.~Cappuzzello {\em et~al.},
\newblock The European Physical Journal A {\bf 54}, 72 (2018).

\bibitem{PhysRevLett.107.062501}
J.~Men\'endez, D.~Gazit and A.~Schwenk,
\newblock Phys. Rev. Lett. {\bf 107}, 062501 (2011).

\bibitem{Menendez:2011zza}
J.~Menendez, A.~Poves, E.~Caurier and F.~Nowacki,
\newblock J.Phys.Conf.Ser. {\bf 312}, 072005 (2011).

\bibitem{Faessler:2011rv}
A.~Faessler, G.~Fogli, E.~Lisi, A.~Rotunno and F.~Simkovic,
\newblock Phys.Rev. {\bf D83}, 113015 (2011), [1103.2504].

\bibitem{Kotila:2012zza}
J.~Kotila and F.~Iachello,
\newblock Phys. Rev. {\bf C85}, 034316 (2012), [1209.5722].

\bibitem{PhysRevLett.120.232502}
O.~Azzolini {\em et~al.},
\newblock Phys. Rev. Lett. {\bf 120}, 232502 (2018).

\bibitem{ARGYRIADES2010168}
J.~Argyriades {\em et~al.},
\newblock Nucl. Phys. A {\bf 847}, 168  (2010).

\bibitem{PhysRevD.92.072011}
NEMO-3 Collaboration, R.~Arnold {\em et~al.},
\newblock Phys. Rev. D {\bf 92}, 072011 (2015).

\bibitem{PhysRevD.98.092007}
A.~S. Barabash {\em et~al.},
\newblock Phys. Rev. D {\bf 98}, 092007 (2018).

\bibitem{Arnaboldi:2002te}
C.~Arnaboldi {\em et~al.},
\newblock Phys. Lett. B {\bf 557}, 167 (2003), [hep-ex/0211071].

\bibitem{collaboration2019improved}
CUORE, D.~Adams {\em et~al.},
\newblock Phys. Rev. Lett. {\bf 124}, 122501 (2020), [1912.10966].

\bibitem{PhysRevLett.117.082503}
KamLAND-Zen Collaboration, A.~Gando {\em et~al.},
\newblock Phys. Rev. Lett. {\bf 117}, 082503 (2016).

\bibitem{PhysRevD.94.072003}
NEMO-3 Collaboration, R.~Arnold {\em et~al.},
\newblock Phys. Rev. D {\bf 94}, 072003 (2016).

\bibitem{deSalas:2020pgw}
P.~de~Salas {\em et~al.},
\newblock 2006.11237.

\bibitem{Alenkov:2019jis}
V.~Alenkov {\em et~al.},
\newblock Eur. Phys. J. C {\bf 79}, 791 (2019), [1903.09483].

\bibitem{Zsigmond:2020bfx}
LEGEND, A.~J. Zsigmond,
\newblock J. Phys. Conf. Ser. {\bf 1468}, 012111 (2020).

\bibitem{RoyChoudhury:2019hls}
S.~Roy~Choudhury and S.~Hannestad,
\newblock JCAP {\bf 07}, 037 (2020), [1907.12598].

\bibitem{Mahajan:2013ixa}
N.~Mahajan,
\newblock Phys. Rev. Lett. {\bf 112}, 031804 (2014), [1310.1064].

\bibitem{Gonzalez:2015ady}
M.~Gonz\'{a}lez, M.~Hirsch and S.~G. Kovalenko,
\newblock Phys. Rev. {\bf D93}, 013017 (2016), [1511.03945],
\newblock [Erratum: Phys. Rev.D97,no.9,099907(2018)].

\bibitem{Pavan:2020ipz}
CUPID, M.~Pavan,
\newblock J. Phys. Conf. Ser. {\bf 1468}, 012210 (2020).

\bibitem{Gando:2020cxo}
KamLAND-Zen, Y.~Gando,
\newblock J. Phys. Conf. Ser. {\bf 1468}, 012142 (2020).

\bibitem{Pocar:2020zqz}
nEXO, A.~Pocar,
\newblock J. Phys. Conf. Ser. {\bf 1468}, 012131 (2020).

\bibitem{Faessler:2014kka}
A.~Faessler, M.~Gonzalez, S.~Kovalenko and F.~Simkovic,
\newblock Phys. Rev. {\bf D90}, 096010 (2014), [1408.6077].

\bibitem{Bolton:2019pcu}
P.~D. Bolton, F.~F. Deppisch and P.~Bhupal~Dev,
\newblock JHEP {\bf 03}, 170 (2020), [1912.03058].

\bibitem{Cai:2017mow}
Y.~Cai, T.~Han, T.~Li and R.~Ruiz,
\newblock Front. in Phys. {\bf 6}, 40 (2018), [1711.02180].

\bibitem{Das:2018hph}
A.~Das,
\newblock Adv. High Energy Phys. {\bf 2018}, 9785318 (2018), [1803.10940].

\bibitem{Mertens:2018vuu}
KATRIN, S.~Mertens {\em et~al.},
\newblock J. Phys. G {\bf 46}, 065203 (2019), [1810.06711].

\bibitem{Krasnov:2019kdc}
I.~Krasnov,
\newblock Phys. Rev. D {\bf 100}, 075023 (2019), [1902.06099].

\bibitem{SHiP:2018xqw}
SHiP, C.~Ahdida {\em et~al.},
\newblock JHEP {\bf 04}, 077 (2019), [1811.00930].

\bibitem{Blondel:2014bra}
FCC-ee study Team, A.~Blondel, E.~Graverini, N.~Serra and M.~Shaposhnikov,
\newblock Nucl. Part. Phys. Proc. {\bf 273-275}, 1883 (2016), [1411.5230].

\bibitem{Pascoli:2018heg}
S.~Pascoli, R.~Ruiz and C.~Weiland,
\newblock JHEP {\bf 06}, 049 (2019), [1812.08750].

\bibitem{Banerjee:2015gca}
S.~Banerjee, P.~S.~B. Dev, A.~Ibarra, T.~Mandal and M.~Mitra,
\newblock Phys. Rev. D {\bf 92}, 075002 (2015), [1503.05491].

\bibitem{Chakraborty:2018khw}
S.~Chakraborty, M.~Mitra and S.~Shil,
\newblock Phys. Rev. D {\bf 100}, 015012 (2019), [1810.08970].

\bibitem{Das:2018usr}
A.~Das, S.~Jana, S.~Mandal and S.~Nandi,
\newblock Phys. Rev. D {\bf 99}, 055030 (2019), [1811.04291].

\bibitem{Nemevsek:2018bbt}
M.~Nemevsek, F.~Nesti and G.~Popara,
\newblock Phys. Rev. D {\bf 97}, 115018 (2018), [1801.05813].

\bibitem{Aaboud:2018spl}
ATLAS, M.~Aaboud {\em et~al.},
\newblock JHEP {\bf 01}, 016 (2019), [1809.11105].

\bibitem{Mandal:2017tab}
S.~Mandal, M.~Mitra and N.~Sinha,
\newblock Phys. Rev. D {\bf 96}, 035023 (2017), [1705.01932].

\bibitem{Deppisch:2012nb}
F.~F. Deppisch, M.~Hirsch and H.~P{\"a}s,
\newblock J.Phys. {\bf G39}, 124007 (2012), [1208.0727].

\bibitem{Dekens:2020ttz}
W.~Dekens, J.~de~Vries, K.~Fuyuto, E.~Mereghetti and G.~Zhou,
\newblock JHEP {\bf 06}, 097 (2020), [2002.07182].

\bibitem{Arbelaez:2016uto}
C.~Arbel\'aez, M.~Gonz\'alez, S.~Kovalenko and M.~Hirsch,
\newblock Phys. Rev. D {\bf 96}, 015010 (2017), [1611.06095].

\bibitem{Gysbers:2019uyb}
P.~Gysbers {\em et~al.},
\newblock Nature Phys. {\bf 15}, 428 (2019), [1903.00047].

\bibitem{Yao:2019rck}
J.~Yao {\em et~al.},
\newblock Phys. Rev. Lett. {\bf 124}, 232501 (2020), [1908.05424].

\bibitem{Belley:2020ejd}
A.~Belley, C.~Payne, S.~Stroberg, T.~Miyagi and J.~Holt,
\newblock 2008.06588.

\bibitem{Novario:2020dmr}
S.~Novario {\em et~al.},
\newblock 2008.09696.

\bibitem{Yao:2020azz}
J.~Yao,
\newblock 2008.13249.

\bibitem{faessler:1997ph}
A.~Faessler, S.~Kovalenko, F.~Simkovic and J.~Schwieger,
\newblock Phys. Rev. Lett. {\bf 78}, 183 (1997), [hep-ph/9612357].

\bibitem{Prezeau:2003xn}
G.~Prezeau, M.~Ramsey-Musolf and P.~Vogel,
\newblock Phys.Rev. {\bf D68}, 034016 (2003), [hep-ph/0303205].

\bibitem{Peng:2015haa}
T.~Peng, M.~J. Ramsey-Musolf and P.~Winslow,
\newblock Phys. Rev. {\bf D93}, 093002 (2016), [1508.04444].

\bibitem{Deppisch:2015yqa}
F.~F. Deppisch, J.~Harz, M.~Hirsch, W.-C. Huang and H.~P{\"a}s,
\newblock Phys. Rev. {\bf D92}, 036005 (2015), [1503.04825].

\bibitem{otsukacode}
T.~Otsuka and N.~Yoshida,
\newblock (1985).

\bibitem{Duval83}
P.~Duval, D.~Goutte and M.~Vergnes,
\newblock Phys. Lett. B {\bf 124}, 297 (1983).

\bibitem{Kaup83}
U.~Kaup, C.~M\"{o}nkemeyer and P.~V. Brentano,
\newblock Z. Physik A {\bf 310}, 129 (1983).

\bibitem{Kaup79}
U.~Kaup and A.~Gelberg,
\newblock Z. Physik A {\bf 293}, 311 (1979).

\bibitem{Shlomo92}
H.~Dejbakhsh, D.~Latypov, G.~Ajupova and S.~Shlomo,
\newblock Phys. Rev. C {\bf 46}, 2326 (1992).

\bibitem{Isacker80}
P.~V. Isacker and G.~Puddu,
\newblock Nucl. Phys. A {\bf 438}, 125 (1980).

\bibitem{Kim96}
K.-H. Kim, A.~Gelberg, T.~Mizusaki, T.~Otsuka and P.~von Brentano,
\newblock Nucl. Phys. A {\bf 604}, 163 (1996).

\bibitem{Giannatiempo91}
A.~Giannatiempo, A.~Nannini, A.~Perego, P.~Sona and G.~Maino,
\newblock Phys. Rev. C {\bf 44}, 1508 (1991).

\bibitem{Sambataro82}
M.~Sambataro,
\newblock Nucl. Phys. A {\bf 380}, 365 (1982).

\bibitem{Iachello96}
S.~Cacciamani, G.~Bonsignori, F.~Iachello and D.~Vretenar,
\newblock Phys. Rev. C {\bf 53}, 1618 (1996).

\bibitem{Puddu80}
O.~S. G.~Puddu and T.~Otsuka,
\newblock Nucl. Phys. A {\bf 348}, 109 (1980).

\bibitem{ScholtenPhD}
O.~Scholten,
\newblock Ph. D. thesis, University of Groningen, The Netherlands (1980).

\bibitem{Bijker80}
R.~Bijker, A.~E.~L. Dieperink and O.~Scholten,
\newblock Nucl. Phys. A {\bf 344}, 207 (1980).

\bibitem{Barfield83}
A.~F. Barfield, B.~R. Barrett, K.~A. Sage and P.~D. Duval,
\newblock Z. Physik A {\bf 311}, 205 (1983).

\bibitem{kot12b}
J.~Kotila, K.~Nomura, L.~Guo, N.~Shimizu and T.~Otsuka,
\newblock Phys. Rev. C {\bf 85}, 054309 (2012).

\end{thebibliography}
\bibliographystyle{h-physrev4}
\end{document}